\documentclass[%
reprint,
superscriptaddress,
preprintnumbers,
nofootinbib,
twocolumn,
amsmath,amssymb,
aps,
prd,
floats,
floatfix,
]{revtex4-2}

\usepackage{dcolumn}
\usepackage[mathlines]{lineno}

\usepackage[T1]{fontenc}
\setcitestyle{authoryear,round}
\usepackage{aecompl}

\usepackage{newtxtext,newtxmath}
\usepackage{mdframed,bm}
\usepackage{graphicx}	
\usepackage{amsmath}	
\usepackage{amssymb}	
\usepackage{color}
\usepackage{enumitem}
\usepackage{eqnarray}
\usepackage[british]{babel} 
\usepackage{placeins}
\usepackage{natbib}
\usepackage{appendix}
\usepackage{algorithm2e}
\usepackage{soul}

\usepackage{hyperref}
\usepackage{cleveref}
\usepackage{xcolor}
\usepackage{float}
\usepackage{physics}

\newcommand{\aap}{A\&A}
\newcommand{\apjs}{ApJS}
\newcommand{\aj}{A.J.}
\newcommand{\mnras}{MNRAS}

\newcommand{\jcap}{JCAP}
\newcommand{\apjl}{ApJ}
\newcommand{\pasj}{PASJ}
\newcommand{\pasp}{PASP}

\newcommand{\ssr}{Space Sci. Rev.}
\newcommand{\physrep}{Phys. Rep.}

\newcommand{\pdd}{\,$P(q_{2D})$}

\newcommand{\qdd}{\,$q_{2D}$}

\newcommand{\eq}[1]{\begin{equation}  #1 \end{equation}}
\newcommand{\eqa}[1]{\begin{eqnarray}   #1 \end{eqnarray}}

\newcommand{\sersic}{S\'{e}rsic}

\newcommand{\galapagos}{{\sc Galapagos}}

\newcommand{\blockfont}[1]{{\textsc{#1}}\xspace}
\newcommand{\metacal}{\blockfont{Metacalibration}}

\begin{document}

\preprint{DES-2022-0689}
\preprint{FERMILAB-PUB-22-474-PPD}

\title{Modeling Intrinsic Galaxy Alignment in the MICE Simulation}

\author{Kai Hoffmann}\email{kai.d.hoffmann@gmail.com}
\affiliation{Institute of Space Sciences (ICE, CSIC), Campus UAB, Carrer de Can Magrans, s/n, 08193 Barcelona, Spain}
\affiliation{Institute for Computational Science, University of Zurich, Winterthurerstr. 190, 8057 Zürich, Switzerland}
\author{Lucas ~F.~Secco}
\affiliation{Kavli Institute for Cosmological Physics, University of Chicago, Chicago, IL 60637, USA}
\author{Jonathan ~Blazek}
\affiliation{Department of Physics, Northeastern University, Boston, MA, 02115, USA}
\author{Martin ~Crocce}
\affiliation{Institute of Space Sciences (ICE, CSIC), Campus UAB, Carrer de Can Magrans, s/n, 08193 Barcelona, Spain}
\affiliation{Institut d'Estudis Espacials de Catalunya (IEEC), 08034 Barcelona, Spain}
\author{Pau Tallada-Cresp\'{i}}
\affiliation{Centro de Investigaciones Energ\'eticas, Medioambientales y Tecnol\'ogicas (CIEMAT), Avenida Complutense 40, 28040 Madrid, Spain}
\affiliation{Port d'Informaci\'{o} Cient\'{i}fica (PIC), Campus UAB, C. Albareda s/n, 08193 Barcelona (Barcelona), Spain}
\author{Simon Samuroff}
\affiliation{Department of Physics, Northeastern University, Boston, MA, 02115, USA}
\author{Judit Prat}
\affiliation{Kavli Institute for Cosmological Physics, University of Chicago, Chicago, IL 60637, USA}
\affiliation{Department of Astronomy and Astrophysics, University of Chicago, Chicago, IL 60637, USA}
\author{Jorge Carretero}
\affiliation{Institut de F\'isica d'Altes Energies (IFAE), The Barcelona Institute of Science and Technology, Campus UAB, 08193 Barcelona, Spain}
\affiliation{Port d'Informaci\'{o} Cient\'{i}fica (PIC), Campus UAB, C. Albareda s/n, 08193 Barcelona (Barcelona), Spain}
\author{Pablo Fosalba}
\affiliation{Institute of Space Sciences (ICE, CSIC), Campus UAB, Carrer de Can Magrans, s/n, 08193 Barcelona, Spain}
\affiliation{Institut d'Estudis Espacials de Catalunya (IEEC), 08034 Barcelona, Spain}
\author{Enrique Gazta\~naga}
\affiliation{Institute of Space Sciences (ICE, CSIC),  Campus UAB, Carrer de Can Magrans, s/n,  08193 Barcelona, Spain}
\affiliation{Institut d'Estudis Espacials de Catalunya (IEEC), 08034 Barcelona, Spain}
\author{Francisco J. Castander}
\affiliation{Institute of Space Sciences (ICE, CSIC),  Campus UAB, Carrer de Can Magrans, s/n,  08193 Barcelona, Spain}
\affiliation{Institut d'Estudis Espacials de Catalunya (IEEC), 08034 Barcelona, Spain}

\date{\today}

\begin{abstract}
The intrinsic alignment (IA) of galaxies is potentially a major limitation in deriving cosmological constraints from weak lensing surveys.
In order to investigate this effect we assign intrinsic shapes and orientations to galaxies in the light-cone output of the MICE simulation, spanning $\sim5000\,{\rm deg}^2$ and reaching redshift $z=1.4$. This assignment is based on a 'semi-analytic' IA model that uses photometric properties of galaxies as well as the spin and shape of their host halos. Advancing on previous work,
we include more realistic distributions of galaxy shapes and a luminosity dependent galaxy-halo alignment.
The IA model parameters are calibrated against COSMOS and BOSS LOWZ observations. The null detection of IA in observations of blue galaxies is accounted for by setting random orientations for these objects.
We compare the two-point alignment statistics measured in the simulation against predictions from the analytical IA models NLA and TATT over a wide range of scales, redshifts and luminosities for red and blue galaxies separately. We find that both models fit the measurements well at scales above $8\,h^{-1}{\rm Mpc}$, while TATT outperforms NLA at smaller scales. The IA parameters derived from our fits are in broad agreement with various observational constraints from red galaxies. Lastly, we build a realistic source sample, mimicking DES Year 3 observations and use it to predict the IA contamination to the observed shear statistics. We find this prediction to be within the measurement uncertainty,
which might be a consequence of the random alignment of blue galaxies in the simulation.
\end{abstract}

\maketitle

\section{Introduction}
\label{sec:introduction}

Weak gravitational lensing, able to directly probe dark-matter-dominated large-scale structures in the Universe,
has become a core cosmological probe \citep{Hikage19, Heymans21, DES21}.
In the coming years, next-generation experiments including Euclid, the Vera C. Rubin Observatory, and the
Nancy Grace Roman Space Telescope, will rely on weak
lensing to provide a substantial part of their overall constraining power. However, weak lensing analyses bring
several challenges, including both measurement methodology and understanding complex astrophysical effects.
One of the main astrophysical effects is the intrinsic alignment (hereafter also referred as IA) of source
galaxies \cite[e.g.][]{Joachimi15, Kiessling15, Kirk15, Troxel15} which contaminates the alignment signal induced
by gravitational lensing. Understanding how IA affects the observed weak lensing statistics is becoming increasingly
important as the statistical errors are decreasing strongly with the larger volumes probed by modern surveys.
It has been shown that ignoring IA can bias the constraints on cosmological parameters from these lensing surveys significantly
\citep{Krause16}. The IA contribution therefore needs to be included in the modeling of the observed data when
deriving cosmological constraints from weak lensing observations.

Analytic IA models \citep[e.g.][]{Catelan01,Crittenden01,Hirata04,Blazek19,fortuna21a} are typically used to mitigate the impact of
IA on lensing measurements. However, it is not yet known which IA models are sufficiently accurate to avoid
biasing cosmological parameter inference. Alternatively, employing overly complex modeling can remove cosmological
constraining power and might introduce parameter degeneracy. It is thus important to test if current IA models 
satisfy the accuracy requirements for the upcoming observations. 
One possibility to do so is provided by direct measurements of IA in spectroscopic surveys, as these surveys enable a clear separation between foreground and background galaxies. Such a separation is not possible with the less accurate photometric redshift estimates that are used in weak lensing surveys. Direct measurements of IA have been made in several spectroscopic surveys, including SDSS, WiggleZ, BOSS, KiDS+GAMA and PAU \citep{Mandelbaum06,Hirata07,Mandelbaum11, Joachimi11, Singh15, Singh16, Johnston19, Johnston21, fortuna21b} and revealed inaccuracies
of the analytic IA models, in particular at small scales. These direct observations further showed that the IA signal
depends strongly on the luminosity and color range probed by a given galaxy sample, indicating that the shapes and
orientations of galaxies are affected by the same evolutionary processes (e.g. merging and cold gas accretion)
that determine the photometric properties of galaxies. This conclusion lines up with results from hydrodynamic
simulations \cite[e.g.][]{Codis18}. The alignment contributions to the lensing signal are therefore
expected to depend strongly on the photometric properties as well as on the redshift of the source samples used in
weak lensing analysis. An assessment of how strongly inaccuracies of analytical IA models may bias the cosmological
constraints derived from lensing surveys can therefore not be derived from the current spectroscopic IA observations,
which are focused mainly on red galaxies at relatively low redshifts ($z\lesssim0.5$).

This lack of observational IA constraints may be filled by cosmological simulations, which can provide insights into
IA behavior and allow for testing of modeling and analysis methods in a realistic setting. Cosmological hydrodynamic
simulations of galaxy formation can predict the alignment of galaxies as a function of color and luminosity up to high redshifts \citep{Chisari15, Tenneti15, Velliscig15, Hilbert17, Samuroff21}.
However, their relatively low resolution as well as the assumptions involved in the implementations of galaxy formation processes may impose a bias on the
IA constraints derived from these simulations, which has not been investigated so far.
In addition, the volumes covered by these simulations
are several orders of magnitudes below those probed by lensing surveys due to computational limitations, which inhibits
investigations at the large scales probed in observations.

The need for simulating IA in large cosmological volumes promoted the development of models which assign intrinsic shapes
and orientations to galaxies that were placed in dark mater-only simulations using approximate methods.
These models (hereafter referred to as 'semi-analytic' IA models) are based on the assumption that
each galaxy can be described either as a discy or as an elliptical object.
Discs are thereby commonly assumed to be perfectly circular and oriented perpendicular to their host halos' angular momentum,
while ellipticals are assumed to have the same projected 2D shape and orientation as their host halo \citep{Croft00,Heavens00}.
While assuming that all galaxies are discs, \cite{Heymans04} added more realism to the IA modeling by introducing
a disc thickness as well as a misalignment between the disc and their host halos' angular momentum, as suggested by
hydrodynamic simulations \citep{vdBosch02}. This misalignment strongly reduced the predicted amplitude
of the IA two-point statistics, bringing it in agreement with COSMOS-17 observations.
\citet{Heymans06} further advanced the semi-analytic IA modeling by considering mixed populations of
discs and ellipticals in their simulation, while applying a galaxy-halo misalignment only to the disc population.
\citet{Okumura09a,Okumura09b} found that a misalignment between ellipticals and their host halo is needed in order
to reproduce the observed alignment signal of luminous red galaxies (hereafter referred to as LRGs) in the
Sloan Digital Sky Survey (hereafter referred to as SDSS).
These different semi-analytic IA models only considered central galaxies, for which information on halo shape
and angular momenta could be obtained from the underlying dark matter simulation.
\citet[][hereafter jointly referred to as J13]{Joachimi13a, Joachimi13b}
were the first to add satellite galaxies to the semi-analytic IA modeling, using constraints on the radial alignment of satellites
with respect to their host halos center from a hydrodynamic simulation \citep{Knebe08}. Considering
both, elliptical as well as disc galaxies, J13 applied their IA model on galaxies from a semi-analytic
model of galaxy formation imposed on the Millennium simulation, which exceeded the N-body simulations used in
previous works in resolution and volume. They showed that variations of the model parameters controlling the
disc thickness and the galaxy-halo misalignment have a significant impact on the predicted IA contamination
in the lensing signal. These authors further pointed out that the ellipticity distribution for late-type galaxies
in their simulation does not reproduce the observed lack of circular face-on disc galaxies.
A more detailed overview on semi-analytic IA models can be found in \cite{Kiessling15}.
More recently, \cite{Wei18} applied the model of J13 on a catalog of galaxies from a semi-analytic model of galaxy
formation that was run on a simulation from the Elucid project, which matched the Millennium simulation in volume
but exceeds its resolution significantly. In contrast to previous works, this IA simulation included not only
intrinsic galaxy ellipticities, but in addition gravitational shear derived from ray tracing, which allowed
for direct predictions of the IA contributions to the lensing signal from the Kilo Degree Survey (KiDS) and the Deep Lens Survey.

Overall these different works predicted small but significant contributions of the IA to future lensing surveys.
However, these predictions may be affected by different shortcomings in the IA implementation, which we aim to
address in this work with the following three steps:
1) We use a new model for the intrinsic galaxy shapes, which reproduces the observed galaxy axis ratio distribution
from the COSMOS survey over wide ranges of redshifts, galaxy luminosities and colors, accounting for the lack of
circular objects;
2) We calibrate the semi-analytic IA model for the first time against observational constraints from the BOSS LOWZ survey, provided by
\citet[][hereafter referred to as SM16]{Singh16}, taking into account the luminosity dependence of the observed signal by introducing a luminosity
dependence in the galaxy-halo misalignment for satellite galaxies;
3) We run this new IA model on the light-cone output of the MICE Grand Challenge simulation \citep{MICEI,MICEII,MICEIII},
which provides lensing information together with mock galaxies generated with a hybrid approach of Halo Occupation Distribution
modeling and Halo Abundance matching that was calibrated to match observational constraints on galaxy luminosity and color
distributions as well as the galaxy clustering. The MICE light-cone covers one octant of the sky and reaches up to
redshift $z=1.4$, which allows us to create the largest IA simulation produced so far.

We use this simulation for a detailed investigation of the accuracy of two analytical IA models that are
applied in current cosmological weak lensing analyses: the Non-Linear Alignment (NLA) model \citep{Catelan01, Hirata04, Bridle07, Hirata07}
and the Tidal Alignment and Tidal Torquing (TATT) model \citep{Blazek19}. We therefore compare these models with measurements in MICE over wide ranges of scales, redshifts, galaxy luminosities and colors.
We further compare constraints on the model parameters derived from the simulation
against observational constraints in luminosity and redshift ranges in which the
simulation was not calibrated.
Lastly, we construct a mock sample resembling \metacal \citep*{Gatti21}, the sample used in the analysis of the  first 3 years of Dark Energy Survey (DES) data, in order to predict the IA contamination in current observations.

The paper is organized as follows. In Section \ref{sec:2pt_shear_statistics} we introduce the different
two-point statistics used in this work together with the two analytical IA models, NLA and TATT.
Section \ref{sec:data} describes the MICE simulation, the spectroscopic mock BOSS LOWZ and the photometric DES-like samples constructed
from the MICE galaxy catalog as well as the COSMOS data that was used in the calibration of the
galaxy shapes in MICE.
Our method for modeling these shapes is described and validated in Section \ref{sec:model_shapes},
while the modeling of galaxy orientations is described and validated in Section \ref{sec:model_orientations}.
In Section \ref{sec:sim_vs_theo} we compare IA two-point statistics measured in MICE using true redshifts against predictions from the NLA and TATT models. In Section \ref{sec:des_aplication} we study the IA contribution to the weak lensing signal in
a DES-like photometric sample, as predicted by our simulation.
We finally summarize and discuss our findings in Section \ref{sec:conclusions}.

\section{correlation functions}
\label{sec:2pt_shear_statistics}

The two-point correlation function of the galaxy shear is the main probe of lensing surveys
and has further been used for the direct detection of IA in spectroscopic data sets. We therefore focus on
this type of statistic for the calibration of the IA signal in MICE and for deriving predictions for the IA contamination in weak lensing observations from the simulation.

Before introducing the specific shear correlations used in this work let us define the shear itself.
In weak lensing studies galaxies are approximated as 2D ellipses. The shapes and
orientations of these ellipses are fully described by the shear, which is commonly defined
as a complex spin-2 vector $\gamma = \gamma_1 + i \gamma_2 = \epsilon \exp(i 2 \phi)$.
The galaxy ellipticity $\epsilon=(1-q_{2D})/(1+q_{2D})$ is defined via the 2D axis ratio
$q_{2D}=B_{2D}/A_{2D}$, where $A_{2D}$ and $B_{2D}$ are the absolute value of the
major and minor axis vectors of the ellipse respectively. The galaxy orientation angle
$\phi$ is defined as the angle between one of the two principle axis and an arbitrary reference axis,
as we will specify later on.

\subsection{Definitions and estimators}

\subsubsection{Projected galaxy-galaxy, galaxy-shear and matter-shear correlations (\texorpdfstring{$w_{gg}$,$w_{g+}$,$w_{m+}$)}{wgg,wg+,wm+}}
\label{sec:projcorr}

The projected galaxy-shear correlation is commonly used for direct measurements
of IA in spectroscopic surveys as it provides a high signal-to-noise ratio compared
to the angular shear statistics that are commonly employed in weak lensing cosmology,
while being only weakly sensitive to redshift space distortions
\citep[e.g.][]{Joachimi10, Kirk15}.
In our work we use this statistic to calibrate the IA model in MICE against observational
constraints derived from the BOSS LOWZ sample by SM16.
In addition we study the projected galaxy-galaxy correlation to validate the
mock BOSS LOWZ samples constructed from MICE that are used for the calibration.
When measuring these correlations we follow SM16 by studying the cross-correlation
between a 'shape' sample $S$, consisting of the galaxies whose IA signal we want to measure
and a 'density' sample $D$, which is used as a tracer for the underlying matter distribution. 

The galaxy-galaxy cross-correlation function is defined as
$\xi_{gg}(r) \equiv \langle \delta_g^S \delta_g^D \rangle(r)$, where $\delta_g^S$ and $\delta_g^D$ are the galaxy density contrasts
of the shape and density samples respectively, separated by
the distance $r$, and  $\langle \ldots \rangle$ is the ensemble average.
We measure this correlation from the data using the estimator from 
\citet{Landy93},
\begin{equation}
{\hat \xi_{gg}} = \frac{(S-R_S)(D-R_D)}{R_S R_D} = \frac{SD - D R_S - R_D S + R_D R_S }{R_S R_D}.
\label{eq:xi_gg_LS}
\end{equation}
Each term in the numerator and denominator on the right-hand side of
this equation stands for the counts of galaxy pairs that are separated
by $r$. $R_S$ and $R_D$ are thereby samples of random points
that are constructed to follow the radial probability distribution $N(d)$ of
the $S$ and $D$ samples respectively, where $d$ is the comoving distance from the observer.
We smooth the $N(d)$ distribution over $20 \ h^{-1}\textrm{Mpc}$ with a top-hat window function to reduce the impact of cosmic variance and tested that reducing the window size
to $10 \ h^{-1}\textrm{Mpc}$ has only a negligible impact on the signal compared to the estimated errors on the signal.

Analogously to the galaxy-galaxy correlation one can define the galaxy - shear correlation as
$\xi_{g+/\times}(r) = \langle \delta^D \gamma_{+/\times} \rangle$(r). The
shear is here defined specifically for each $\delta^D$ - $\gamma$ pair considered in the average
$\langle \ldots \rangle$ such that the orientation angle is the angle between the galaxies' major axis
and the distance vector $\bf r$, i.e. $\phi' = \phi - \phi_{r}$. In this coordinate system the shear
components are denoted as $\gamma = \gamma_+ + i \gamma_\times$. Radial ($\phi'=0$) and tangential ($\phi'=\pi/2$)
alignment then leads to $\gamma_+=1$ and $\gamma_+=-1$ respectively, with $\gamma_\times=0$.
An alignment of $\phi'=\pi/4$ and $\phi'=-\pi/4$ leads to $\gamma_\times=1$ and $\gamma_\times=-1$ respectively, with $\gamma_+=0$.
Following SM16 we focus our analysis on $\xi_{g+}$, which we measure using
a variation of Equation (\ref{eq:xi_gg_LS}) given by \citet{Mandelbaum06},
\begin{equation}
\hat{\xi}_{g+} = \frac{S_+ (D - R_D)}{R_S R_D}  = \frac{S_+ D - S_+ R_D}{R_S R_D},
\label{eq:xi_g+_LS}
\end{equation}
with
\begin{equation}
S_+ X = \sum_{i,j} \gamma_+ (i|j),
\end{equation}
where $\gamma_+ (i|j) = \Re \left[ \gamma \exp{-i2\phi_r^{ij}} \right]$
is the $(+)$ component of the shear of a galaxy $i$ in sample $S$, defined with respect to
the vector ${\bf r}$ pointing to position $j$ in sample $X$, where $\phi_r^{ij}$
is the orientation angle of ${\bf r}$ at the position $i$ and $X$ refers to either $D$ or $R_D$.

So far we introduced $\xi_{gg}$ and $\xi_{g+}$ (jointly referred to $\xi_{gx}$ in the following) as isotropic quantities, that are
averaged over all orientations of ${\bf r}$. In order to obtain the projected
correlations we measure $\xi_{gx}$ first as a two-dimensional quantity by
separating the distance vector ${\bf r} = {\bf r_2} - {\bf r_1}$ between
two points at position ${\bf r_1}$ and ${\bf r_2}$ into a line-of-sight and a
transverse (or projected) component.
The line-of-sight vector is thereby defined as
${\bf n} \equiv ({\bf r}_1 + {\bf r}_2)/2$.
The line-of-sight and transverse components are then
obtained as $r_{\Pi} = {\bf r} \cdot \hat{\bf n}$ and
$r_p = (r_\Pi^2 - r^2)^{1/2}$ respectively.
For the measurements we follow SM16 by using $25$
logarithmic bins in the interval $0.1 < |r_p| < 200 \ h^{-1} \textrm{Mpc}$
and $20$ linear bins in the interval $0 < |\Pi| < 60 \ h^{-1} \textrm{Mpc}$.
The projected correlation is then given by
\begin{equation}
w_{gx}(r_p) = \int_{-\Pi_{max}}^{\Pi_{max}} \xi_{gx}(r_p, \Pi) \ d\Pi,
\label{eq:w_theo}
\end{equation}
where $w_{gx}$ stands for $w_{gg}$ and $w_{g+}$. Note that
the projected matter-shear correlation $w_{m+}$, which we investigate in Section \ref{sec:sim_vs_theo} is defined analogously to $w_{g+}$, while the
density sample in Equation (\ref{eq:xi_g+_LS}) is replaced by
a random sub-sample of the dark matter particle distribution in the simulation.

Errors on the measurements of $w_{gg}$, $w_{g+}$ and $w_{m+}$ are estimated using jackknife resampling.
The MICE octant is therefore split into $N_{JK} = 88$ angular sub-regions, which are defined as healpix
pixels with $N_{side}=8$ (see Fig. \ref{fig:jk_samples}). The covariance is then estimated as
\begin{equation}
C^{JK}_{ij} = (N_{JK}-1) \langle \Delta_i \Delta_j \rangle,
\label{eq:jk_cov}
\end{equation}
with $\Delta_i = w^{JK}_i - w_i$, where $w_i$ is the projected correlation measured in the $r_p$ bin $i$ on the full area,
$w^{JK}_i$ is the same measurement, but neglecting one 
jackknife sub-region and $\langle \ldots \rangle$
is the average over the $N_{JK}$ measurements of $\Delta_i$.
When measuring the projected correlations we use the same healpix sub-regions to organize the data in a
one-dimensional tree-structure in order to accelerate the search of galaxy pairs that enter the estimators
in Equation (\ref{eq:xi_gg_LS}) and (\ref{eq:xi_g+_LS}). We have verified that the angular correlations
measured by our code match corresponding measurements from the public
code TreeCorr\footnote{\url{https://github.com/rmjarvis/TreeCorr}} \citep{TreeCorr}.

\subsubsection{Angular shear-shear correlation}
\label{sec:ang_corr}

The real-space angular shear-shear cross-correlation between galaxy samples in
different redshift bins $A$ and $B$
is one of the main observables used for weak lensing tomography in current surveys such as the DES.
The shear field gives rise to a pair of two-point correlations that preserve parity invariance, defined as 
$\xi_+^{AB} \equiv \langle \gamma^A \gamma^{*B} \rangle$ and
$\xi_-^{AB} \equiv \langle \gamma^A \gamma^{B} \rangle$,
where $\langle \dots \rangle$ is the average over the products of all galaxy pairs that are separated
by an angle $\theta$. With the ($+/\times$) decomposition of the complex shear
these correlations can be written as
\begin{equation}
\xi^{AB}_{\pm}=\left\langle{\tilde \gamma}^A_+{\tilde \gamma}^{B}_+\right\rangle \pm\left\langle {\tilde \gamma}^A_{\times}{\tilde \gamma}^{B}_{\times}\right\rangle.
\label{eq:xip}
\end{equation} 
\noindent
We define ${{\tilde \gamma} \equiv - \gamma}$, following the literature convention in weak lensing cosmology,
according to which ${\tilde \gamma}_+$ (often denoted as $\gamma_t$) $=1$ indicates perfect tangential alignment
(see \citet{Kiessling15} for a discussion of differences between shear definitions in weak lensing and IA studies).
We further follow literature conventions for the notations of the correlations $\xi^{AB}_{\pm}$ and $\xi_{g+}$.
Note here that the subscript $+$ has different meanings in both cases.
For $\xi^{AB}_{\pm}$ the $+$ refers to the addition of the $\times$ term in Equation (\ref{eq:xip})
whereas for $\xi_{g+}$ it refers to the radial shear $\gamma_+$.

We measure $\xi^{AB}_\pm$ in our mock DES-like source sample using a similar estimator as in the analysis
of DES Y3 data \citep*[e.g.][]{Secco21}, but with significant simplifications which can be made because of
the absence of observational effects in MICE. In detail,  the response factor and
weight associated to each galaxy's shear are set to unity while the mean shear of each sample is negligible.
This simplified estimator can be written as
\begin{equation}
{\hat \xi}_{\pm}^{AB}(\theta) =
\frac{S_+^A S^B_+ \pm S_\times^A S^B_\times}{S^A S^B},
\label{eq:xipm_estimator}
\end{equation} 
with
\begin{equation}
S^A_+S^B_+ =
\sum_{i \neq j} \gamma^{A}_+ (i|j) \ \gamma^B_+ (j|i).
\label{eq:xipm_estimator_num}
\end{equation} 
$S^A_\times S^B_\times$ is defined analogously and ${S^A S^B}$ is the number of galaxy pairs
between the sample $A$ and $B$ that are separated by $\theta$. Note that this estimator 
does not use pair counts between random samples in the denominator in contrast to the $\xi_{g+}$
estimators. 
The sums are taken over pairs for which the angular separation is in the range
$|\boldsymbol{\theta}-\Delta\boldsymbol{\theta}|$ and $|\boldsymbol{\theta}+\Delta\boldsymbol{\theta}|$.
Both, $\xi^{AB}_+$ and $\xi^{AB}_-$ are measured using 20 logarithmically spaced angular bins between $2.5'$ and $250'$, using TreeCorr.
The data covariance matrix estimate and cosmology inference are described in Sec. \ref{sec: results of likelihood analysis}.

\subsection{Analytical modeling}

\subsubsection{NLA and TATT models for IA}
\label{sec:nla_tatt}

In weak lensing analyses the observed shear is described as the superposition
of a component induced by gravitational lensing ($\gamma^G$) and a component
related to the galaxies intrinsic ellipticity ($\gamma^I$), i.e. $\gamma = \gamma^G + \gamma^I$.
The contribution of the intrinsic shear to the observed shear correlations is most commonly described
analytically using the 
{\it Non Linear Alignment} model \citep["NLA",][]{Hirata04, Bridle07} and the more recent
{\it Tidal Alignment and Tidal Torquing} model \citep["TATT",][]{Blazek19}.
Both models are based on the assumption that the galaxy alignment is induced by the tidal tensor of the
large-scale matter distribution,
\begin{equation}
    s_{ij}({\bf k}) = \left(\hat{k}_i \hat{k}_j + \frac{1}{3}\delta_{ij}\right)\delta(k).
    \label{eq:tidal_tensor}
\end{equation}
The TATT model uses a perturbative approach in which the intrinsic galaxy
shear is expressed via the tidal tensor as,
\begin{equation}
    \bar{\gamma}^{\rm I}_{ij} \simeq A_1 s_{ij} +A_{1\delta} \delta s_{ij} + A_2 s_{ik}s_{kj} .
    \label{eq:tatt_model}
\end{equation}
The free parameters of the model, $A_1$, $A_2$ and $A_{1\delta}$ are effective parameters that capture the total response of galaxy shape to the corresponding combination of cosmic tidal and density fields. In this framework, $A_1$ and $A_2$ capture the direct impact of "tidal alignment" and "tidal torquing", respectively, as well as contributions from any small-scale astrophysical effects that produce the corresponding response. Similarly, $A_{1\delta}$ includes the impact of "density weighting" -- the fact that we observe IA only at the location of galaxies -- as well as other potential effects that can change its value from what we would expect if only density weighting contributed.
All three of these IA parameters can depend on galaxy redshift, luminosity,
and potentially other properties.
The NLA model corresponds to the TATT model without contributions from
tidal torquing, i.e. $(A_2,A_{1\delta})=(0,0)$.
Finally, because we always use the galaxies shape sample, the TATT model can be applied to describe measurements of $w_{m+}$, even though this statistic correlates with the unbiased matter density field. We note that this statistic would not capture the impact of higher-order biasing and correlations of these bias terms with IA. However, these contributions are expected to be small and are currently not included in TATT implementations applied to weak lensing data (e.g.\ \cite{DES21}).

\subsubsection{Prediction for \texorpdfstring{$w_{m+}$}{wg+}}
\label{sec:wmp_predictions}

One goal of our work is to obtain predictions for the IA model parameters
by fitting a model for the projected matter-intrinsic shear correlation $w_{m+}$
against measurements in the MICE simulation. By studying the matter-shear instead of the
galaxy shear correlation we circumvent the modeling of galaxy bias, which would 
add uncertainties to our analysis. Besides the bias the gravitational shear
$\gamma^G$ can also thereby be neglected since we can separate it out in the simulation signal.
In any case its effect on $w_{m+}$ would be negligible by  construction since the
correlations are studied for pairs with line-of-sight distances $|r_\Pi|<60 \ h^{-1} \textrm{Mpc}$
over which gravitational lensing contributions should be weak.

We model $w_{m+}$ using a Hankel transformation of the position - intrinsic galaxy shear power spectrum
and the Limber approximation
\begin{equation}
    w_{m+} = -
    \int_0^\infty
    \frac{dk_\perp k_\perp}{2 \pi}
    J_2 (k_\perp r_p) P_{m I}(k_\perp r_p),
    \label{eq:wgp_prediction}
\end{equation}
where $J_2$ is the second-order Bessel function of the first kind.
We compute this transformation by using the code Fast-PT\footnote{https://github.com/JoeMcEwen/FAST-PT} \citep{McEwen16, Fang17}.
The matter position - intrinsic galaxy shear power spectrum $P_{m I}=\langle \delta_m \gamma^{I}\rangle$
is thereby set by the intrinsic alignment parameters as detailed in \citet{Blazek19}.

\subsubsection{Prediction for \texorpdfstring{$\xi_\pm$}{xi+/-}}

The modeling of the $\xi_\pm$ measurements in our mock DES-like catalog in MICE is more complex
than in the case of $w_{m+}$ since we now need to take into account the gravitational as well
as the intrinsic component for the "observed" shear, which are superposed as $\gamma_{\mathrm{obs}} = \gamma^G + \gamma^I$.
Inserting this superposition into the definition of the shear-shear
correlation between two redshift bins $A$ and $B$ leads to the emergence
of several terms in $\xi_\pm$,
%
\begin{equation}
\xi^{AB}_{\mathrm{obs}} = \xi^{AB}_{GG} + \xi^{AB}_{GI} + \xi^{AB}_{IG} + \xi^{AB}_{II}.
\label{eq:xi_GI}
\end{equation}
Observationally, these terms cannot be separated from each other, and hence, they need to be modeled
when extracting cosmological information from the measurements.
However, the MICE simulation allows us to measure each of these terms separately to investigate
their contribution to the observed signal in mock surveys constructed from the simulation.
In general the predictions for the different terms of the angular shear correlation
are obtained as
\begin{equation} \label{eq: xipm Hankel transform}
\begin{split}
    \xi_{\pm}^{AB}(\theta)= & \sum_{\ell}\frac{2\ell+1}{2\pi\ell^{2}\left(\ell+1\right)^{2}}
    \bigl[G_{\ell,2}^{+}\left(\cos\theta\right) \\ 
    & \pm G_{\ell,2}^{-}\left(\cos\theta\right)\bigr]C^{AB}(\ell),
\end{split}
\end{equation}
where
$G^\pm_\ell(x)$ are related to Legendre polynomials $P_\ell(x)$ and averaged over angular bins \citep[see for instance][]{y3-generalmethods}.
The $GG$, $II$, $GI$ and $IG$ terms from Equation (\ref{eq:xi_GI}) enter via the
2D convergence power spectrum,
$C^{AB} = C^{AB}_{GG} + C^{AB}_{GI} + C^{AB}_{IG} + C^{AB}_{II}$ and are obtained
from the 3D power spectra $P$ again under the Limber approximation as
\begin{equation}\label{eq:Limber}
C_{GG}^{AB}(\ell)=\int_{0}^{\chi_\mathrm{H}} d\chi\frac{W^{A}(\chi)W^{B}(\chi)}{\chi^{2}}P_{\delta\delta} \left(\frac{\ell+1/2}{\chi},z(\chi)\right),
\end{equation}
\noindent
%
\begin{equation}\label{eq:Limber_GI}
C_{\rm GI}^{AB}(\ell)=\int_{0}^{\chi_\mathrm{H}} d\chi\frac{W^{A}(\chi)n^{B}(\chi)}{\chi^{2}}P_{\rm \delta I}\left(\frac{\ell + 1/2}{\chi},z(\chi)\right),
\end{equation}
\noindent
and
\begin{equation}\label{eq:Limber_II}
C_{\rm II}^{AB}(\ell)=\int_{0}^{\chi_\mathrm{H}} d\chi\frac{n^{A}(\chi)n^{B}(\chi)}{\chi^{2}}P_{\rm II}\left(\frac{\ell + 1/2}{\chi},z(\chi)\right),
\end{equation}
\noindent
where
$P_{\delta \delta}\equiv \langle \delta \delta\rangle$ is the matter-matter power spectrum,
$P_{\delta I}\equiv \langle \delta \gamma^I\rangle$ is the same matter-intrinsic shear power spectrum
which enters the $w_{m+}$ prediction in Equation (\ref{eq:wgp_prediction}) and
$P_{I I}\equiv \langle \gamma^I \gamma^I\rangle$ is the intrinsic shear-intrinsic shear power spectrum.
Both, $P_{\delta I}$ and $P_{I I}$ are obtained from the NLA and the TATT model, as detailed in
\citet{Blazek19}. Additionally, $n^{A/B}$ is the normalized source galaxy redshift distribution
in redshift bins $A$ or $B$,
\begin{equation}\label{eq: lensing kernel W}
W^{A/B}(\chi)=\frac{3H_{0}^{2}\Omega_{\mathrm{m}}}{2c^{2}}\frac{\chi}{a(\chi)}\int_{\chi}^{\chi_{\mathrm{H}}}d\chi'\,n^{A/B}\left(z(\chi')\right)\frac{dz}{d\chi'}\frac{\chi'-\chi}{\chi'}
\end{equation}
is the lensing efficiency kernel, $\chi$ is the comoving distance, $\chi_\mathrm{H}$ is the comoving distance at the horizon,
$a$ is the scale factor, $H_0$ is the Hubble constant and $\Omega_{\mathrm{m}}$ is the matter density.

\section{Data}
\label{sec:data}

\subsection{COSMOS}

    \label{sec:data:cosmos}
    
    We use observed galaxy magnitudes, redshifts and shapes from the COSMOS survey to
    calibrate the color cut and the parameters of the galaxy axes ratio distribution
    in our IA model. These galaxy properties are obtained from two public catalogs,
    the {\it COSMOS2015} \footnote{\url{https://www.eso.org/qi/}} \citep{Laigle16} and
    the {\it Advanced Camera for Surveys General Catalog} \citep[ACS-GC \footnote{\url{vizier.u-strasbg.fr/viz-bin/VizieR-3?-source=J/ApJS/200/9/acs-gc}},][]{Griffith12}.
    In the following we briefly described the main properties of these data sets.
    Details on quality cuts and the matching between both catalogs are described in \cite{Hoffmann22}.
    The COSMOS2015 catalog comprises photometry in $30$ bands and provides redshift estimates,
    which were derived by fitting templates of spectral energy distributions to the photometric
    data \citep{Ilbert06, Ilbert09}.
    We discard objects which are classified as i) residing in regions flagged as "bad" ii)
    saturated, and iii) not classified as galaxies. After these cuts the sample contains $521,935$
    objects which are used to calibrate the color cut employed in our IA model. 
    
    In order to constrain the galaxy shape parameters as a function of redshift we 
    further impose the recommended cuts on the $3 \sigma$ limiting AB magnitudes in the near-infrared
    $K_s$-band of $24.0$ and $24.7$ in the {\it deep} and {\it ultra-deep} fields, respectively \citep{Laigle16}.
    The ACS-GC is based on {\it Hubble Space Telescope} (HST) imaging in the optical red $I_{AB}$
    broad band filter F814W. The absence of atmospheric distortions allows for an excellent image
    resolution, which is mainly limited by the width of the HST point spread function (PSF) of $0.085"$
    in the F814W filter and the pixel scale of $0.03"$. Sources were detected using the \galapagos\
    software \citep{Haeussler11}. Galaxy shapes are described by the two-dimensional major over minor
    axes ratios \qdd, which are derived from fits of a single \sersic\ model and corrected for
    PSF distortions. We select objects from the catalog which were classified as galaxies with good
    fits to the \sersic\ profile. After applying the quality cuts, the two catalogs are matched based
    on galaxy positions and magnitudes as described in \citet{Hoffmann22}. The final matched
    catalog contains $98,604$ objects.

\subsection{MICE}\label{section:mice_intro}

The MICE Grand Challenge (MICE-GC) simulation \citep{MICEI} is a large N-body run which evolved $4096^3$ particles in a volume of $(3072 \ h^{-1} \textrm{Mpc})^3$ using the \textsc{gadget-2} code \citep{Springel05}. It assumes a flat $\Lambda$CDM cosmology with $\Omega_m=0.25$, $\Omega_{\Lambda}=0.75$, $\Omega_b=0.044$, $n_s=0.95$, $\sigma_8=0.8$ and $h=0.7$. This results in a particle mass of $2.93\times10^{10} h^{-1} M_\odot$. The initial conditions were generated at $z_i=100$ using the Zel'dovich approximation and a linear power spectrum generated with \textsc{camb}\footnote{\texttt{http://camb.info}}. 

The dark-matter light-cone is decomposed into a set of concentric all-sky spherical shells of a given width $\Delta_r$ around the observer, following the approach introduced in \citet{Fosalba08} \citep[see also][]{MICEIII}. Given the size of the simulation box, the resulting light-cone outputs show negligible repetition along any line of sight up to $z=1.4$.  A set of 265 maps of the projected mass density field with $\Delta_r = 35$ mega-years in look-back time, and angular Healpix resolution $N_{side}=8192$ (i.e, 0.43 arcmin pixels) were used to discretize the light-cone volume. These maps were then used to derive the all-sky convergence field $\kappa$ in the Born approximation by integrating them along the line of sight weighted by the appropriate lensing kernel (see \cite{Fosalba08} for details). The convergence was transformed to harmonic space, where a simple relation to the shear field holds (for which the B-mode exactly vanishes), and transformed back to angular space to obtain the  $(\gamma_1,\gamma_2)$ components of the shear field. In this way discretized 3D lensing properties (kappa and shear) were produced across the 3D volume covered by the light-cone.

Halos in the ligh-cone were identified using the Friends-of-Friends (FoF) algorithm with linking length $b=0.2$ down to the limit of two particles per halo \citep{MICEII}. Following \citet{Carretero15}, a combination of Halo Occupation Distribution (HOD) and Sub Halo Abundance Matching (SHAM) techniques were then implemented to populate halos with galaxies
in one octant of the light-cone, covering $5156.6$ $deg^2$. Galaxy positions, velocities, luminosities and colors were thereby assigned, such that the catalog reproduces
SDSS observations of the luminosity function, the color-magnitude distribution
and the clustering as a function of color and luminosity \citep{Blanton03,Zehavi11}.
Spectral energy distributions (SEDs) were then assigned to the galaxies re-sampling from the COSMOS catalog of \cite{Ilbert09} galaxies with compatible luminosity and (g-r) color at the given redshift. Once the SEDs are assigned,
magnitudes can be computed in any desired filter. In particular,
DES $griz$ magnitudes are generated by convolving the SEDs with
the DES pass bands.

In order to reproduce with high fidelity the distribution of colors and magnitudes of the
DES Year 3 (hereafter Y3) data, we remap the MICE photometry into the observed photometry using an N-dimensional probability density transfer method \citep{remap}, which preserves the correlation among colors. Once we have remapped the photometry (i.e. distributions of magnitudes and colors) to the one of DES Y3, we compute photometric redshift estimates using the Directional Neighborhood Fitting \citep[DNF,][]{DeVicente16} training-based algorithm. DNF is one of the algorithms used to compute photometric redshifts in DES albeit not the default one for the source sample.  As a training sample for DNF we consider the same sample used to run DNF on the Y3 data, which is a compilation of spectra from spectroscopic surveys that overlap with the DES footprint (see \cite{DESY3GOLD} for details). We will use the remapped photometry and the DNF photometric redshifts in Section \ref{sec:data:des_samples}.

\subsubsection{Halo orientations and angular momenta}
\label{sec:halo_orientations}
    The orientations and angular momenta of the FoF halos are main components of our IA model.
    The orientations are obtained from the eigenvectors of the reduced moment of inertia
    \eq{
        I_{i,j} = \frac{1}{N_p} \sum_{n}^{N_p} \frac{r_{n,i} r_{n,j}}{r_n^2},
     \label{eq:rmi}
    }
    where $N_p$ is the number of FoF particles,
    $r_{n,i}$ are the components of the 3D position vector of the $n^{th}$ particle
    with respect to the FoF center of mass and 
    $r_n = \sqrt{r_{n,1}^2+r_{n,2}^2+r_{n,3}^2}$ is the particle distance to that center.
    The angular momentum vectors are given by
    \eq{
        {\bf J} = \sum_{n}^{N_p} {\bf r}_n \times {\bf v}_n,
    }
    where ${\bf v}_n$ is the 3D velocity vector of the $n^{th}$ particle, defined
    with respect to the average velocity vector of all halo particles.
    
    The orientations and angular momenta where measured for FoF groups with down to
    $10$ particles. Using such low numbers of particles can be problematic
    since noise in the measurements could decrease the halo alignment to a degree that inhibits
    the induction of a galaxy alignment signal that is sufficiently high to match
    observational constraints. We therefore investigate the impact of noise on the
    alignment of halo orientations and angular momenta in Appendix \ref{app:halo_alignment}. 
    For that purpose we compute these quantities from subsets of random
    particles of massive halos in MICE and measure a 3D alignment statistics
    for different subset sizes. We find that
    even with $10$ particles we are still able to detect a clear signal,
    although with a significantly decreased amplitude. The dependence of noise
    in the halo orientations and angular momenta on the number of halo particles
    will affect the mass dependence of the halo alignment and
    therefore potentially also the luminosity dependence of the galaxy alignment in the simulation.
    However, in Section \ref{sec:model_orientations} we argue that we can compensate
    for such systematic effects when calibrating the galaxy-halo misalignment
    as a function of galaxy luminosity.

    Note further that the FoF particle positions have not been stored in MICE for halos
    with less than $10$ particles,
    while the HOD model uses halos containing as few as two particles. The $10$ particle limit
    therefore imposes a luminosity cut in the simulation, which we discuss in Appendix \ref{app:cmz_distributions}.
    
    A common alternative to the reduced moment of inertia is the standard moment of inertia, which is
    defined as in Equation (\ref{eq:rmi}), but with $r_n=1$ in the denominator.
    By using the reduced instead of the standard moment of inertia we hence
    assign more weight to the inner regions of the halos when measuring
    their orientations. This choice is motivated by the assumption that the
    central galaxy orientation should be more closely related to the
    orientation of the host halo center than to the orientation of the host halos' outer regions.
    Furthermore, FoF particles in the outer regions are more likely to be spuriously
    linked by the FoF algorithm \citep[e.g.][]{Springel01}, which may bias the measured orientations.
    The halo properties measured for this work are part of a public halo catalog
    that has been presented by \citet{Gonzalez22}.

\subsection{Color cuts}
\label{sec:color_cuts}
    
\begin{figure}
        \begin{center}
        \includegraphics[width=0.45\textwidth]{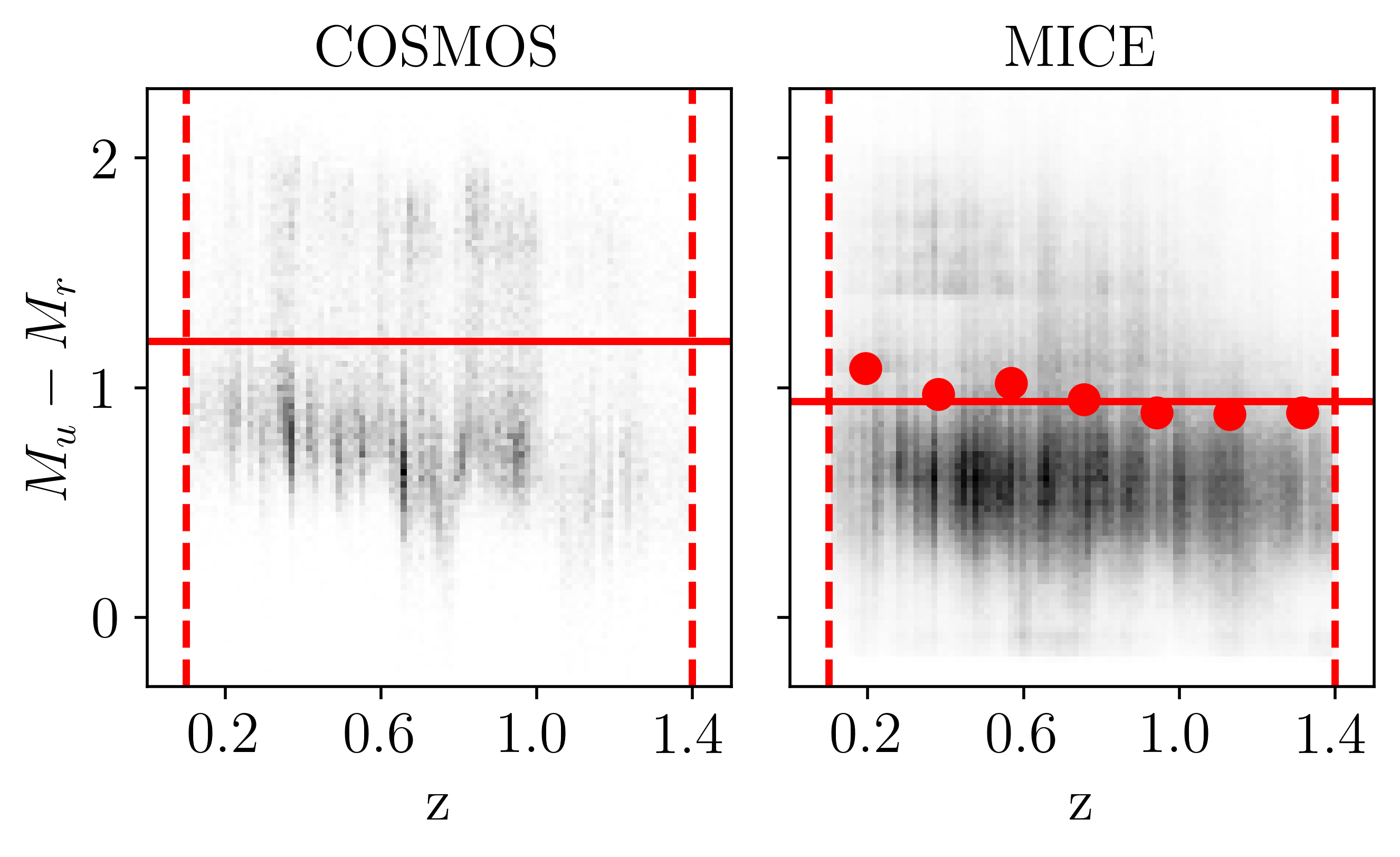}
        \caption{
        Absolute restframe color index versus redshift
        for galaxies with $m_i<24$ in COSMOS and MICE. Horizontal
        solid lines mark the redshift independent cuts used for selecting red and
        blue sub-samples. Dots in the right panel indicate the color cuts
        in MICE that would reproduce the fraction of red and blue galaxies in COSMOS
        in different redshift bins.
        }
        \label{fig:cz_cosmos_mice}
        \end{center}
    \end{figure}

\begin{figure}
        \begin{center}
        \includegraphics[width=0.45\textwidth]{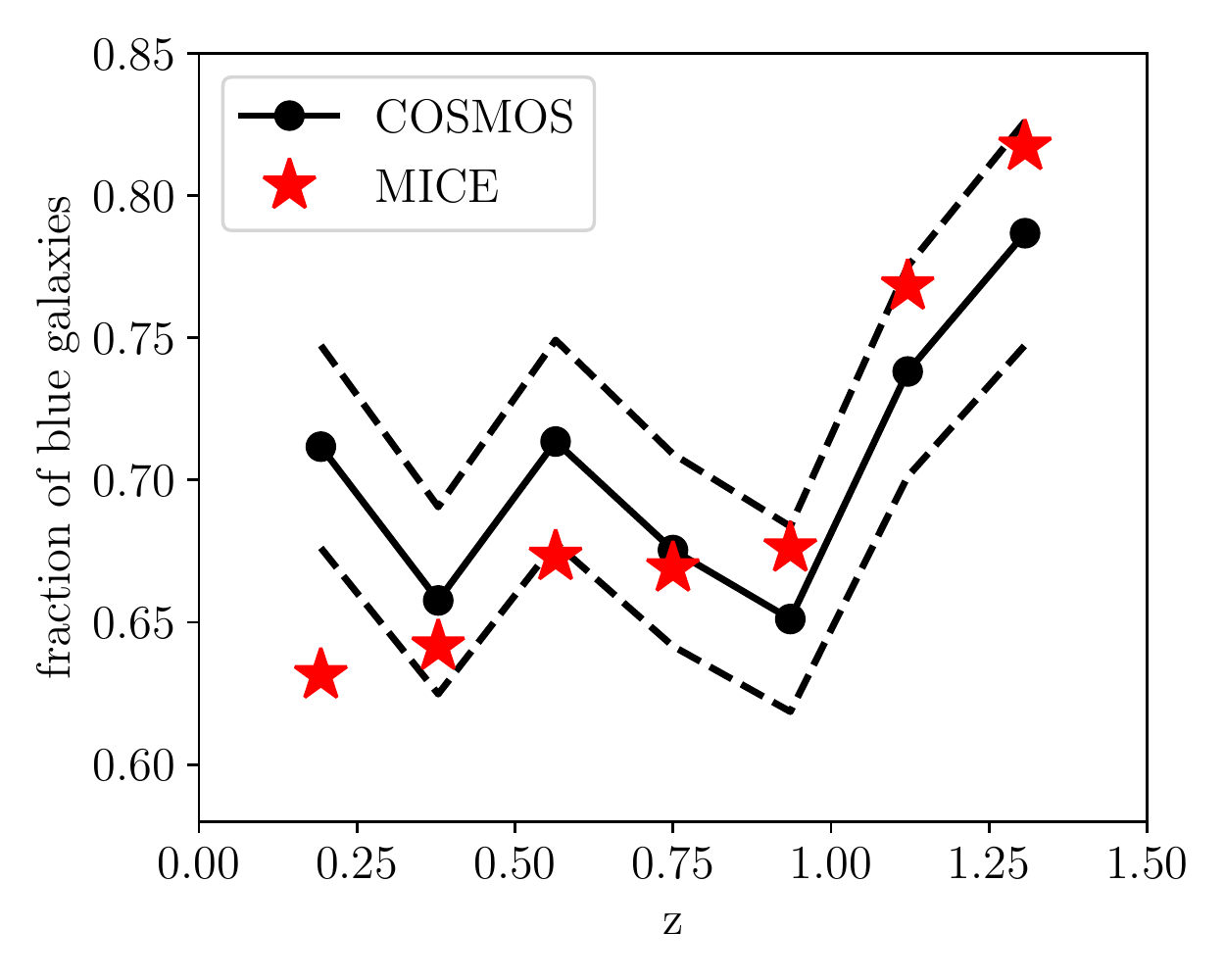}
        \caption{Fraction of blue galaxies in MICE and COSMOS, selected by the redshift
        independent color cuts, shown in Fig. \ref{fig:cz_cosmos_mice}.
        The dashed lines mark $\pm 5\%$ deviations from the COSMOS data.}
        \label{fig:frac_blue}
        \end{center}
    \end{figure}

    Observations have shown that the shapes as well as the intrinsic alignment signal depend strongly on galaxy color
    \citep[e.g.][]{Joachimi15,Guo20}. We incorporate such a color dependence in our model by using different
    model parameters for red and blue galaxies. The color type is set by a cut in the $u-r\equiv M_u-M_r$ color index,
    where $M_u$ and $M_r$ refer to the absolute rest frame magnitudes in the CFHT $u$-band and the Subaru $r$-band respectively.
    We infer the value of this cut by comparing the $u-r$ distributions from MICE and COSMOS in Fig. \ref{fig:cz_cosmos_mice},
    focusing on galaxies within the redshift and apparent $i$-band magnitude range covered by MICE, i.e. $0.1 < z < 1.4$ and $m_i<24$
    \footnote{The fluctuations in the redshift distribution,
    noticeable as vertical stripes in Fig. \ref{fig:cz_cosmos_mice},
    result from cosmic variance. This variance is expected to be high since the data used
    for this figure was sampled in narrow light-cones of a few square degree in COSMOS as well as in MICE.}.
    We find that a cut at $u-r=1.2$, shown as horizontal solid line in the 
    left panel of Fig. \ref{fig:cz_cosmos_mice}  separates the red and the blue sequences in
    COSMOS reasonably well at all considered redshifts. The global fraction of blue galaxies in COSMOS defined by this cut
    is $f_{blue}=0.69$. We adjust this cut in the MICE simulation to $u-r=0.94$ to obtain the same
    global fraction of blue galaxies, as shown in the right panel of
    Fig. \ref{fig:cz_cosmos_mice}. The red dots indicate the color cut which would
    reproduce the exact fraction of blue galaxies from COSMOS in different redshift bins. We find
    that these redshift dependent cuts lie close to the globally defined cut, which confirms that using a
    redshift independent cut in MICE is an appropriate choice.
    As an additional validation we compare the fractions
    of blue galaxies in the redshift bins
    to results in COSMOS in Fig. \ref{fig:frac_blue}. The blue fractions in MICE lie within
    $5 \%$ of the COSMOS results, except for the lowest redshift bins at $z\simeq0.2$,
    where we find a $\simeq 10\%$ deviation.
    
    Note here that a simple color cut does not separate morphological types very well, in particular because
    a significant fraction of disc galaxies is red due to dust extinction when seen edge-on
    \cite[e.g.][]{Graham08,Hoffmann22}. A more robust selection of morphological types based on photometric
    properties could be done using a color-color cut, based on two different color indices
    \cite[e.g.][]{Joachimi13b}. However, it is less obvious how
    to adjust such a color-color selection in the simulation to match the relative abundance
    of the different morphological types in an observational reference sample. For the sake of simplicity
    we therefore proceed using a simple color-cut, leaving more sophisticated cuts as improvements for
    future updates of our model.
    
\subsection{Mock BOSS LOWZ samples}
\label{sec:data:mice:lowz}

For the calibration of our IA model against observed IA statistics of LRGs from the BOSS LOWZ survey from
SM16,
we construct a mock LOWZ catalog from the MICE simulation.
We therefore select galaxies from MICE in the redshift range analyzed by SM16 ($0.16 < z < 0.36$) and apply the LOWZ
selection in color-magnitude space, given by
\begin{eqnarray}
    \label{eq:lowzcuts1}
    m_r < 13.5 + c_\parallel/0.3 + \Delta m_r \\
    16.0  < m_r < 19.6 + \Delta m_r \nonumber \\
    |c_\perp| < 0.2 \nonumber \\
    0.16 < z < 0.36,  \nonumber
\end{eqnarray}
where
\begin{eqnarray}
    \label{eq:lowzcuts2}
    c_\parallel = 0.7(m_g-m_r) + 1.2[(m_r-m_i) - 0.18] \\
    c_\perp = (m_r-m_i) - (m_g-m_r) / 4.0 -0.18 \nonumber,
\end{eqnarray}
and $m_g$, $m_r$, $m_i$ are the apparent magnitudes in the corresponding SDSS broad-band
filters
\footnote{
Note that the BOSS target selection is based on model magnitudes \citep{Dawson13},
while MICE magnitudes were assigned to match the \citet{Blanton03} SDSS luminosity function derived
from Petrosian magnitudes. However, the latter authors find only a weak change of the luminosity function
when using model magnitudes. We therefore do not expect the differences in the magnitude definition
to be relevant for the construction of the mock BOSS LOWZ samples.}.

$\Delta m_r$ is a constant that is zero in the observational LOWZ selection and 
adjusted to a value of $0.085$ in MICE to obtain the observed galaxy number density of the LOWZ sample.

In order to study the luminosity dependence of the IA signal we
follow SM16 by splitting the mock LOWZ sample into four luminosity sub-samples,
called L1-L4 (from bright to dim), which are selected as quantiles of the absolute
SDSS $r$-band magnitude distribution, containing $20\%$, $20\%$, $20\%$ and $40\%$
of the objects respectively. The number of galaxies in each sub-sample is given
in Table \ref{tab:mag_samps_mice_lowz} together with the corresponding
magnitude ranges, mean magnitudes and mean redshifts.
More details on the selection of the MICE LOWZ sample and its luminosity sub-samples
are given in Appendix \ref{app:cmz_distributions}.

\begin{figure*}
    \includegraphics[width=1\textwidth]{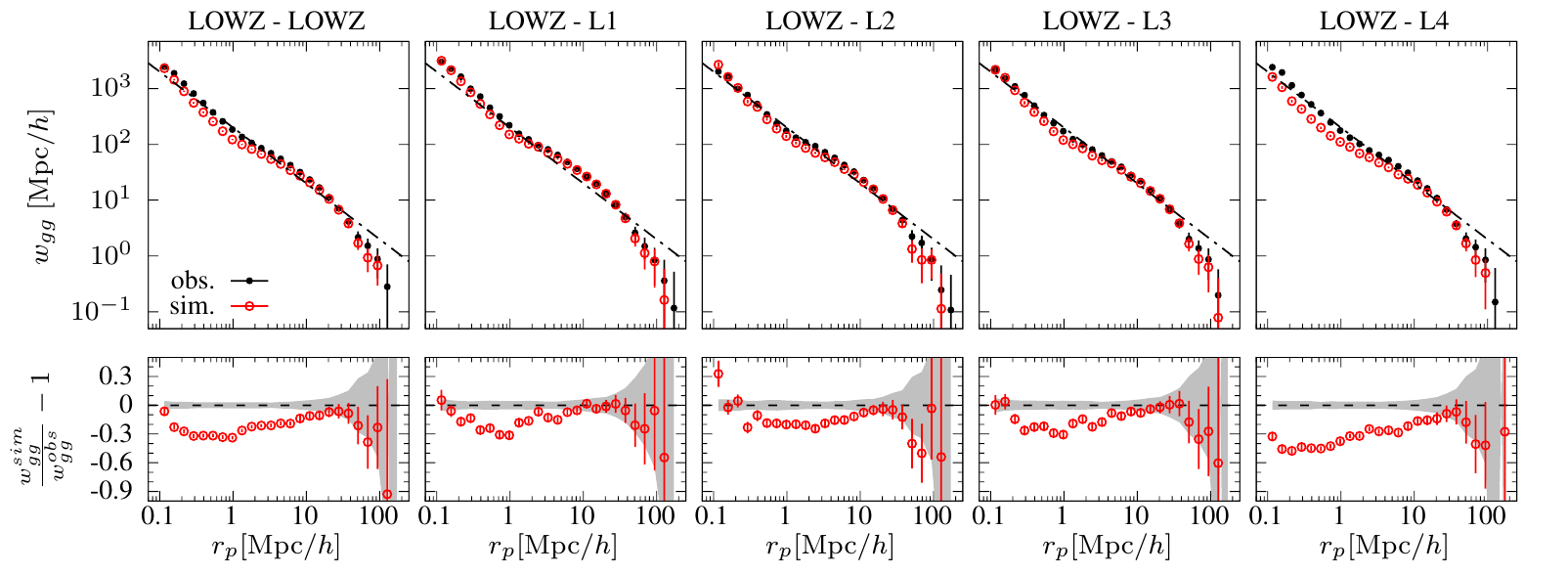}
    \caption{{\it Top}: Projected galaxy correlation functions from the BOSS LOWZ survey
    (SM16) and a mock catalog constructed from the MICE simulation (black dots and red open circles respectively).
    The left panel shows the auto-correlation measured in the full LOWZ sample,
    the panels on the right show the cross-correlation between the full LOWZ sample and
    four luminosity sub-samples L1-L4 (from bright to dim, see Section \ref{sec:data:mice:lowz}
    for details on the sample selection).
    Error bars show estimates of the $1 \sigma$ uncertainty on the measurements. The dashed-dotted
    black lines shows the same $200 \ r_p^{-1}$ power law in all panels as a visual guidance for
    comparing the variation in the amplitude across different samples.
    {\it Bottom}: Relative deviations between measurements in observations and in MICE. The gray shaded
    areas indicate the $1 \ \sigma$ errors on the observations.
    }
    \label{fig:wgg}
\end{figure*}

We validate our sample selection by comparing the $w_{gg}$ auto-correlation of the full MICE LOWZ sample
and its cross-correlation with the sub-samples L1-L4 to the corresponding measurements in BOSS observations
from SM16 in Fig. \ref{fig:wgg}.
We find that the simulation reproduces the overall scale dependence of the observed $w_{gg}$ signal as well as the relatively
weak dependence on luminosity. At scales between $20<r_p<40 \ h^{-1} \textrm{Mpc}$ the amplitudes for
the simulated and the observed samples are in good agreement as well with deviations of $\lesssim 10 \%$.
At smaller and larger scales the $w_{gg}$ measurements in MICE are up to $\sim 60\%$ and $\sim 90\%$ below
the observations respectively. For scales $r_p \gtrsim 10 \ h^{-1} \textrm{Mpc}$ the deviation between observation
and simulations are consistent with the $1 \sigma$ error estimates, while the deviations at small scales are
highly significant as the errors are smaller. These small scale deviations are similar for the samples L1-L3,
and highest for the dimmest sample L4, which could be related to the over-density artefict
at $z\simeq0.25$ that is discussed in Appendix \ref{app:cmz_distributions}.

When interpreting these deviations it is important to keep in mind that the MICE HOD-SHAM model has been calibrated
against the clustering statistics of the SDSS main sample, which covers lower redshifts and dimmer magnitudes
than those probed by the BOSS LOWZ survey \citep{Carretero15}. Deviations of the galaxy clustering statistics in our
mock LOWZ samples from observational results are therefore not unexpected.
Furthermore, the cosmological parameters used to simulate the matter distribution in MICE differ significantly
from recent constraints, while we expect these deviations to have a weak effect on the clustering
compared to the HOD-SHAM parameters.
However, given the implications of these deviations on the IA model
calibration that we discuss in Section \ref{sec:wgp_mice_lowz}, it might be worth trying a more sophisticated
mock construction by adjusting the LOWZ cuts in MICE, such that the mock samples match the observed
clustering instead of the observed number density.

\begin{table}
    \centering
    \begin{tabular}{c | c | c | c | c | c}
        Sample & $M_r^{min}$ & $M_r^{max}$ & $\langle M_r \rangle$ & $\langle z \rangle$ & $N_g$ \\\hline
        L1 & $-23.61$ & $-22.21$ & $-22.43$ & $0.29$ & $30924$ \\
        L2 & $-22.21$ & $-21.98$ & $-22.08$ & $0.28$ & $30923$ \\
        L3 & $-21.98$ & $-21.76$ & $-21.87$ & $0.27$ & $30923$ \\
        L4 & $-21.76$ & $-19.53$ & $-21.41$ & $0.24$ & $61847$
    \end{tabular}
    \caption{Characteristics of the luminosity sub-sample from the mock LOWZ catalog constructed from MICE.
    The columns (from second left to right) show the minimum, maximum and mean values of the
    absolute rest-frame SDSS $r$-band magnitude ($M_r$), the mean redshifts and the number of galaxies
    for each sub-sample.}
    \label{tab:mag_samps_mice_lowz}
\end{table}

%
%
\subsection{DES-like source sample}
\label{sec:data:des_samples}

In order to predict the IA signal for a galaxy population that approximates a realistic weak lensing sample, we construct a mock tomographic catalog utilizing photometric redshift estimates in MICE.
The mock sample resembles the one used in the DES Y3 analysis \citep[\metacal,][]{Gatti21} in its overall magnitude and redshift distribution, as well as in constraining power in the cosmological parameter space as described below. 

Firstly, we verify in Appendix \ref{app:cmz_distributions}
that the (non-tomographic) magnitude distributions in the $r$, $i$ and $z$ DES broad bands from the Y3 data are in good
agreement with the distributions of remapped magnitudes in MICE (described in Section \ref{section:mice_intro}).
The resulting MICE DES-like mock contains over $130$ million galaxies,
which is slightly more than the  $100$ million galaxies in the DES Y3
catalog, but is roughly consistent in number density given the $\sim25\%$ greater
area of the MICE octant compared to DES. Additionally, where cosmology inference is carried out, we adapt the
per-galaxy shape noise in order to match the DES Y3 small-scales covariance (see Section \ref{sec: results of likelihood analysis}).

\begin{figure}
	\includegraphics[width=\columnwidth]{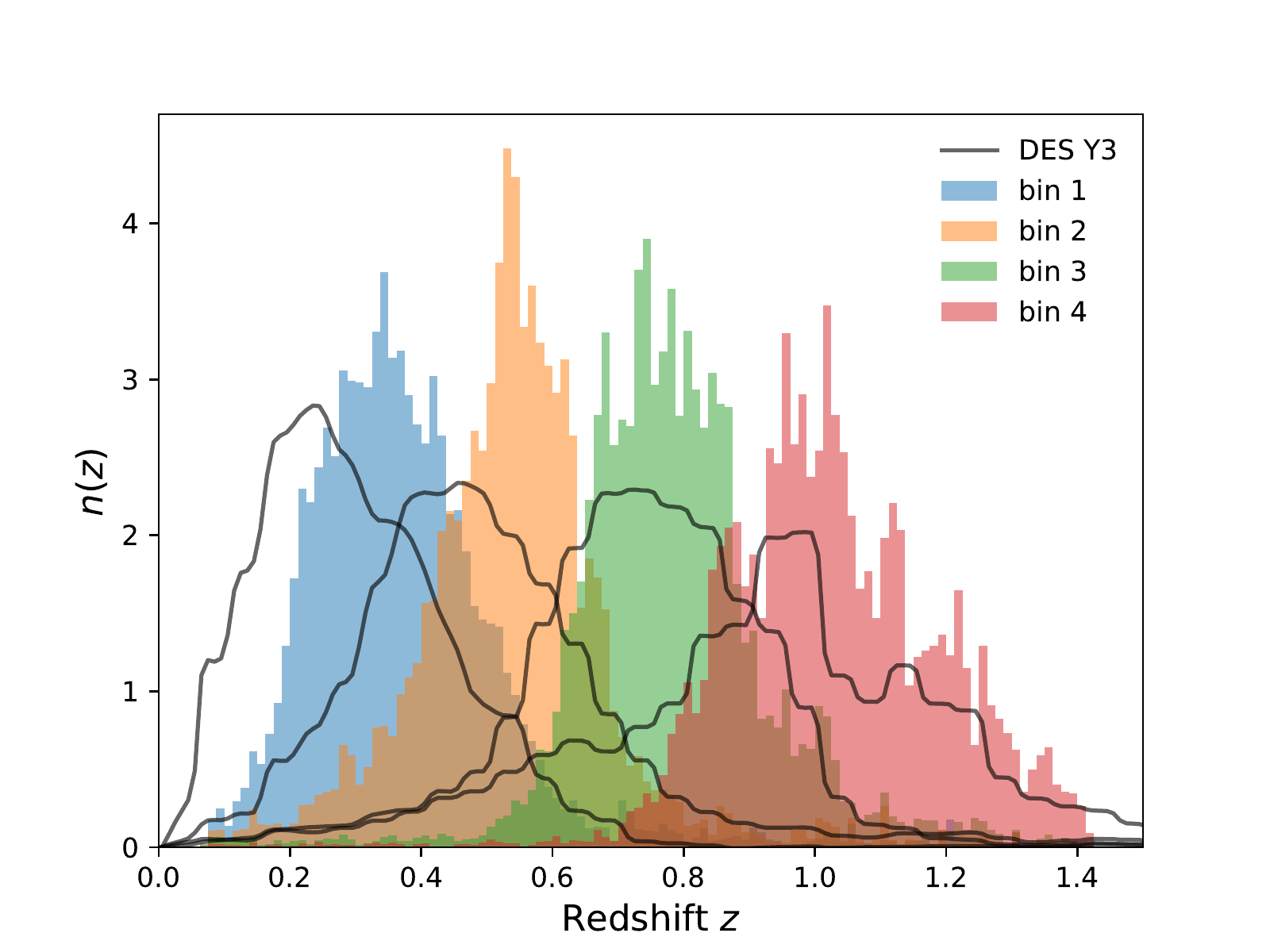}
   \caption{Arbitrarily normalized redshift distribution of the DES-like MICE source sample (solid histograms) and real DES Y3 data (black lines). Redshift point estimates in MICE are obtained with DNF and qualitatively resemble the $n(z)$ distributions in DES Y3 (see text for further details).}
    \label{fig:redshift binning}
\end{figure}

Secondly, we split our mock sample into tomographic bins along the line of sight using photometric redshifts estimated for MICE galaxies with the
DNF algorithm described in Section \ref{section:mice_intro}. We sort galaxies into four bins defined by hard nominal edges that match DES Y1: [$0.2$, $0.43$, $0.63$, $0.90$, $1.30$] \citep{Troxel18}.
While this procedure is different than the methodology employed in DES Y3, based on Self-Organizing Maps (SOMPZ) \citep{Myles20}, it suffices for our goal to create a set of realistic redshift distributions that approximate a DES selection. We show histograms of the true redshifts of the galaxies binned via
DNF point-estimates in Fig. \ref{fig:redshift binning}, with an overall mean redshift of $z=0.6$, along with the binned DES Y3 distributions \citep{Myles20} for a visual comparison. We note that the MICE redshift distributions are generally narrower and peak higher redshifts than their DES Y3 counterparts. The methodological differences between MICE and DES Y3 redshifts exist for practical purposes and imply that the testing presented here should be taken as an additional piece of evidence that the IA modeling in DES Y3 is sound, though not as a final proof.

We find that a significant fraction of central galaxies are defined as blue by
the color cut used in our modeling (see Table \ref{tab:fblue_des}). Since the orientations of blue galaxies are highly randomized in our model
(see Section \ref{sec:model_orientations}), we can already expect from this finding
that the IA signal in the DES-like samples predicted by our model will be weak,
which is indeed the case (see Section \ref{sec:des_aplication}).
A more detailed discussion on the galaxy color distribution in the DES-like samples
can be found in Appendix \ref{app:cmz_distributions}.

\begin{table}
    \centering
    \begin{tabular}{c|c c c}
        z-bin & centrals & satellites & centrals+satellites\\\hline
        $1$ & $0.526$ & $0.386$ & $0.467$\\
        $2$ & $0.583$ & $0.332$ & $0.467$\\
        $3$ & $0.589$ & $0.298$ & $0.488$\\
        $4$ & $0.667$ & $0.358$ & $0.594$
    \end{tabular}
    \caption{Fraction of blue galaxies in the DES-like from MICE samples, defined by our $u-r=0.94$ color cut.}
    \label{tab:fblue_des}
\end{table}

\subsection{Volume limited samples}
\label{sec:data:zmc_samples}

We construct two sets of volume-limited color samples. The first set is used to derive predictions
for the two-point IA statistics up to high redshifts where observational constraints are currently not available.
It covers three redshift bins that are centered around $z=0.2$, $z=0.4$ and $z=0.6$ and have a width of
$\Delta z=0.2$. Galaxies in each redshift bin are separated into six bins by their absolute restframe SDSS
$r$-band magnitude ($M_r$) which have a width of $\Delta M_r=1.0$. The faint limit of these magnitude samples 
is set to $M_r<-20$ to ensure that host halo shape measurements are available for all central galaxies
in the sample (see Appendix \ref{app:cmz_distributions}).
Each of the resulting volume limited samples is further split into a red and a blue
sub-sample at the same $u-r=0.94$ color cut which we use in our IA model (Section \ref{sec:color_cuts}).
The selection of the resulting $36$ samples is illustrated Fig. \ref{fig:mice_zmc_samples}.

A second set of volume-limited color samples is constructed to calibrate the parameters of our
shape model against COSMOS observations. Each of these samples has a width of $\Delta z=0.2$ and $\Delta M_r=1$
in redshift and absolute Subaru $r$-band magnitude respectively. The samples are equally spaced on a regular grid
in the $z-M_r$ space with an overlap of $\Delta z / 2$ and $\Delta M_r/2$. This overlap allows for an increased
sampling resolution while keeping the number of galaxies per sample large enough to allow for statistically meaningful
measurement of the 2D axis ratio distribution over a wide range in magnitude and redshift. We discard samples that
contain less than $100$ galaxies, which leads to $115$ and $196$ samples for red and blue galaxies respectively.
The positions of these samples in magnitude-redshift space are shown as dots in the right panels of Fig. \ref{fig:shape_parameters}, where
each dot's color indicates the number of galaxies in the corresponding sample. 
We use $10$ samples from the second set as examples to validate if
the distribution of galaxy axis ratios in the final MICE IA simulation
matches the reference observations from COSMOS.
The areas spanned in the magnitude-redshift space by these example samples are shown in Fig. \ref{fig:zmc_samples_cosmos_mice}.
Note that red and blue sub-samples are selected in MICE by the same cut at $u-r=0.94$ used in the first set of samples
while we cut the COSMOS samples at $u-r=1.2$ as explained in Section \ref{sec:color_cuts}.

\section{Modeling galaxy shapes}
\label{sec:model_shapes}

Our model for galaxy shapes is based on the assumption that each galaxy's shape can be approximated as a 3D ellipsoid whose shape
is fully described by two of the three axis ratios
\begin{equation}
q_{3D} \equiv \frac{B_{3D}}{A_{3D}}, ~~\mbox{ }~~
r_{3D} \equiv \frac{C_{3D}}{B_{3D}}, ~~\mbox{ }~~
s_{3D} \equiv \frac{C_{3D}}{A_{3D}},
\label{eq:3D_axes_ratios}
\end{equation}
where $A_{3D}$, $B_{3D}$, $C_{3D}$ are the 3D major, intermediate and minor axis respectively.
This modeling choice is motivated by findings reported in the literature, which show
that randomly oriented populations of such 3D ellipsoids can lead to distributions of
projected 2D axes ratios,
\begin{equation}
    q_{2D} \equiv B_{2D} / A_{2D}
\end{equation}
which match those from observed ensembles of early- as well as late-type galaxies
with high accuracy \citep[e.g.][]{Sandage70, Binney78, Noerdlinger79, Lambas92,Ryden04}.
In particular this model describes successfully the lack of circular face-on galaxies
(i.e. $q_{2D} \simeq 1$) found in observations. Achieving such a match was shown
to be problematic in previous work in which discs were modeled as as flat coin-like cylinders \citep{Joachimi13a}.
However, whether this lack is physical, a result of observational limitations or both
remains an open question \citep[e.g.][]{Bertola91, Huizinga92, Rix95, Bernstein02, Joachimi13a}.

Besides the ellipsoidal model for each galaxy's shape, matching the observed 2D axis ratio
distribution further requires a model for the distribution of 3D axes ratios.
Several models of such distributions have been presented in the literature
\citep[see][for an overview]{Hoffmann22}. In this work we employ a simple Gaussian model,
\eq{
\label{eq:P3d}
\tilde P(q_{3D},r_{3D}) =
exp \Biggl\{
-\frac{1}{2}
\left[
\left( \frac{q_{3D} - q_0}{\sigma_{qr}} \right)^2 +
\left( \frac{r_{3D} - r_0}{\sigma_{qr}} \right)^2
\right]
\Biggr\},
}
where $q_0$, $r_0$ and $\sigma_{qr}$ are the free model parameters.
The normalized truncated distribution is then given by
\eqa{
\label{eq:P3dnorm}
    P =
    \begin{cases}
     \tilde P_{3D} / \mathcal{N} & \text{if} \ q_{3D},r_{3D} \in (0,1] \\
    0 & \text{else}
    \end{cases}
\label{eq:qr_pdf}
}
with $\mathcal{N} = \int_{0}^1 \int_{0}^1 \tilde P_{3D}(q_{3D},r_{3D}) dr_{3D} dq_{3D}$.
This model is motivated by the model proposed by \citet{Hoffmann22}, which we simplify by
assuming the same width $\sigma_{qr}$ for $q_{3D}$ and $r_{3D}$ to reduce the numbers of free
parameters in our simulation.

To model the shape of a specific galaxy in the simulation we first draw the two 3D axis ratios
$q_{3D}$ and $r_{3D}$ randomly from the distribution in Equation (\ref{eq:qr_pdf}).
The observed 2D axis ratio is obtained later on by projecting the 3D ellipsoid on a tangential plane that
is oriented perpendicular to the observers line of sight, following the methodology presented in J13.
Note that this projection requires not only the 3D axis ratios as input, but
also each galaxy's 3D orientation. The modeling of the latter is described in Section \ref{sec:model_orientations}.
An important aspect for producing realistic mock observations is to incorporate the
dependence of the galaxy shapes on photometric properties and redshift.
We introduce such a dependence in our model by adjusting the parameter
vector  ${\bf p} \equiv (q_0$, $r_0$, $\sigma_{qr}$), according to each galaxy's
redshift, absolute magnitude and color before drawing its 3D axes ratios.

\subsection{Parameter calibration}
\label{sec:shape_model_calibration}

We determine the dependence of the parameter vector ${\bf p}$
on redshift and absolute $r$-band magnitude for a given color
(red or blue) from the observed distribution of 2D axes ratios \pdd\ in COSMOS.
The \pdd\ distribution is therefore measured for red and blue galaxies
(defined via the $u-r$ color index as detailed in Section \ref{sec:color_cuts})
in the volume limited COSMOS samples that are described as the 'second set' in
Section \ref{sec:data:zmc_samples}. For each of these samples we determine the values of
${\bf p}$ for which the corresponding $P(q_{2D})$ prediction fits the observations.
We obtain this prediction for a given candidate ${\bf p}$ by first generating a set
of $N$ 3D ellipsoids, whose 3D axis ratios are drawn randomly from the distribution
in Equation (\ref{eq:P3dnorm}). For each 3D ellipsoid in this set we then compute
$q_{2D}$ following J13 while assuming a random 3D orientation.
The \pdd\ prediction is then measured from the resulting set of $N$ projected 2D axis
ratios and compared to the reference measurement from the observed sample.
The observed as well as the predicted distributions are thereby measured using the
same binning in $q_{2D}$. The number of bins is adjusted to the number of galaxies in
each COSMOS sub-sample, following the Freedman–Diaconis rule for optimal binning \citep{Freedman81}.
We derive the best fit values ${\bf p}$ by maximizing the likelihood which is 
computed from the $\chi^2$ deviation between the predicted and the observed $P(q_{2D})$ distribution.
For the measurements we assume shot-noise errors, while neglecting errors on the predictions
since those are generated using much higher number of axis ratios (i.e. $1000$ points per bin on average).
The posterior of the parameter space is estimated using the Markow-Chain-Monte-Carlo algorithm
\verb|emcee|\footnote{emcee.readthedocs.io} \citep{Foreman13} with flat priors in the ranges
$0.01 < q_0 < 0.99$, $0.01 < r_0 < 0.99$ and $0.01 < \sigma_{qr} < 0.35$.
The upper limit for $\sigma_{qr}$ is set to an arbitrary value that is chosen to be well above the typical
best fit values found for this parameter.
We define the best fit parameters as the position of the maxima of the marginalized posterior distribution.

The distribution of the fitted ${\bf p}$ components
($q_0$, $r_0$, $\sigma_{qr}$) in the redshift-magnitude plane, interpolated between
the positions of the volume limited samples, is shown for red and blue galaxies
in the three left panels Fig. \ref{fig:shape_parameters}. 
The second panel from the right shows the corresponding $\chi^2$ per $q_{2D}$ bin,
which correlates with the number of galaxies per sample, shown on the right of
Fig. \ref{fig:shape_parameters}. This correlation means that deviations between best
fit model and reference measurements become more significant as the shot-noise errors
on the measurements decrease. This indicates that our shape model is too simple to
capture the details of the observed 2D shape distributions. An improvement on that
aspect might be possible by using more flexible extensions for the 3D axis ratio
distribution model \citep[e.g.][]{Hoffmann22}. However, such an extension would
introduce additional parameters in our modeling, while we find the model employed
here to be sufficiently accurate for the purpose of this work, as detailed in the
following.

\subsection{Shape mock construction and validation}

\begin{figure*}
    \includegraphics[width=1.\textwidth]{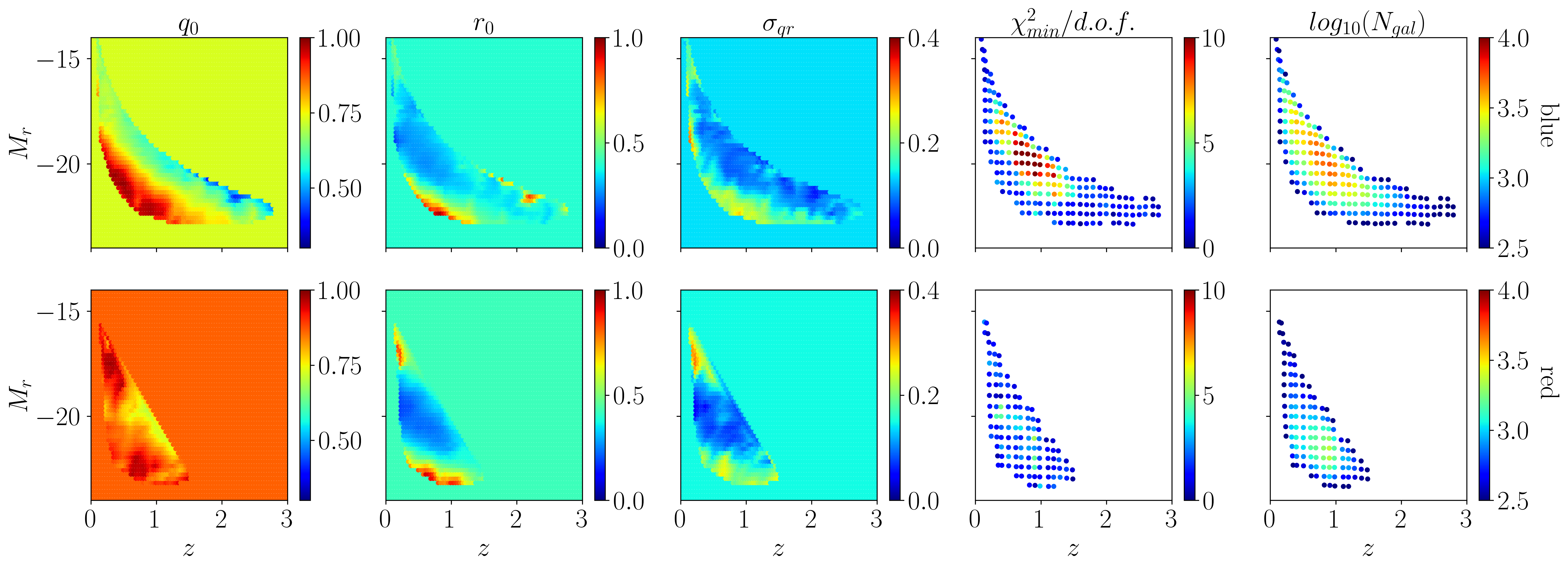}
    \caption{
    {\it Left, central left, central panels}: Interpolated distribution of the parameters in our Gaussian model
    for the 3D galaxy axis ratio distribution, $q_0$, $r_0$ and $\sigma_r$, given in Equation (\ref{eq:P3d})
    as a function of galaxy redshift and absolute $r$-band magnitude for blue and red galaxies (top and bottom panels respectively).
    The parameters were derived from the 2D galaxy axis ratio distributions, measured in overlapping volume limited samples in the COSMOS survey (see Section \ref{sec:model_shapes} for details). 
    {\it Central right panel}: $\chi^2$ per degree of freedom ($d.o.f.$) of the fits to the observed 2D axis ratio distribution for each volume limited sample.
    {\it Right panel}: Number of galaxies per volume limited sample.
    The dots in the right panels are located at the mean redshifts and $r$-band magnitudes of
    the volume limited samples.
    }
    \label{fig:shape_parameters}
\end{figure*}

We assign 3D axes ratios to a given galaxy in the simulation by linearly
interpolating the constrained values of
$q_0$, $r_0$ and $\sigma_{qr}$ for red and blue
samples at the galaxy's position in the magnitude-redshift space.
For galaxies in the simulation which lie outside of the magnitude-redshift
range covered by the COSMOS data we assign the average values of the parameters
over all volume limited sub-samples within each red and blue sample,
shown as homogeneously colored areas in Fig. \ref{fig:shape_parameters}.
Note that a more sophisticated extrapolation of the observational constraints is not
trivial due to the complex dependence of the parameters on magnitude and redshift.
However, in practice this problem is not relevant as most galaxies used in our
analysis lie within the magnitude and redshift ranges covered by COSMOS.

\begin{figure}
    \includegraphics[width=0.45\textwidth]{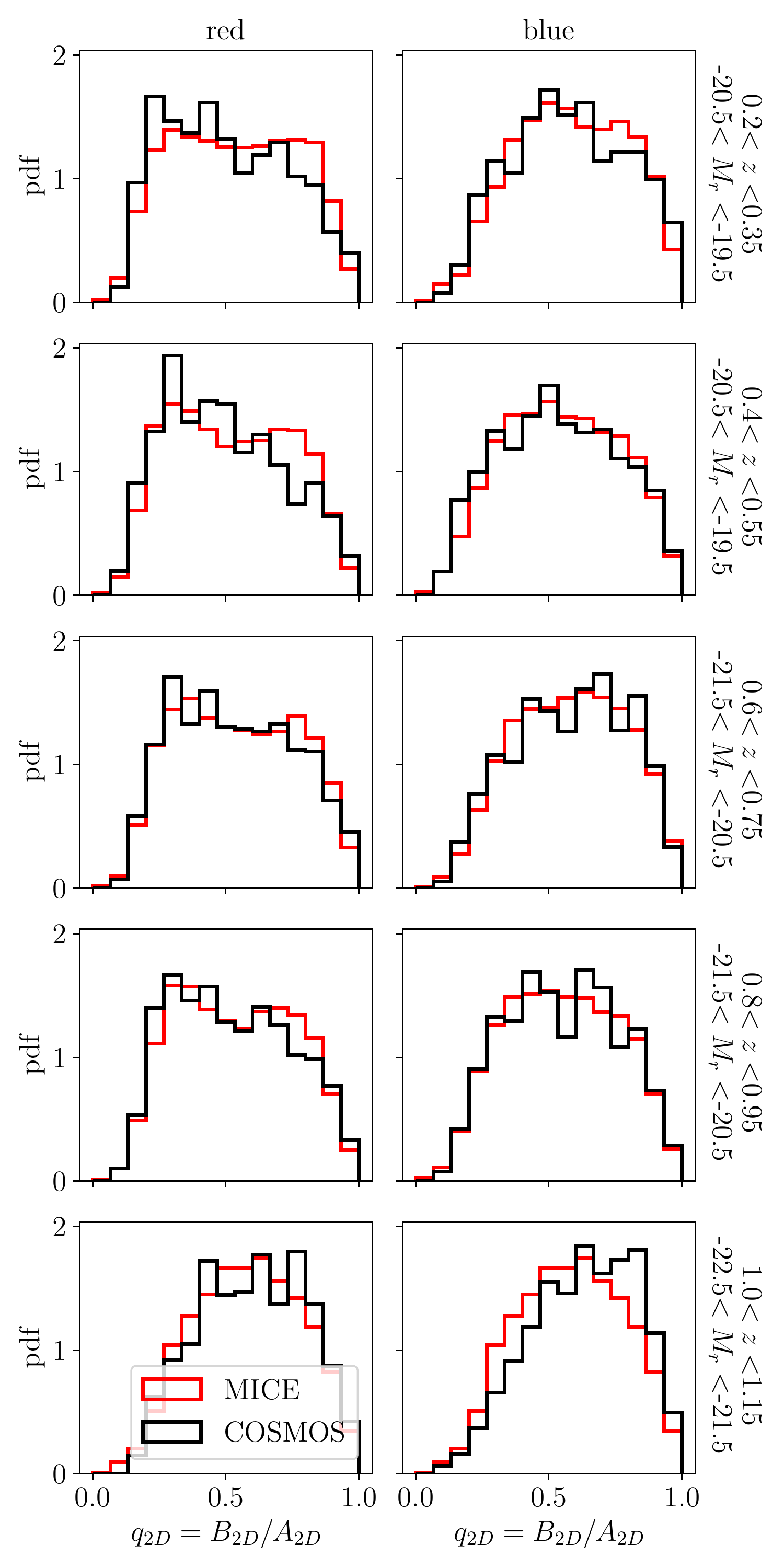}
    \caption{2D galaxy axes ratios measured for red and blue galaxies in different
    volume limited samples.
    The redshift and absolute Subaru $r$-band limits of each sample are indicated on the right. 
    Red and black histograms show results from the MICE simulation and COSMOS observations respectively.}
    \label{fig:q2d_cosmos_mice}
\end{figure}

We validate the performance of our model by comparing the \pdd\ distributions
from MICE against measurements from COSMOS in Fig. \ref{fig:q2d_cosmos_mice}
for the set of $10$ volume limited samples described in Section \ref{sec:data:zmc_samples}
and displayed in Fig. \ref{fig:zmc_samples_cosmos_mice}. We find an overall good
agreement between the simulated an observed data. Deviations are most noticeable
for the brightest sample of blue galaxies at $z\simeq1.1$. They may result from
the shortcomings of the modeling that we discussed in the previous sub-section,
from potential inaccuracies in the linear interpolation of the model parameters as
well as from differences between the redshift-magnitude distributions of observed and
simulated galaxies within a given sample.
It is interesting to note that the \pdd\ distributions for red and blue galaxies
deviate significantly from those expected for discy and elliptical galaxies respectively.
The observed distributions for disc galaxies show typically a plateau in the
center (at $q_{2D}\simeq0.5$) with two knee-like cut-offs on each side. Those for ellipticals
have typically the shape of a skew Gaussian distribution with a maximum
close to unity and a long tail towards low axis ratios \citep[e.g.][]{Rodriguez13}.
The reason that the axis ratio distributions of our color sub-samples do not follow this
expectation may result from the fact that a single color cut does not
separate different morphological types very well as detailed in Section \ref{sec:color_cuts}.

\section{Modeling galaxy orientations}
\label{sec:model_orientations}

\begin{figure}
    \centering
    \includegraphics[width=0.4\textwidth]{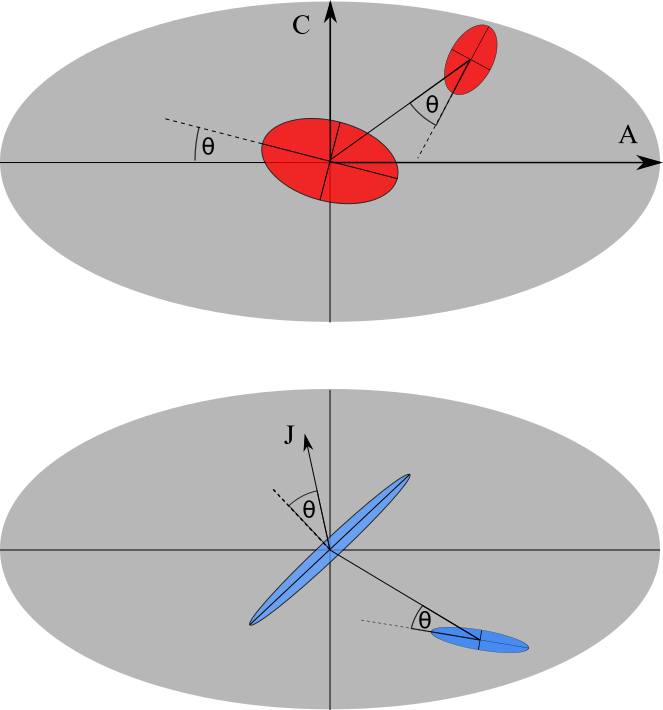}
    \caption{2D Illustration of the model used for assigning 3D galaxy orientations
    in the MICE simulation for red and blue galaxies (top and bottom sub-figures respectively). 
    The major and minor axes of red centrals are aligned with those of their host halo.
    The minor axis of blue centrals is aligned with the angular momentum vector of the
    host halo. The major axes of red and blue satellites are pointed towards the 
    halo center. In a subsequent step of the modeling the orientations
    are distorted by a random angle $\theta$ that depends on galaxy type, magnitude and color.
    }
    \label{fig:ia_halo_model}
\end{figure}

We implement 3D galaxy orientations using methodology from \citet{Joachimi13b},
with some modifications.
Galaxies are thereby separated into three groups: red centrals, blue centrals and satellites,
where the latter include red as well as blue objects. Red and blue galaxies are selected as
described in Section \ref{sec:color_cuts}.

\paragraph*{Red centrals}
have their 3D principle axes aligned with those of their host halo,
i.e. $(\hat A_{3D}, \hat B_{3D}, \hat C_{3D})^{gal} = (\hat A_{3D}, \hat B_{3D}, \hat C_{3D})^{halo}$.
This alignment is based on the assumption that all red galaxies are pressure supported ellipticals
whose shape and orientation is set by the same tidal stretching that determines the shape and
orientation of the host halo.

\paragraph*{Blue centrals}
are assumed to be rotationally supported discs, whose minor axis is aligned with the
angular momentum vector of the host halo, i.e. $\hat C_{3D}^{gal} = \hat J_{3D}^{halo}$
while the major axis $\hat A_{3D}$ is oriented randomly on a plane that is perpendicular
to the minor axis.

\paragraph*{Satellites}
Red and blue satellites are assumed to have their major axes pointed towards the host halo center
while the minor axis is oriented randomly on a plane which is perpendicular to the major axis.
This model assumption is motivated by evidence for a preferred orientation towards the center that has been found in observations as well as simulation.\mbox{}\\
An illustration of the model is shown in Fig. \ref{fig:ia_halo_model}.
Within the framework of the analytical IA models described in Section \ref{sec:2pt_shear_statistics}
the alignment between central ellipticals and their host halo can be associated
with the tidal alignment terms, while the alignment between central discs and the
host halo's angular momentum can be associated to the tidal torquing terms.
The combined effects of tidal alignment, tidal torquing, and the impact of galaxy density weighting
are captured by the parameters $A_1$, $A_2$ and $A_{1\delta}$.

Note here that our assumption that all blue galaxies are discs and all red galaxies are ellipticals
is motivated by the observed correlation between morphological and photometric galaxy properties.
However, the simple color cut used in this work may lead to an inaccurate discrimination between
the two morphological types, as we discuss in Section \ref{sec:color_cuts}. Future updates of
our model may therefore employ more complex photometric cuts to define discs and ellipticals.
For detailed discussions of how the different model assumptions are motivated by observations,
hydrodynamic simulations and analytical models, we refer the reader to the reviews of
\citet[][]{Joachimi15, Kiessling15, Kirk15} and references therein.

\subsection{Misalignment}
\begin{figure}
    \centering
    \includegraphics[width=0.4\textwidth]{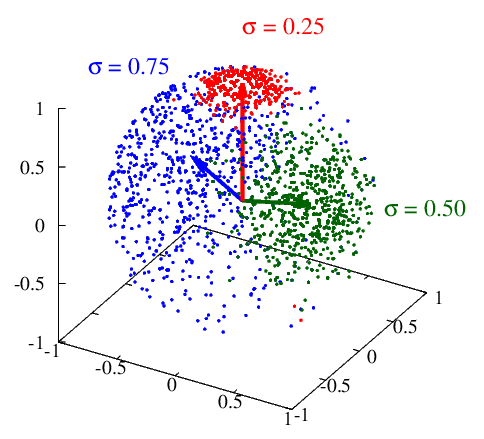}
    \caption{Mises-Fisher distribution for different misalignment parameters $\sigma_{mf}$.
    Points mark randomized orientations on the surface of a unit sphere
    with respect of an input vector. The Mises-Fisher distribution is
    used for randomizing galaxy orientations in our IA model.}
    \label{fig:mf_dist}
\end{figure}

Deviations from this simplistic model are accounted for by randomizing the galaxy orientations
in a subsequent step. Such a randomization has been shown to be an effective way to calibrate semi-analytic IA simulations
against observed alignment statistics \citep[e.g.][]{Heymans04, Okumura09b, Joachimi13b}.
In this work we randomize
the galaxy orientations in 3D before projection along the observers line of sight. This approach
allows for extracting constraints on the 3D galaxy alignment from calibrating the model against 2D observations.
In addition it opens up the possibility to calibrate the model against 3D alignment statistics measured at high redshifts
in hydrodynamic simulations in future studies.
For the randomization we draw misalignment angles $\theta$ from the Misis-Fisher distribution,
\begin{equation}
    P(\cos(\theta)) = \frac{1}{2 \sigma_{mf}^2 sinh(\sigma_{mf}^{-2})}exp\left( \frac{cos(\theta)}{\sigma_{mf}^2} \right),
    \label{eq:ms_dist}
\end{equation}
where the width $\sigma_{mf}$ is a free parameter of our model. Higher values of $\sigma_{mf}$ lead to a higher randomization
of the original orientation vector (see Fig. \ref{fig:mf_dist}) and therefore to a lower alignment signal.
\citet{Bett12} showed that in hydrodynamic simulations the distribution of misalignment
angles between the galaxy spin vector and the host halos minor axis is well approximated by Equation (\ref{eq:ms_dist}).
It has therefore been used by \citet{Joachimi13b} to model the 3D misalignment of the circular discs in their model.
In our model we use Equation (\ref{eq:ms_dist}) to 
model the 3D misalignment for all types of galaxies (including
discs, ellipticals, centrals and satellites) with respect to
their initial orientations.
Assuming that this modeling is valid not only for discs, but also for ellipticals, we are
able to successfully reproduce observed alignment statistics of LRG samples which consist mainly of ellipticals
(see Section \ref{sec:wgp_mice_lowz}). However, it would be worthwhile to validate this assumption
with measurements of galaxy-halo misalignment of ellipticals from
hydrodynamic simulations \citep[similar to those presented for instance by][]{Tenneti15, Chisari17, Bhowmick20}.

Since we model all galaxies as 3D ellipsoids which are in general rotationally asymmetric we need to
randomize their orientations in two directions. We thereby start by randomizing the orientation of the
minor and major axes $A_{3D}$ and $C_{3D}$, using
two misalignment angles $\theta_A$ and $\theta_C$, which are drawn from the Misis-Fisher distribution with the same
value of $\sigma_{mf}$ for both angles. The randomized orientation vectors
${\hat{\bf A}}_{3D}^r$ and ${\hat{\bf C}}_{3D}^{r,\prime}$ are constructed
such that
${\hat{\bf A}}_{3D} \cdot {\hat{\bf A}}_{3D}^r = \cos(\theta_A)$
and ${\hat{\bf C}}_{3D} \cdot {\hat{\bf C}}_{3D}^{r,\prime} = \cos(\theta_C)$.
${\hat{\bf C}}_{3D}^{r,\prime}$ is thereby a temporary vector which is in general not perpendicular to ${\hat{\bf A}}_{3D}^r$.
The final randomized minor axis orientation is therefore obtained as
${\bf C}_{3D}^r = ({\hat{\bf A}}_{3D}^{r} \times {\hat{\bf C}}_{3D}^{r,\prime}) \times {\hat{\bf A}}^r_{3D}$ and is then normalized
to ${\hat{\bf C}}_{3D}^r = {\bf C}_{3D}^r / |{\bf C}_{3D}^r|$.
In order to control the dependence of the alignment on galaxy magnitude and color, we introduce
simple dependencies of $\sigma_{mf}$ on these properties. We thereby assume a 
linear relation between $\sigma_{mf}$ and the absolute r-band magnitude $M_r$,
\begin{equation}
    \sigma_{mf} (M_r) = a + b \left(\frac{M_r}{M_0}-1\right),
    \label{eq:sigMr}
\end{equation}
where $a$ and $b$ are free model parameters and $M_0=-22$
is an arbitrarily chosen normalization constant.
The color dependence of the alignment is introduced in the model by using different values of $a$ and $b$ for
red and blue galaxies, where the colors are defined as described Section \ref{sec:color_cuts}.
When adjusting these parameters we further separate between central and satellite galaxies, which
provides control over the scale-dependence of the IA signal in the simulation.
The parameters used in our model are summarized in Table \ref{tab:IA_params}. They are
obtained from calibrating the model by hand, as outlined in the next subsection.
\begin{table}
    \centering
    \begin{tabular}{c|c|c}
         & a & b \\\hline
        red centrals & 0.65 & 0.0 \\
        red satellites & 0.7 & -7.7 \\
        blue centrals & 2.0 & 0.0 \\
        blue satellites & 2.0 & 0.0 \\
    \end{tabular}
    \caption{Parameters describing the magnitude dependence of the
    misalignment parameter $\sigma_{mf}$ in Equation (\ref{eq:sigMr}).}
    \label{tab:IA_params}
\end{table}
%
%
The final output of the IA model are the two 3D axis ratios $q_{3D}$ and $s_{3D}$
as well as the orientations of the three principle axes for each
galaxy in the simulation. In order to compare this output to observations
we project these ellipsoids along the observer's line of sight who is located
at $z=0.0$ in the MICE light-cone and obtain the intrinsic shear components,
as described in Section \ref{sec:model_shapes}.

\subsection{Parameter calibration against observed IA statistics}
\label{sec:wgp_mice_lowz}

We calibrate the parameters for controlling the randomization of galaxy orientations,
$a$ and $b$ in Equation (\ref{eq:sigMr}), for red and blue galaxies separately.
For blue galaxies, including centrals as well as satellites, we set $[a,b] = [2,0]$
such that $\sigma_{mf}=2$, independent of the galaxy magnitude. The randomized orientations
for blue galaxy are consequently close to a uniform distribution on a sphere (see Fig. \ref{fig:mf_dist}).
This choice is motivated by the non-detection of intrinsic alignment for blue galaxies 
in the surveys WiggleZ, SDSS, DES and PAUS \citep{Mandelbaum11, Samuroff19, Johnston19, Johnston21}.
However, achieving such a non-detection in the simulation may also be possible
with much lower values of $\sigma_{mf}$, since the halos' angular momentum
alignment is relatively weak compared to the alignment of the halos'
principle axes, as we show in Appendix \ref{app:halo_alignment}.

For red galaxies we adjust $a$ and $b$ such that the simulation reproduces the observed
scale and magnitude dependence of the alignment statistics, measured for LRGs in the BOSS LOWZ
sample by SM16. 
The alignment is thereby quantified with the projected cross-correlation between
positions of galaxies in a 'density' sample and the intrinsic shear of galaxies in a 'shape' sample, $w_{g+}$, as detailed in Section \ref{sec:2pt_shear_statistics}.

\begin{figure}
    \includegraphics[width=0.45\textwidth]{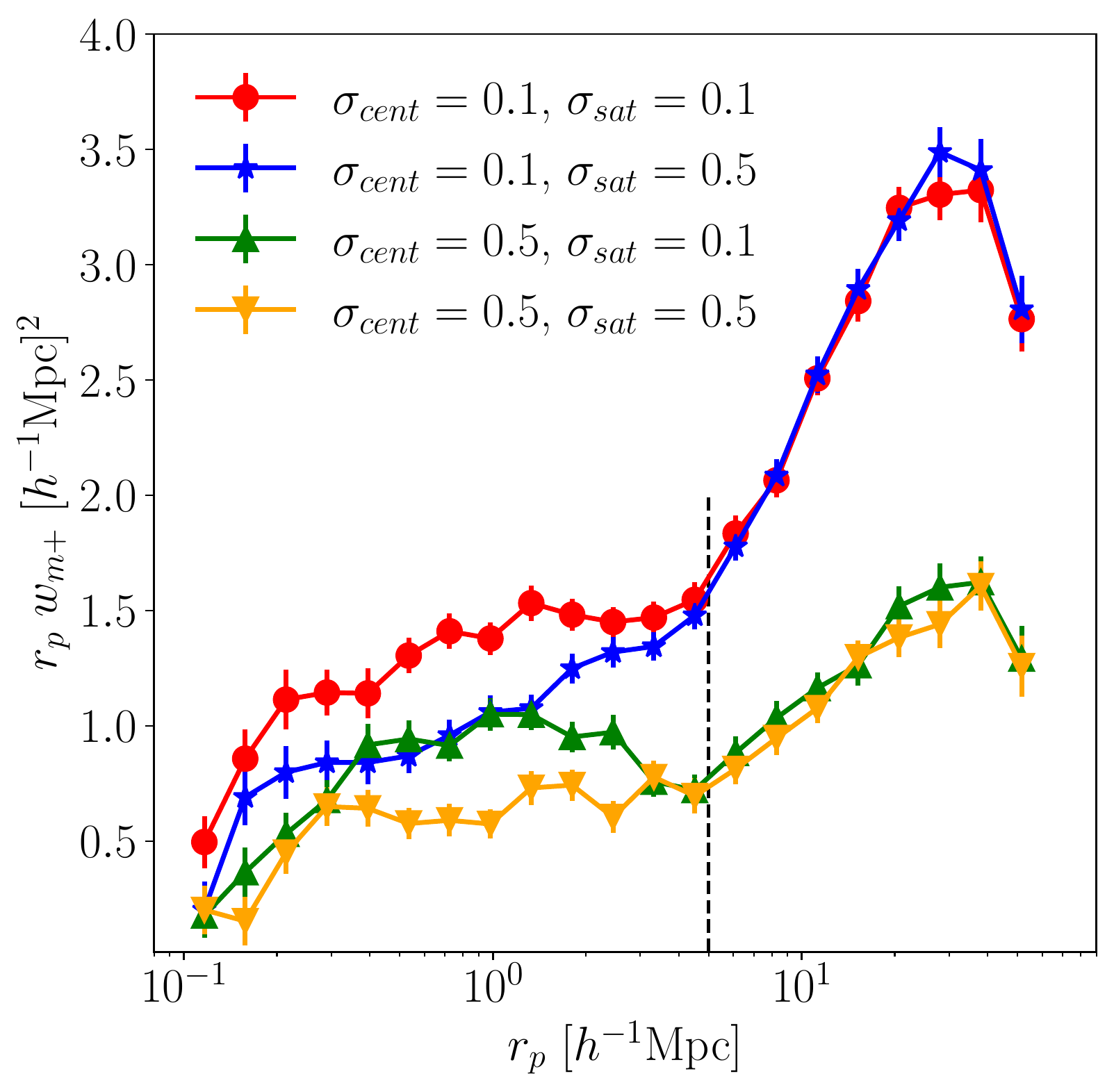}
    \caption{Projected matter - intrinsic shear correlation $w_{m+}$ for test runs of the semi-analytic IA model
    on a catalog of MICE galaxies with $0.1<z<0.3$ and $M_r<-21$, using one misalignment parameter
    for all centrals ($\sigma_{cent}$) and one for all satellites ($\sigma_{sat}$), independently from
    galaxy luminosity and color. Results are shown for different values of $\sigma_{cent}$ and $\sigma_{sat}$.
    Increasing $\sigma$ decreases the amplitude of $w_{m+}$. Misalignment of centrals affects the signal at all scales,
    while the impact of satellite misalignment is limited to $r_p \lesssim 5 \ h^{-1} \textrm{Mpc}$ (vertical dashed line).}
    \label{fig:wmp_fixed_sigma}
\end{figure}

Before discussing the calibration in more detail we show in Fig. \ref{fig:wmp_fixed_sigma} how the IA correlation
reacts to variations of $\sigma_{mf}$ for a test sample of galaxies that are brighter than $M_r=-21$ in the redshift range $0.1<z<0.3$.
When computing the IA correlation we use the matter distribution of the simulation as the density sample in order to minimize
noise on the measurement. The corresponding correlation is hereafter referred to as $w_{m+}$.
In this test case we set the same $\sigma_{mf}$ for galaxies of all luminosities and all colors.
We find in Fig. \ref{fig:wmp_fixed_sigma} that the overall amplitude decreases by roughly a constant factor when
increasing the misalignment by increasing $\sigma_{mf}$ for satellites as well as for centrals
from $0.1$ to $0.5$ (comparing red and yellow lines). When increasing only
the misalignment of satellites we find the signal to decrease only at small scales ($r_p < 5 \ h^{-1} \textrm{Mpc}$),
while it remains unaffected at large scales (comparing red and blue lines).
Increasing the misalignment only for centrals on the other hand has an effect
on all scales, while the impact is stronger on scales larger than $r_p > 5 \ h^{-1} \textrm{Mpc}$
(comparing red and green lines). In practice, the fact that satellite alignment does not affect 
the alignment statistics in the simulation at large scales simplifies the model calibration, as we can first
calibrate $\sigma_{mf}$ for centrals focusing on the large scales, before calibrating the parameters for satellites,
focusing on small scales.
%

In order to calibrate the parameters $a$ and $b$ in the $\sigma(M_r)$ relation from Equation (\ref{eq:sigMr})
we measure $w_{g+}$ in our mock LOWZ sample
(described in Section \ref{sec:data:mice:lowz}), where we take the full sample as density
sample and the four luminosity sub-samples $L1-L4$ as shape samples, following SM16
\footnote{
Note that SM16 showed that different shape measurement methods can lead to $\sim \sigma$ variations of
the observed signal, which introduces additional uncertainties in our modeling. In this work we calibrate
the simulation against their results based on re-gaussianized shapes \citep{Hirata03, Reyes12}.
Further note that SM16 apply cuts based on the quality galaxy shape measurements, which we cannot
mimic in the construction of mock catalogs from MICE.}

\begin{figure*}
    \includegraphics[width=1\textwidth]{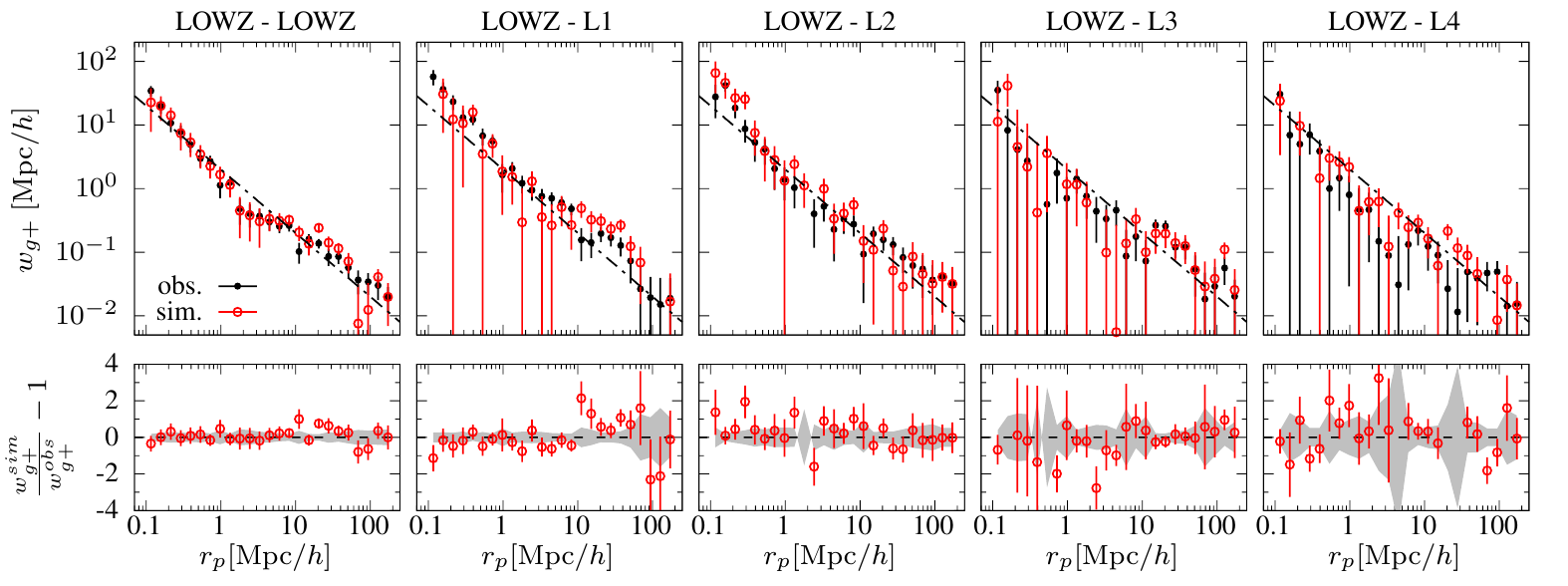}
    \caption{Similar to Fig. \ref{fig:wgg}, but for the projected galaxy-shear cross-correlation
    between a "density sample" and different "shape samples". The left panel shows results
    using the full LOWZ sample as shape as well as  density sample. The panels on the right
    show results based on shapes from the four luminosity sub-samples L1-L4, while using the full LOWZ sample as density sample.
    }
    \label{fig:wgp}
\end{figure*}

As a first step in the calibration we then set magnitude independent values of $\sigma_{mf}$ for centrals and
satellites for each LOWZ luminosity sample separately. These values are chosen such that
$\chi^2$ deviation between $w_{g+}$ measurements in MICE and the observational reference is minimized.
We thereby obtain a relation between $\sigma_{mf}$ and the mean $r$-band magnitude $\langle M_r \rangle$
of each sample, from which we can infer a first guess of the parameters $a$ and $b$.
Starting from this first guess we then vary $a$ and $b$ by hand until the simulation matches
the observed $w_{g+}$ measurements for the different luminosity samples $L1-L4$ simultaneously.
Note that this match is quantified purely by eye.
In future work we plan to improve the calibration technique, using quantitative measures for IA model
performance and an automated calibration pipeline.
Fig. \ref{fig:wgp} shows the comparison between $w_{g+}$ from the calibrated MICE
simulation together with the observational reference measurements from SM16.
The simulation reproduces the observed dependence of $w_{g+}$ on scale as well as on magnitude
as the deviations from the observations are consistent with the $1 \sigma$ jackknife error estimates.
The errors on the MICE results are overall larger than those on the observations,
which can be expected from the smaller area covered by the MICE octant. In addition, differences
in the errors can result from differences in the size and geometry of the jackknife samples.

When calibrating $\sigma_{mf}$ on the different LOWZ luminosity samples
we compensate automatically for systematic effects in the mass dependence of the host halo alignment
(see Appendix \ref{app:halo_alignment}), at least within the luminosity and redshift ranges
covered by the LOWZ sample. It is not obvious that this compensation also works
for magnitudes and redshifts that are not considered in the calibration. However, our
results in Section \ref{sec:sim_vs_theo} indicate that this might be the case, since 
the IA amplitudes predicted by MICE for luminosities and redshifts that are not covered
by the LOWZ sample are consistent with various observational constraints.

A shortcoming in our calibration based on $w_{g+}$ results from the fact that this
statistics is not only sensitive to the alignment, but also to the clustering of galaxies.
Since the clustering, quantified by $w_{gg}$ for the MICE LOWZ samples, is predicted 
to be $\lesssim 30\%$ below the reference observations from BOSS (Fig. \ref{fig:wgg}),
we are setting the IA signal in the simulation too high, when trying to match the observed $w_{g+}$
signal. However, we expect the error on the $w_{g+}$ amplitude to be significantly
smaller than $30\%$ based on the following consideration. At large scales
we can approximate $w_{gg} \propto b_1^2$ and $w_{g+} \propto b_1$,
where $b_1$ is the linear clustering bias. Assuming that the
difference in $w_{gg}$ between MICE and BOSS is mainly driven by differences in
$b_1$, a $30\%$ inaccuracy in $b_1^2$ would propagate into a $17\%$ inaccuracy on $b_1$
and hence on $w_{g+}$. This inaccuracy is well below the dependence of $w_{m+}$ on luminosity,
color and redshift, which we will study later on.

Another potential source of bias in our calibration may result from the fact that
the galaxy shapes in our simulation are calibrated against observed axis ratio distributions
that were derived from \sersic\ model fits (Section \ref{sec:data:cosmos}). Using a reference
distribution based on a different shape measurement method may
change the galaxy ellipticity distribution and hence lead to a change in $w_{g+}$ (e.g. SM16).
However, since in our simulation the orientations and shapes are calibrated independently,
a bias in the ellipticities can be compensated by adjusting the galaxy misalignment, such that
$w_{g+}$ still matches the observational constraints.

\subsection{Distribution of misalignment angles}
\label{sec:gh_misalignment}

The distribution of misalignment angles between galaxies and their host halos has been investigated
in several previous studies. It therefore provides an opportunity to validate our simulation
in a way that is independent of the $w_{g+}$ comparison against the BOSS LOWZ constraints, used
for the calibration of the simulation parameters.
Observational constraints on the distribution of misalignment angles of LRGs have been derived by
\citet{Okumura09a} and \citet{Okumura09b} (jointly referred to as OO9 in the following). Using a methodology
similar to the one presented in this work, these authors randomized the orientations of dark
matter halos from an N-body simulation such that the simulation reproduces the observed alignment statistics
of LRGs in the SDSS. In contrast to our approach of randomizing galaxy orientations in 3D
before projection, OO9 randomized the 2D orientations after projection, assuming a Gaussian
distribution of misalignment angles with zero mean and a variance $\sigma_{\phi}$.
In order to compare their results to predictions for the LRGs in the mock BOSS LOWZ sample from MICE, we compute the
2D misalignment angles as the difference between the 2D orientation angles before and after
randomizing the galaxies in the simulation. We find the variance of the distribution of 2D misalignment
angles in the LOWZ sample to be $\sigma_{\phi}=32.64^{\circ}$, which deviates by just $\sim7\%$ from the $\sim35^{\circ}$ degree variance reported by OO9.
This finding is interesting, given that LRGs in SDSS and
those in the BOSS LOWZ sample probe different ranges in color, luminosity and redshift. Furthermore, the simulations
employed to interpret the observations are based on N-body simulations which differ in their resolution and cosmology,
the definition of halo shapes and orientations as well as in the HOD model and the IA model used to produce
the mock catalogs that are compared to the observations.

\begin{figure}
    \includegraphics[width=0.45\textwidth]{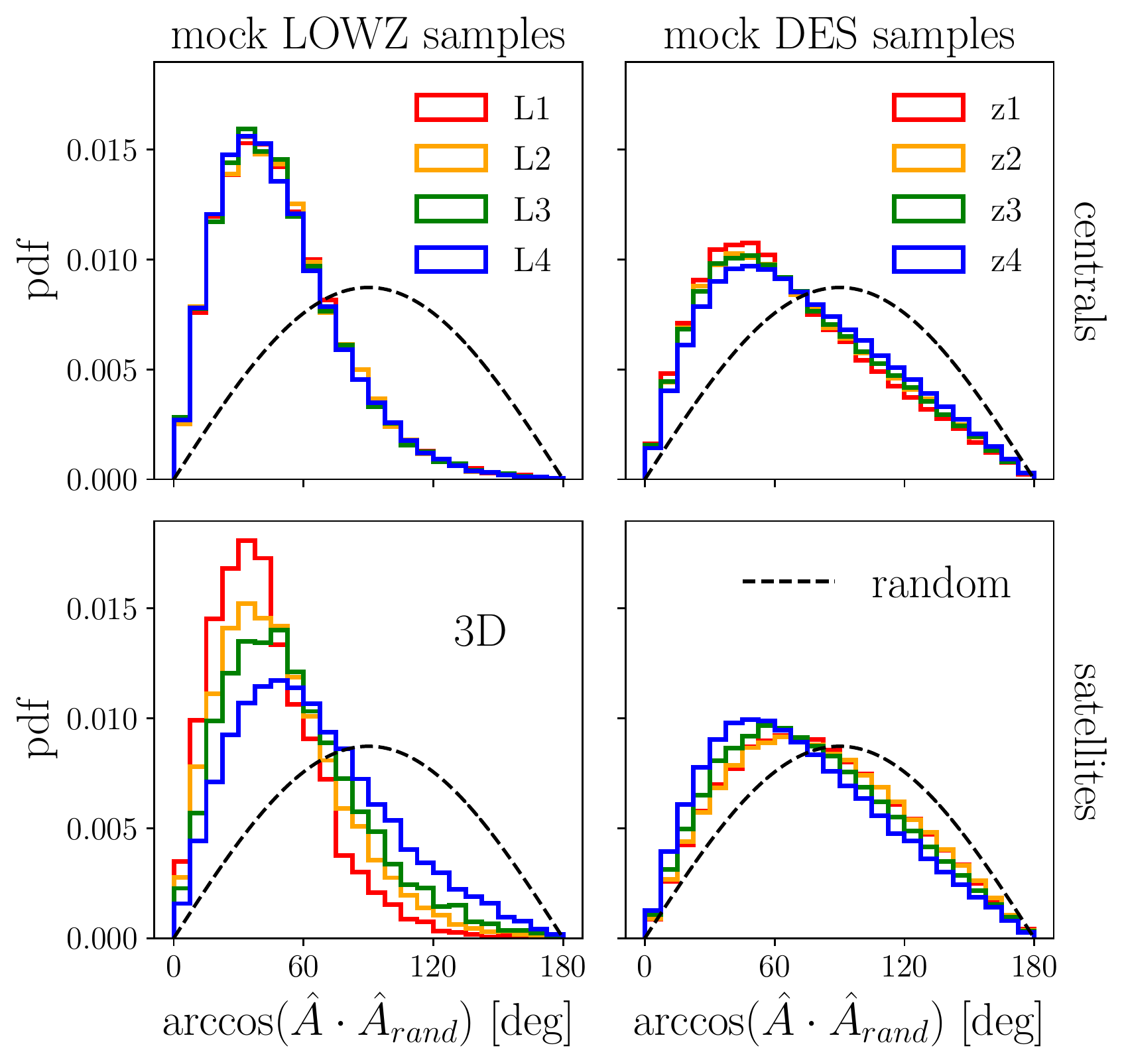}
    \caption{Distribution of misalignment angles between the 3D major axis $A$ of galaxies in MICE before and after randomization. Results are shown for the different luminosity and redshift sub-samples in the mock BOSS LOWZ and DES catalogs respectively.  
    The black dashed line shows the theoretical expectation for completely randomized orientations.
    }
    \label{fig:alpha_rand_3d}
\end{figure}

In addition to the constraints on the 2D misalignment, the MICE simulation provides predictions for the distribution
of 3D misalignment angles. In Fig. \ref{fig:alpha_rand_3d} we show the distribution of these misalignment angles, defined
as the angle between the 3D major axes ${\bf{A}}$ before and after randomization for the mock samples of the BOSS LOWZ and the DES surveys.
The results for the LOWZ sample demonstrate that our IA model implementation for red galaxies works as expected.
In our model the misalignment of red centrals is independent of the galaxy magnitude (Table \ref{tab:IA_params})
which leads to almost identical distributions of misalignment angles for the different luminosity sub-samples.
The misalignment angles for satellites are increasing significantly
for dimmer samples, as expected from the modeling. Similar trends can be seen for the mock DES samples, although less clearly
since these samples consist to $\sim 50\%$ of blue galaxies (Table \ref{tab:fblue_des}), which are almost completely randomized in our model.
The distributions of misalignment angles in the DES samples lie therefore closer to a distribution expected for completely
randomly oriented objects (shown as dashed line in Fig. \ref{fig:alpha_rand_3d}) than the distributions of misalignment angles
in the LOWZ samples.

It is further interesting to note here that the 3D misalignment angles that we obtain from calibrating the IA model in MICE
are significantly higher than those found for the galaxy-halo misalignment in the MassiveBlack-II simulation
by \citet{Tenneti15}. These authors report an average misalignment of $\sim 13$ degree for galaxies
in halos with masses larger than $10^{13} h^{-1}\textrm{M}_\odot$. This prediction is significantly below the values that
we find in the MICE LOWZ samples, which reside mostly in halos of that mass range. One potential explanation could be that
these authors study the alignment of galaxies with respect to their host subhalo, whereas our results refer to the
alignment of galaxies with respect to their host FOF group, which may present a weaker alignment with the central galaxies.

\section{Predictions for two-point IA statistics}
\label{sec:sim_vs_theo}
After having validated that MICE is consistent with observational IA constraints
from LRGs in SDSS and BOSS, we now proceed by using the simulation to derive predictions
at redshift and luminosity ranges that are not covered by these surveys. We are thereby
interested in the following three questions.
1) How well do the analytical IA models NLA and TATT
fit the IA statistics measured in MICE at the redshifts covered
by current photometric weak lensing surveys, such as DES?
2) How do the parameters of these models depend on galaxy color, luminosity, and redshift,
and how well do these dependencies agree with observational constraints from surveys other than
BOSS, to which the simulation has not been calibrated?
3) How strong is the IA contribution to the observed shear statistics predicted by the simulation
in mock DES observations? We address these questions in the following.

\subsection{Dependence of \texorpdfstring{$w_{m+}$}{wm+} on galaxy magnitude, color and redshift}
\label{sec:wmp_mz_dependence}

In order to test the accuracy of the NLA and TATT models we fit corresponding predictions
for the projected matter-intrinsic shear cross-correlation, $w_{m+}$
(introduced in Section \ref{sec:projcorr}), to the measurements in MICE.
Note that using matter instead of galaxies as the density field facilitates the interpretation
of our results, as we do not need to take into account inaccuracies of galaxy clustering bias models. 
However, before discussing these fits we would like to point out some interesting aspects
of the $w_{m+}$ measurements themselves.
In Fig. \ref{fig:wmp_zmc_samples} we show these measurements for the $36$ volume
limited color samples described in Section \ref{sec:data:zmc_samples}.
We find that the measurements for samples of blue galaxies are consistent with zero, which confirms
that the galaxy-halo misalignment set for these galaxies in the simulation (see Table \ref{tab:IA_params})
is high enough to eradicate a statistically significant signal at all scales, magnitude and redshift ranges
covered in our analysis.
For the red galaxy samples the measured signal is clearly present, showing dependencies on
scale, magnitude and redshift.
At a given redshift the amplitude of $w_{m+}$ increases with the brightness of the sample.
At small scales ($r_p \lesssim 5 \ h^{-1}\textrm{Mpc}$), such an increase can be expected from
our IA model, since the misalignment of red satellites is set to decrease for brighter magnitudes
by the corresponding parameters in Table \ref{tab:IA_params}.
At large scales ($r_p \gtrsim 5 \ h^{-1}\textrm{Mpc}$) the $w_{m+}$ alignment signal
is dominated by central galaxies (see Fig. \ref{fig:wmp_fixed_sigma}) for which the
galaxy-halo misalignment is set to be independent of the galaxy magnitude. The 
luminosity dependence of $w_{m+}$ at large scales is hence induced by a change of the
host halo alignment. According to the SHAM technique employed for the production of
the MICE galaxy catalog, the brightness of central galaxies increases with the mass
of their host halos (see Section \ref{section:mice_intro}).
The increase of the alignment of central galaxies with luminosity is therefore
induced by an increase of the host halo alignment with halo mass, which we
study in Appendix \ref{app:halo_alignment} \citep[see also][]{Piras18}.

In addition to the magnitude dependence, we find in Fig. \ref{fig:wmp_zmc_samples}
a decrease of the $w_{m+}$ amplitude with redshift for red galaxies within a fixed magnitude
range. This redshift dependence is most clearly apparent at large scales. Since our model
does not include a redshift dependence of the galaxy-halo misalignment, the decrease of
$w_{m+}$ with redshift at large scales is presumably induced by the decrease of the host halo
alignment with redshift , which we find in Appendix \ref{app:halo_alignment}.
Furthermore one could expect the $w_{m+}$ signal to decrease, even if halo alignment was redshift independent,
due to the decrease of the matter power spectrum amplitude with redshift. We conclude that the
interpretation of the luminosity and redshift dependence of IA statistics in terms of galaxy-halo
misalignment relies on a detailed understanding of the mass and redshift dependence of halo
alignment which we investigate in Appendix \ref{app:halo_alignment}.

For the comparison between the theory predictions for $w_{m+}$ and the corresponding measurements,
which we discuss in the next section, it is interesting to inspect how strongly measurements on
different scales are correlated with each other, which is described by the covariance $C_{ij}$.
In Fig. \ref{fig:cov_wmp} we show examples of the normalized covariance for four of our $36$ volume
limited samples. We find that the covariances are dominated by the diagonal elements, indicating
that the errors on $w_{m+}$ are dominated by noise that originates from the dispersion of intrinsic
galaxy ellipticities, which are spatially uncorrelated.

\subsection{NLA and TATT fits to \texorpdfstring{$w_{m+}$}{wm+} measurement}
\label{sec:fitting}

In order to examine the accuracy of the NLA and TATT models we fit the corresponding
predictions for $w_{m+}$ (Section \ref{sec:wmp_predictions}) to the
measurements in MICE (hereafter referred to as the data vector {\bf d}) by maximizing the
posterior probability $P(\boldsymbol{\theta}|{\bf d})$ for the parameter vector
$\boldsymbol{\theta}$, given ${\bf d}$, where
$\boldsymbol{\theta}$
is given by $A_1$ and $(A_1, A_2, A_{1\delta})$ in the case of the NLA and the TATT model respectively.
$P(\boldsymbol{\theta}|{\bf d})$ is inferred from the likelihood
$\mathcal{L}({\bf d}|\boldsymbol{\theta})$ of measuring {\bf d} given $\boldsymbol{\theta}$,
using Bayes' theorem.
We estimate the likelihood from the data, assuming that it is well
described by a multivariate normal distribution, i.e.
\eq{
\ln \mathcal{L}({\bf d}|\boldsymbol{\theta}) =
-\frac{1}{2}\chi^2({\bf d}|\boldsymbol{\theta}) + const.
\label{eq:lnL}
}
with
\eq{
\chi^2({\bf d}|\boldsymbol{\theta}) = [{\bf d} - {\bf m(\boldsymbol{\theta})}]^{\mathsf T}C^{-1}[{\bf d} - {\bf m(\boldsymbol{\theta})}].
\label{eq:chisq}
}
The model $m(\boldsymbol{\theta})$ is the NLA or TATT prediction for $w_{m+}$ from Equation (\ref{eq:w_theo})
for a given $\boldsymbol{\theta}$.
The covariance $C$ is estimated from measurements of $w_{m+}$ in jackknife samples
as detailed in Section \ref{sec:projcorr}.
The posterior is given by Bayes' theorem as
\eq{
P(\boldsymbol{\theta}|{\bf d}) \propto \mathcal{L}({\bf d}|\boldsymbol{\theta})\Pi(\boldsymbol{\theta}).
\label{eq:def_likelihood}
}
In our analysis we set the prior $\Pi(\boldsymbol{\theta})$ flat to unity in the interval $[-25,25]$
and zero elsewhere for all parameters, covering the range of parameter values expected
from observations with a high margin.
We estimate $P(\boldsymbol{\theta}|{\bf d})$ by sampling the parameter space with
the Markov-Chain-Monte-Carlo algorithm \verb|emcee|
(introduced in Section \ref{sec:shape_model_calibration}).
For each posterior we run $16$ chains with $300$ steps each. The best fit parameters are defined
as the sampling point with the highest posterior probability. The confidence intervals for each
parameter are derived from the corresponding marginalized posterior distribution.

We fit the NLA as well as the TATT model up to scales of $r_p<60 \ h^{-1}{\textrm{Mpc}}$,
which corresponds to the projection length $\Pi$, used for the $w_{m+}$  measurements.
\citet{Blazek15} pointed out that the Limber approximation that enters the $w_{m+}$ prediction
requires $r_p<<\Pi$ to be valid. However, we do not
find a significant change in our parameter constraints when reducing the upper limit to
$r_p<30 \ h^{-1}{\textrm{Mpc}}$ (see Appendix \ref{app:tatt_params_scale}), presumably because
the likelihood is dominated by small scales measurements, as the measurement errors increase with scale.
The lower scale limit of the fitting range is set to $1$ and $8 \ h^{-1} {\textrm{Mpc}}$ for the
TATT and the NLA model respectively. The choice of these lower limits is motivated in
Appendix \ref{app:tatt_params_scale}. Since we limit the NLA fits to large scales
at which satellite galaxy alignment does not affect the $w_{m+}$ amplitude measured in MICE
(Fig. \ref{fig:wmp_fixed_sigma}), we expect the $A_1$ constrains from the NLA fits to be set
solely by the large-scale alignment of central galaxies in the simulation.

\begin{figure*}
    \centering
    \includegraphics[width=0.8\textwidth]{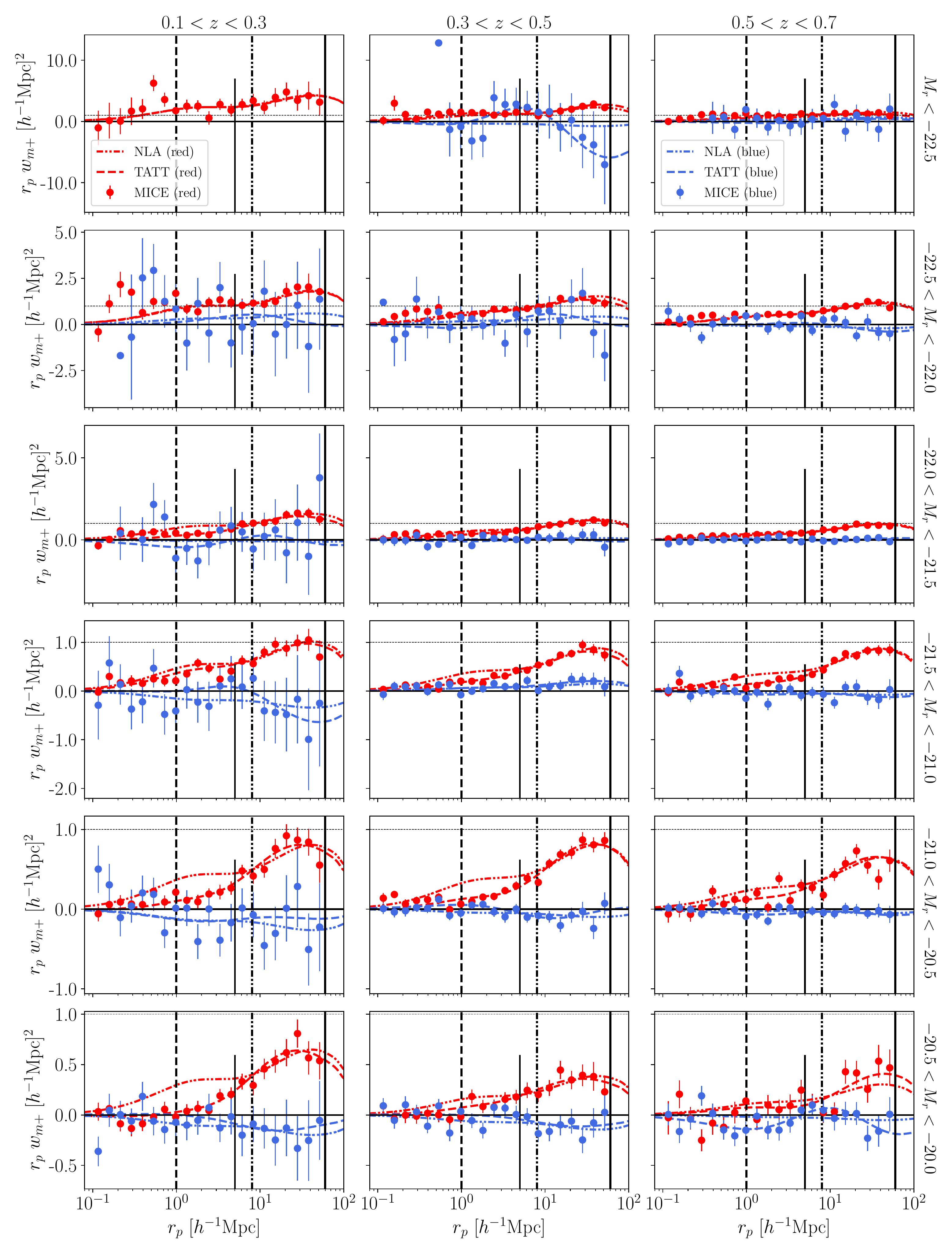}
    \caption{
    Projected matter-intrinsic shear two-point correlation functions
    measured in the $18$ volume limited samples shown in Fig. \ref{fig:mice_zmc_samples}.
    Each sample's range in redshift and absolute $r$-band magnitude is 
    indicated on the top and right respectively. Red and blue symbols show measurements in the MICE simulation
    for red and blue galaxies. Dashed-dotted and dashed lines show fits of the NLA and the TATT model respectively,
    where the line color is matched to the corresponding color sample.
    The lower limits of the fitting ranges are shown for each model as vertical black lines
    in the corresponding line-style. The upper limit is indicated by a vertical black solid line.
    The dotted horizontal line at $r_p \ w_{m+}=1.0$ facilitates the visual comparison of the
    amplitudes in different samples.
    Error bars indicate $1 \sigma$ uncertainties, which can be smaller than the symbol size.
    The significance of deviations between fits and measurements is shown for various fitting ranges
    in Fig. \ref{fig:delta_wmp_zmc_samples}. Note that results for blue galaxies with $M_r<-22.5$ and
    $0.1<z<0.3$ (top left panel) are not shown because the low number of objects did not allow for meaningful
    measurements. The absence of the small scale signal in the same
    magnitude range at higher redshifts results from a lack of bright blue satellites in these samples (see Fig. \ref{fig:mice_zmc_samples}).
    The short vertical solid lines at $r_p = 5 \ h^{-1} \textrm{Mpc}$ indicate the scale above which the
    correlation function is dominated by central galaxies (see Fig. \ref{fig:wmp_fixed_sigma}).
    }
    \label{fig:wmp_zmc_samples}
\end{figure*}

We compare the best fits of the NLA and the TATT predictions against $w_{m+}$ measurements
in Fig. \ref{fig:wmp_zmc_samples} and find that both models fit the data with similar
$\sim 1\sigma$ accuracy at scales above $8 \ h^{-1}{\textrm{Mpc}}$.
At smaller scales the fit of the TATT model stays within the $1 \ \sigma$ uncertainties of the data down
to the lower limit of the fitting range of $1 \ h^{-1}{\textrm{Mpc}}$. The NLA model tends to
lie above the measurements below $8 \ h^{-1}{\textrm{Mpc}}$, which is also the case when
reducing the lower limit of the NLA fitting range (see Appendix \ref{app:tatt_params_scale}).
These results indicate that the TATT model provides accurate predictions of galaxy
alignment statistics over a wide range of scales, redshifts and luminosities.

\begin{figure}
    \centering
    \includegraphics[width=0.45\textwidth]{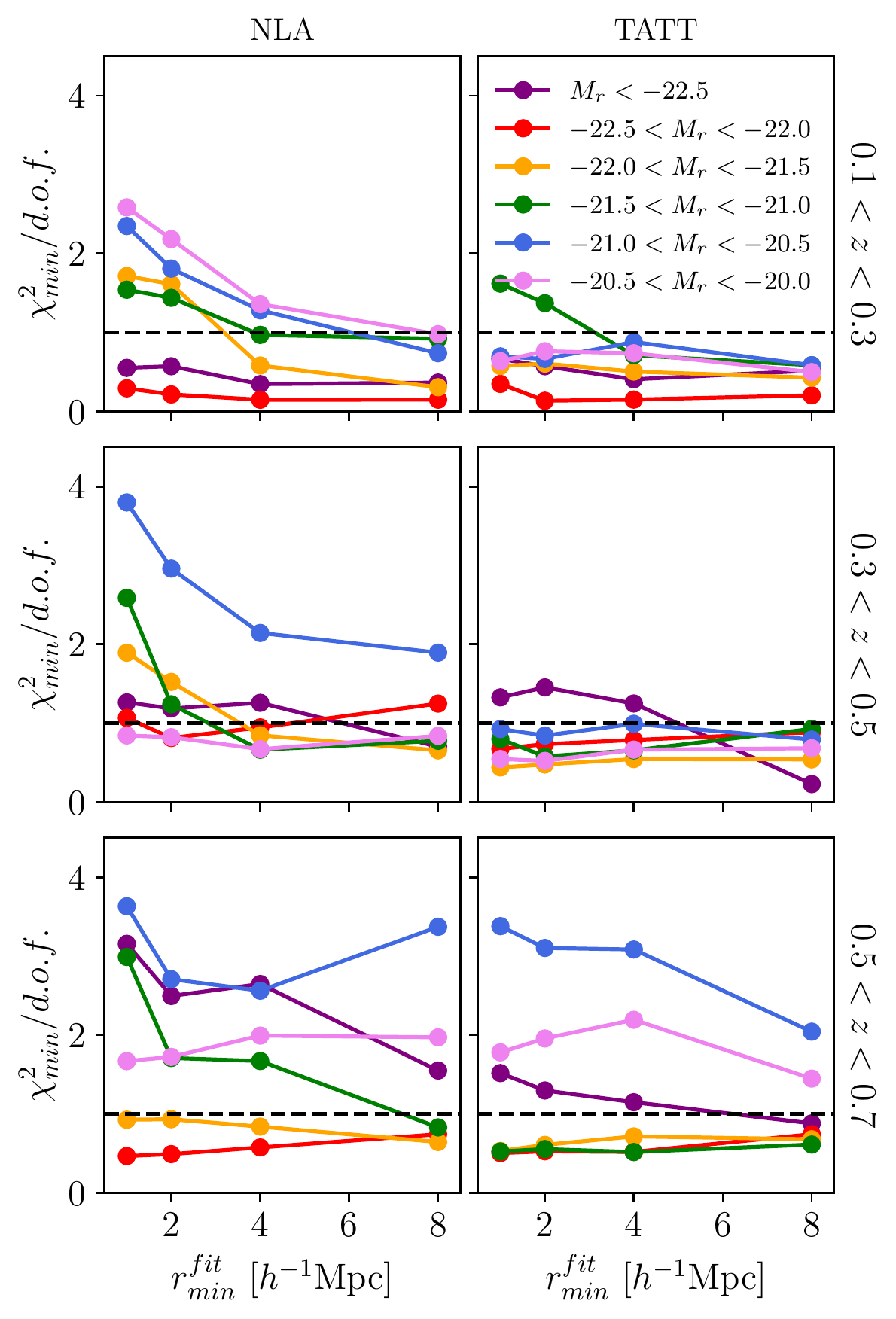}
    \caption{$\chi^2$ deviation between the best fits of the NLA and TATT model predictions for $w_{m+}$
    and the corresponding measurements (as shown in Fig. \ref{fig:wmp_zmc_samples}), versus the smallest
    scale of the fitting range. Results are shown for the $18$ volume limited samples of red galaxies in MICE.}
    \label{fig:chisq_rmin}
\end{figure}

We assess the fitting performance of the IA model predictions for $w_{m+}$ in a more quantitative way
in Fig. \ref{fig:chisq_rmin}, where we show the minimum $\chi^2$ deviation between
measurements and predictions (corresponding to the maximum of the likelihood)
per degrees of freedom versus the smallest scale used in the fit.
The degrees of freedom are given by $d.o.f. = n-m$, where $n$ is the number of $w_{m+}$ bins
within the fitting range and $m$ is the number of model parameters ($m=1$, $3$ for
NLA, TATT respectively).
The figure confirms the TATT model predictions fit the $w_{m+}$ measurements better than those
based on the NLA model as the $\chi^2/d.o.f.$ values tend to be lower, in particular at small scales.
We further find the fits to perform better at lower redshifts, which could result from larger errors
on the measurements at low redshifts, due to the smaller volume of the light-cone
(see Fig. \ref{fig:wmp_zmc_samples}). Interestingly we
do not find a clear dependence of the fitting performance on the magnitude range probed by the 
different samples and hence on the amplitude of $w_{m+}$. This finding might as well be
related to the fact that the different samples have distinct errors, even if they cover the same redshift bin,
due to their different number densities and noise properties. Inspecting the absolute $\chi_{min}^2/d.o.f.$ values
shown in Fig. \ref{fig:chisq_rmin}, we notice that various results are well below unity.
This finding could indicate overfitting of the data by the models. However, the fact that these low values
are also present for the NLA model (which has only one free parameter) as well as over a wide range of scales
(i.e. different numbers of scale bins $m$) could be an indication for shortcomings in the $\chi_{min}$ estimation.
Such shortcomings may result from the approximations implied in the jackknife method for estimating the covariance,
leading, for instance, to an overestimation of the variance.
Note that the noise in the covariance estimates, which we see in the off-diagonal elements in the examples
shown in Fig. \ref{fig:cov_wmp}, is expected to reduce the amplitude of the inverse covariance
by $\sim 25 \%$ for the number of bins and samples used on this analysis \citep{Hartlap07}. Correcting for
this effect of noise would further reduce the $\chi^2$ values by the same factor.
When assessing the fitting performance of the IA model predictions, one
must further bear in mind that the reference data from MICE is based on a simple model for IA, which provides a good match with the alignment signal of $w_{g+}$ in BOSS LOWZ,
but so far has not been validated against corresponding constraints from other
surveys covering different ranges in redshift, luminosity and color ranges.

\subsection{Dependence of IA model parameters on galaxy magnitude, color and redshift}
\label{sec:IA_params_zmc_dependence}
We proceed by investigating the dependence of the NLA and TATT model parameters
on galaxy redshift, luminosity and color by comparing the posterior
distributions that we derived from the $w_{m+}$ fits in the different volume limited
samples from MICE.

\begin{figure}
    \centering
    \includegraphics[width=0.45\textwidth]{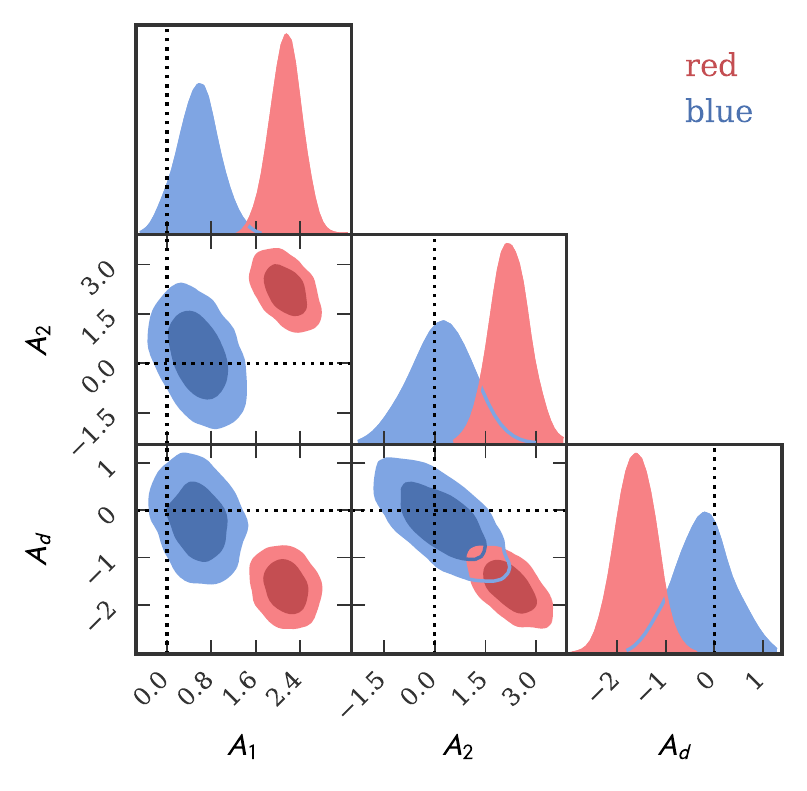}
    \caption{Posterior probability distribution of the TATT model parameters,
    derived from fits to the $w_{m+}$ measurements from red and blue galaxies 
    in a volume limited sample,
    selected by $0.3<z<0.5$ and $-21.5<M_r<-21.0$ (shown in Fig. \ref{fig:wmp_zmc_samples}).
    }
    \label{fig:posterior_tatt}
\end{figure}

\begin{figure*}
    \centering
    \includegraphics[width=0.95\textwidth]{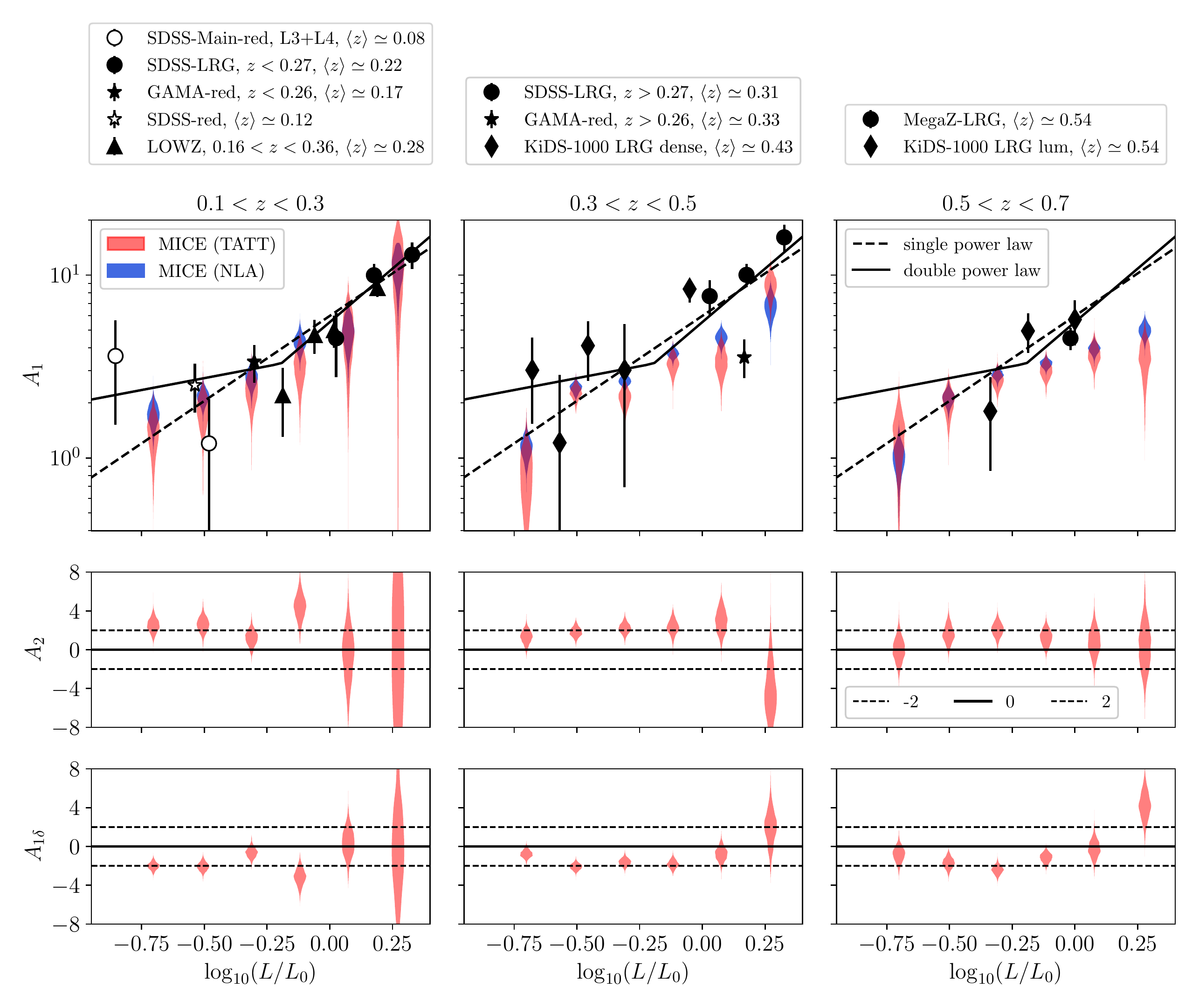}
    \caption{Marginalized posterior distributions of the NLA and TATT model parameters, derived
    from fits to $w_{m+}$ measurements for red galaxies in different volume limited samples of the MICE simulation,
    shown in Fig. \ref{fig:wmp_zmc_samples}.
    The posteriors are displayed at each samples logarithmic mean $r$-band luminosity, normalized by a pivot
    luminosity $L_0$ that corresponds to $M_r=-22$. The top panel shows results for the $A_1$ parameter,
    derived from fits of the NLA and the TATT model together with constraints from fits of the NLA model
    to different $w_{g+}$ measurements in observational samples of red galaxies, provided in the literature
    (circles: \citet{Joachimi11}, stars: \citet{Johnston19}, triangles: \citet{Singh15}, diamonds: \citet{fortuna21b}). Dashed and solid lines show power law fits to the observed data from 
    \citet{fortuna21b}.
    The central and bottom panels show results for the parameters
    $A_2$ and $A_{1\delta}$ from the TATT model. The horizontal lines at $\pm2$ facilitate the
    comparison of the parameter amplitudes by eye.
    }
    \label{fig:loglum_a1a2a3}
\end{figure*}

We start by inspecting the joint posterior distributions for the
TATT parameters ($A_1$, $A_2$, $A_{1\delta}$), derived from red and blue sub-samples 
at intermediate redshifts and magnitudes ($0.3 < z < 0.5$, $-21.5 < M_r < -21.0$)
in Fig. \ref{fig:posterior_tatt}.
We find the constraints on the parameters for the blue sub-sample to be consistent with 
$(A_1$, $A_2$, $A_{1\delta}) = (0,0,0)$ at the $1\sigma$ level. This is expected from the
null detection of the $w_{m+}$ (see Fig. \ref{fig:wmp_zmc_samples})
which results from the highly randomized orientations of blue galaxies in MICE.
Since we find similar results for all blue sub-samples we focus in the following
discussion on the parameter constraints from samples of red galaxies.
The parameters for the red sub-sample, shown in Fig. \ref{fig:posterior_tatt}, differ significantly
from zero which lines up with the significant signal of the corresponding $w_{m+}$ measurements.
We further find the parameter $A_2$ to correlate weakly with $A_1$ and slightly stronger with $A_{1\delta}$,
while $A_1$ and $A_{1\delta}$ appear to be uncorrelated. The joint constraints on the TATT parameters from
the other volume limited samples (not shown) exhibit a similar behaviour.
The dependence of the NLA and TATT parameter constraints on luminosity is shown for the red sub-samples in
three redshift bins in Fig. \ref{fig:loglum_a1a2a3}. In that figure we
display the marginalized posterior distributions as violins at the logarithm of
each samples luminosity $L$, normalized by a pivot luminosity $L_0$,
i.e. $log_{10}(L/L_0) = (\langle M_r \rangle - M_0)/(-2.5)$,
where $\langle M_r \rangle$ is each sample's mean SDSS $r$-band magnitude and $M_0 = -22$,
according to literature conventions.
We find that the marginalized posteriors of the $A_1$ parameter in the TATT model are mostly consistent with
those from the NLA model, shown as red and blue violins in the top panels of Fig. \ref{fig:loglum_a1a2a3}
respectively.
However, for certain samples, in particular in the central redshift bin
($0.3<z<0.5$), deviations between the NLA and TATT constraints are significant, while the general trends
of how $A_1$ changes with redshift and luminosity are the same for both models. These deviations
remain significant when varying the scale range over which the TATT model is fitted to $w_{g+}$
(see Fig. \ref{fig:loglum_a1a2a3_scale_cuts}). We further note that the $A_1$ constraints are tighter
for the NLA than for the TATT model, which can be expected from the higher number of free parameters in TATT.

The $A_1$ constraints from MICE are compared to observational constraints from various
samples of red galaxies from different spectroscopic surveys, that were presented in the literature
(shown as black symbols in Fig. \ref{fig:loglum_a1a2a3}). These observational constraints have been
derived analogously to those from MICE from fits of the NLA model to $w_{g+}$ measurements.
Circles show results from \citet{Joachimi11} for two luminosity samples
(L3 and L4) of dim red galaxies in the SDSS-Main sample, two redshift samples of SDSS LRGs (with $z\lessgtr0.27$)
and the MegaZ-LRG sample. Triangles show $A_1$ constraints from \citet{Singh15} for four luminosity
sub-samples of the BOSS LOWZ sample \footnote{These sub-samples are similar, although somewhat brighter than
the luminosity sub-samples from \citet{Singh16} to which we calibrate the MICE simulation}. Stars show results
from \citet{Johnston19} for red galaxies in the SDSS main sample and the high and low redshift sub-samples
from the combined KiDS+GAMA survey. Diamonds show results from \citet{fortuna21b} from the KiDS survey
for dim and bright sub-samples (denoted by the authors as 'dense' and 'luminous' respectively),
that cover different ranges of redshifts. Note that the observational results are displayed in Fig. \ref{fig:loglum_a1a2a3}
across the three redshift bins in which we analyzed the MICE simulation, according to each samples mean redshift
$\langle z \rangle$.

We find in Fig. \ref{fig:loglum_a1a2a3} that most observational constraints on $A_1$ are consistent with those
derived from the volume limited samples in MICE within the estimated errors. This finding is remarkable, given that the MICE IA model has been calibrated
only against constraints from LRGs in the BOSS LOWZ sample. Predictions beyond the color-magnitude-redshift
range covered by the LOWZ sample rely on the simple assumptions of the IA model. Furthermore, the
observational constraints are based on samples of
red galaxies from various surveys and have been selected
with different cuts on color, magnitude and redshift.
These differences in the selection may
contribute to the deviations between observations and simulation as well as to the variation across the
observational constraints. A more meaningful comparison between observations and simulations would require
the construction of mock catalogs, which is beyond the scope of this work.

In addition to the observational $A_1$ constraints from separate surveys, we compare in Fig. \ref{fig:loglum_a1a2a3}
the MICE results with two power laws fitted to the combined observational constraints on $A_1$, provided by \citet{fortuna21b}. We find that the MICE results
roughly follow the single power
law fit in the lowest redshift bin over the full luminosity range. At higher redshifts
the MICE constraints are still consistent with the same single power law at low luminosities ($log_{10}(L/L_0)<-0.25$),
but decrease with redshift for brighter sub-samples. This behaviour is apparent in the $A_1$ constraints from
the TATT as well as from the NLA model. Since the latter was fitted at ($r_p > 8 \ h^{-1} {\textrm{Mpc}}$),
we attribute the magnitude and redshift dependence of $A_1$ to be set by the large-scale amplitude of $w_{m+}$.
We argued previously that the
large-scale alignment signal is dominated by the alignment of central galaxies (Fig. \ref{fig:wmp_fixed_sigma}).
Given that the misalignment between central galaxies and their host halos is modeled
independently of redshift and luminosity, we expect the dependence of $A_1$ on these
quantities to be driven by the redshift and mass dependence of the host halo alignment,
which we discuss in Appendix \ref{app:halo_alignment}.
Due to a lack of observational constraints on $A_1$ for luminous, high redshift samples, it remains an open question
if the redshift evolution of $A_1$ for red galaxies as predicted by MICE is supported by observation.
However, our results line up with predictions from the Horizon AGN hydrodynamic simulation, according to which
the alignment between massive elliptical galaxies and their surrounding tidal field decreases with redshift \citep{Bate2020}.

The central and bottom panels of Fig. \ref{fig:loglum_a1a2a3} show the marginalized posteriors for
the parameters $A_2$ and $A_{1\delta}$ of the TATT model. For dim samples ($L<L_0$) we find both parameters
to be roughly constant, taking values of $A_2 \sim 2$ and $A_{1\delta} \sim -2$. When approaching brighter
luminosities the parameters switch their sign and reach higher amplitudes for the samples at $z>0.3$, while
constraints at lower redshifts are too noisy to reveal any trend.  We further do not find a clear
dependence of the $A_2$ and $A_{1\delta}$ constraints on redshift. Note here that $A_{1\delta}$ in our modeling
depends not only on galaxy alignment but also on galaxy clustering, as detailed in Section
\ref{sec:2pt_shear_statistics}.

\section{Application to DES Y3}
\label{sec:des_aplication}

After having investigated the IA signal predicted by MICE in volume limited samples,
we now study how strongly IA contaminates the lensing signal in the simulated DES-like samples constructed from MICE (see Section \ref{sec:data:des_samples}). This predicted contamination was used in the DES Y3 cosmic shear analysis
to show that both the NLA and TATT models  recover the input cosmology in MICE within the 1$\sigma$ uncertainties of parameter posteriors. This process provides a valuable test of the modeling, since the simulated data is not analytically generated with either model. Note that this investigation becomes
possible due to the large area and redshift range covered by the MICE simulation without repetition, while including both
the lensing and the IA signal, allowing for separate measurements of the GG, GI, IG and II terms introduced
in Section \ref{sec:2pt_shear_statistics}.

\subsection{Projected matter-intrinsic shear correlation (\texorpdfstring{$w_{m+}$}{wm+})}

\begin{figure}
    \centering
	\includegraphics[width=0.45\textwidth]{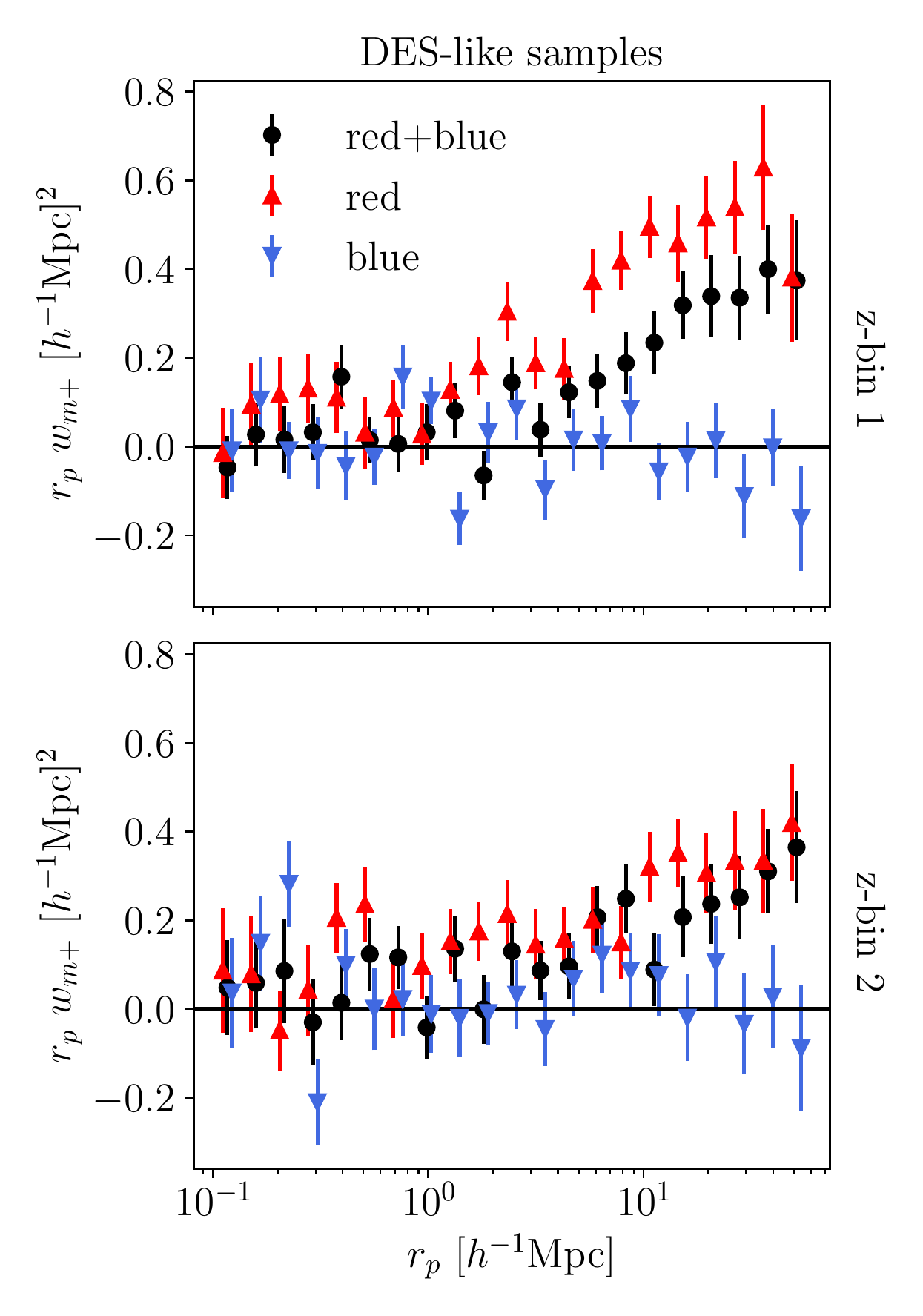}
   \caption{Projected matter-intrinsic shear correlation, measured in the two lowest
   redshift bins of the DES-like sample constructed from MICE. 
   Results are shown for the full sample as well as for sub-samples of red and blue galaxies.
   The amplitudes are comparable to those from our measurements in the dimmest volume
   limited samples (i.e. $r_p w_{m+} \lesssim 0.5 \ h^{-2} \textrm{Mpc}^2$, see
   Fig. \ref{fig:wmp_zmc_samples}), showing that the IA contamination in the DES-like
   samples is predicted to be weak. Symbols for the red and blue sub-samples are
   slightly shifted along the $r_p$-axis for clarity.}
    \label{fig:wmp_des}
\end{figure}

In order to compare the IA signal in the DES-like samples with the signal in the volume limited samples
we measure the $w_{m+}$ statistics, focusing on the lowest two of the four photometric redshift bins,
which cover a similar redshift range. The measurements are shown in Fig. \ref{fig:wmp_des} for the full sample in each bin as well
as for the red and blue sub-samples which are defined by our $u-r=0.94$ color cut, as shown in Fig. \ref{fig:cmd_mice_des_samples}.
We find the strongest signal for the red sub-sample and no significant signal for the blue sub-sample, while the signal for the
full sample lies between those of the two sub-samples. The color dependence is expected from our modeling as well as from the $w_{m+}$ measurements in the volume limited samples.
The amplitudes of the full samples and the red sub-samples are comparable to those derived from red galaxies in our dimmest volume limited
samples (Fig. \ref{fig:wmp_zmc_samples}), showing that the IA contamination in our DES-like samples is relatively weak.
This finding is consistent with the large galaxy-halo misalignment angles for the DES-like samples, shown in Fig. \ref{fig:alpha_rand_3d}.

\subsection{Angular shear-shear correlation (\texorpdfstring{$\xi_{\pm}$}{xi+/-}) }

\begin{figure*}
	\includegraphics[width=\textwidth]{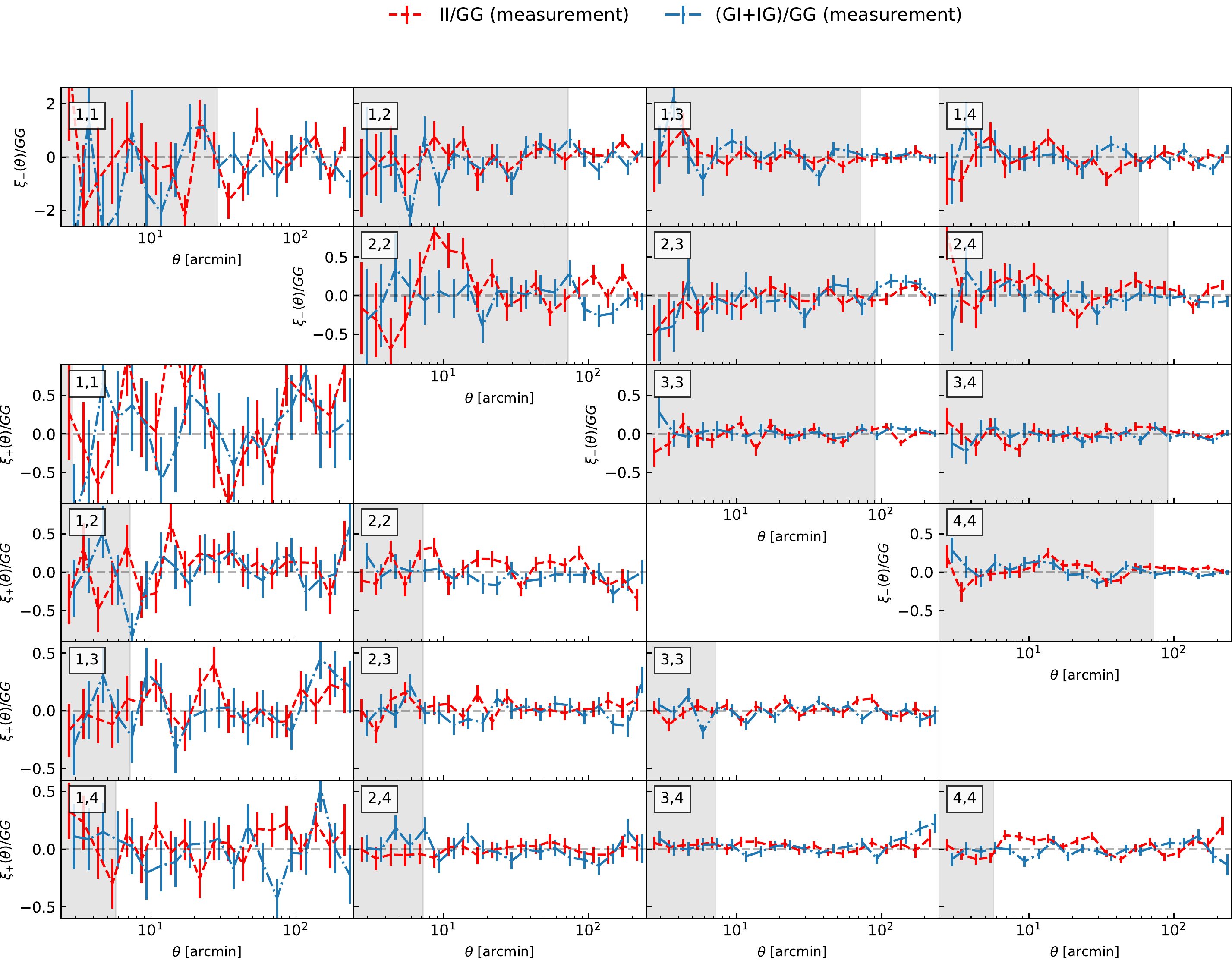}
   \caption{Intrinsic alignment correlation functions as measured from combinations of the individual $G$ and $I$ ellipticities in MICE, divided by the theory GG signal computed at the MICE cosmology and with the mock redshift distributions. Different panels show cross-correlations of redshift bins in $\xi_+$ and $\xi_-$, and shaded bands correspond to the scales that are removed from the inference of cosmological and nuisance parameters. Error bars indicate the estimated $1 \sigma$ shape noise uncertainties.
   }
    \label{fig:xipm}
\end{figure*}

We focus now on the angular shear-shear correlation $\xi_{\pm}$
which was introduced in Section \ref{sec:2pt_shear_statistics} and used in the cosmological weak lensing analysis of the
 DES Y3 data release.
In Figure \ref{fig:xipm} we show the $\xi_{\pm}$ measurements of the IA terms that contribute to the cosmic shear signal, GI+IG and II, divided by the theoretical GG signal computed at the cosmology and redshift distributions of the MICE mock. Each panel corresponds to a different cross-correlation between
redshift bins, and the shaded bands correspond to angular scales that are removed from the cosmic shear cosmology inference, mainly due to the effect
of baryonic feedback (see \cite{Amon21}, \citet*{Secco21} for a justification).

We find that the amplitudes of the IA signals compared to GG is small and consistent overall with shape noise fluctuations.
This finding indicates that the alignment signal, which we found to be weak but significantly above the
noise level in the $w_{m+}$ measurements for the DES-like samples in Fig. \ref{fig:wmp_des}, falls below
the noise level when probing it in the same samples with angular statistics. We attribute this decrease in signal-to-noise to the projection over large line-of-sight
distances that is implied in the definition of angular correlations.

\subsection{Results of Likelihood Analysis}\label{sec: results of likelihood analysis}

In order to produce a likelihood analysis with the measured data vector, we construct an analytic Gaussian covariance using \textsc{CosmoCov}\footnote{\url{https://github.com/CosmoLike/CosmoCov}} \citep{Fang_2020} for the present DES-like MICE sample, matching the statistical power of DES Y3
\citep*[see the description in Appendix A of][]{Secco21}. The theory prediction in the likelihood evaluation is summarized in Equation \ref{eq: xipm Hankel transform} (with $C(\ell)$ spectra containing IA contributions), and we ultimately infer the posteriors on cosmological and nuisance parameters of the $\xi_\pm(\theta)$ model.

For this inference, we sample the parameter space utilizing \textsc{Polychord} \citep{Handley2015}, with similar performance settings as those used in the DES Y3 analyses. Our flat $\Lambda$CDM  cosmology model has 6 free parameters: the matter density $\Omega_\textrm{m}$, the baryon density $\Omega_\textrm{b}$, the primordial amplitude $A_\textrm{s}$, the spectral index $n_\textrm{s}$, the Hubble parameter $h$ and the neutrino density $\Omega_\nu h^2$. We additionally vary 8 calibration parameters (a shear multiplicative bias $m_i$ and a photo-$z$ shift parameter $\Delta z_i$ for each redshift bin $i$). Freeing these nuisance parameters is not strictly required in our present analysis, which does not include shear and photo-$z$ systematics, but are important to guarantee a roughly similar figure-of-merit between this exercise and the actual DES Y3 analysis. In addition, we also vary 2 IA parameters in the NLA model ($A_1$ and its redshfit evolution power-law $\eta_1$) and 5 parameters for the TATT model (both NLA parameters plus a galaxy bias parameter $b_{\mathrm{TA}}$ and the amplitude and redshift evolution of torquing terms, $A_2$ and $\eta_2$). We note that this choice follows closely the DES Y3 analysis: we explicitly parameterize the impact of the linear bias of the source galaxies contributing to the tidal signal as  $A_{1\delta}=b_{\mathrm{TA}}A_1$, allowing for extra freedom in the model.  This "fiducial" setting thus includes 19 (22) free parameters. The explicit priors on these cosmological and nuisance parameters are generally uninformative and can be found in e.g. \citet*{Secco21}, Table I.

In addition to this fiducial setting, we also run inferences with cosmological parameters fixed at the known MICE truth values and shear/redshift calibration parameters fixed at zero (thus freeing only the IA model parameters), and a "baseline" analysis in which no IA signals are added, in which case we expect IA constraints to be compatible with zero.      
In the baseline scenario, we find IA posteriors to be consistent with zero, in line with our expectations: 
\begin{eqnarray*}
A_{1} & =-0.32_{-0.25}^{+0.57}\qquad\textrm{(TATT baseline)}\\
A_{2} & =-0.10_{-0.59}^{+0.74}\qquad\textrm{(TATT baseline)}
\end{eqnarray*}
where the central values are the maxima of the marginalized posterior distributions and the
upper and lower values are the distances to the bounds of the corresponding
$68\%$ confidence intervals.
The main IA constraints as probed by $\xi_\pm$ with this DES-like source sample come from the simplified, fixed-cosmology scenario. In this reduced parameter space we find, when fitting NLA:
\begin{eqnarray*}
 \,\,\,\,\,A_{1}\,\,\,\,\,=-0.11_{-0.27}^{+0.21}\qquad\,\,\textrm{(NLA with fixed cosmology)}
\end{eqnarray*}
and similarly we find, for the TATT model:
\begin{eqnarray*}
A_{1} & =-0.30_{-0.51}^{+0.49}\qquad\textrm{(TATT with fixed cosmology)}\\
A_{2} & =0.67_{-0.41}^{+1.23}\qquad\,\,\,\,\textrm{(TATT with fixed cosmology)}
\end{eqnarray*}
In all of the cases above, we find $A_1$ parameters that are consistent with zero within one standard deviation, and only find a marginal preference for positive $A_2$ in the case of TATT at a fixed cosmology. These results are not unexpected given the overall small impact on the DES-like source sample as a fraction of the GG signal, as seen in the correlation functions shown in Fig. \ref{fig:xipm} .

Finally, for the cases where we additionally vary cosmological and nuisance parameters, when fitting the data vector with the NLA model we find:
\begin{eqnarray*}
\quad\,\,\,\,\, A_{1} & =-0.51_{-0.27}^{+0.62}\quad(\textrm{NLA with free cosmology \& nuis.})
\end{eqnarray*}
and similarly when employing the TATT model:
\begin{eqnarray*}
A_{1} & =-0.55_{-0.33}^{+0.71}\quad(\textrm{TATT with free cosmology \& nuis.})\\
\qquad A_{2} & =-0.37_{-1.65}^{+1.68}\,\quad(\textrm{TATT with free cosmology \& nuis.}).
\end{eqnarray*}
These results are also in line with our expectations based on Fig. \ref{fig:xipm}: the overall IA amplitudes are consistent with zero within a standard deviation, and the constraining power on the individual 1D parameters is suppressed with respect to the fixed cosmology analogs since there is a greater number of parameters in the likelihood analysis, weakening those constraints. Apart from IA, the cosmological parameters of main importance for this type of analysis are the amplitude $S_8$ and the matter density $\Omega_\mathrm{m}$, which are presented for this simulated MICE sample in \citet*[][Appendix A]{Secco21} and found to be unbiased with respect to the MICE input cosmology.

In conclusion, utilizing a cosmic shear measurement obtained from the DES-like simulated source sample, we find that IA amplitudes are subdominant as a contributor to the tomographic $\xi_\pm(\theta)$ data vector and that, accordingly, posterior distributions on IA parameters for both TATT and NLA models are largely consistent with zero. In tandem, the analysis presented in \cite*{Secco21} also shows, using the same simulated mock catalog, that inferring cosmological parameters with both IA models recovers the input MICE simulation cosmology without biases. 

We emphasize that these results are specific to our simulation, which was calibrated to match the IA signal of LRGs at relatively low redshifts,
while assuming effectively no alignment for blue galaxies, which constitute a significant fraction of the DES-like samples
(see Section \ref{sec:data:des_samples}).

\section{Summary and conclusions}
\label{sec:conclusions}

We implemented intrinsic galaxy alignment (IA) in the light-cone output of the cosmological simulation MICE to study it as a contamination in measurements of two-point correlation functions from weak
lensing observations of the cosmic large-scale structure. The simulation was thereby used for two purposes:
a) to investigate the accuracy of analytical models that describe the IA contamination
at luminosities and redshifts for which observational constraints from spectrospcopic
surveys are currently not available and b) to predict the IA contamination in the weak lensing
observations of the Dark Energy Survey (DES). We thereby take advantage of the fact that
MICE provides both, the intrinsic as well as the gravitational shear components.
For the IA implementation we use a semi-analytic model to assign a shape and an orientation to each
galaxy of the HOD-SHAM catalog of the MICE simulation, taking into account the galaxy's brightness and
color as well as the orientation and angular momentum of its host halo. Our model is inspired
by semi-analytic IA models presented previously in the literature \citep[e.g.][]{Joachimi13a,Joachimi13b},
but includes substantial advancements in three aspects.

\begin{enumerate}
    \item We developed a new method for assigning 3D galaxy shapes, assuming a simple ellipsoidal morphology
    for each object. The parameters of this shape model were calibrated such that the distribution of
    projected 2D axis ratios matches observational constraints from the COSMOS survey for different ranges
    of galaxy color, absolute magnitude and redshift (Section \ref{sec:model_shapes}).
    \item The misalignment between the orientations of galaxies and those of their host halos was calibrated such that
    the projected galaxy-intrinsic shear correlation ($w_{g+}$), measured in a
    mock BOSS LOWZ sample of LRGs from MICE, matches
    direct IA measurements in the corresponding observations from \citet{Singh16}
    in four different magnitude
    bins over a large range of scales ($0.1<r_p<200 \ h^{-1}\textrm{Mpc}$, see Section \ref{sec:model_orientations}).
    We found that the galaxy-halo misalignment for LRGs in MICE is consistent with constraints derived
    by \citet{Okumura09b}.
    \item The MICE light-cone covers one octant of the sky ($\sim 5000 \ deg^2$) and reaches up to redshift $z=1.4$.
    The simulated IA catalog is therefore the largest presented in the literature so far, which allows us
    to construct realistic mock catalogs of current weak lensing surveys and measure the IA signal with high significance.
\end{enumerate}

In our investigation of the accuracy of analytical IA models we focus on the NLA model and the TATT model.
We assess the models' accuracy by comparing their predictions for the projected matter-intrinsic shear correlation
($w_{m+}$) against corresponding measurements in MICE (Section \ref{sec:sim_vs_theo}). The latter are derived
for a set of volume limited samples of red and blue galaxies that span over the redshift range $0.1<z<0.7$
and probe absolute magnitudes down to $M_r=-20$.
In contrast to observations, the simulation allows us to access the matter field directly,
which significantly reduces the impact of galaxy bias on the IA statistics. 
As discussed in Section \ref{sec:2pt_shear_statistics}, we can therefore study the accuracy of IA modeling with less sensitivity to the details of nonlinear galaxy bias than when using the observable $w_{g+}$.

Our $w_{m+}$ measurements in MICE show strong dependencies on galaxy color, magnitude and redshift,
which allow to test the analytical models in a wide
range of possible alignment scenarios (Fig. \ref{fig:wmp_zmc_samples}). We find that the NLA and the TATT model
fit the $w_{m+}$ measurements with similar accuracy when restricting the fit to
scales larger than $8 h^{-1} \textrm{Mpc}$ as deviations from the measurements are consistent with the
$\sim 1 \sigma$ error estimates. When including smaller scales the NLA model breaks down,
while the TATT model retains a $\sim 2 \sigma$ accuracy down to the smallest scale considered of
$1 h^{-1} \textrm{Mpc}$ (Fig. \ref{fig:delta_wmp_zmc_samples}). It is important to keep here in mind
that the IA signal predicted in MICE is based on assumptions employed in the HOD and  semi-analytic IA modeling,
which might be too simplistic. However, the fact that the $w_{g+}$ signal in MICE matches the BOSS
observations, even in the 1-halo regime below $1 h^{-1} \textrm{Mpc}$, is an indication that these simplistic 
assumptions provide reasonably effective descriptions of the true galaxy alignment.

As an additional validation of the simulation we compare the constraints on the NLA and TATT parameter
$A_1$, which is sensitive to the IA signal at large scales, 
to constraints from the literature that were derived from various observed samples of red galaxies
to which the MICE simulation has not been calibrated (Fig. \ref{fig:loglum_a1a2a3}). We find that the
$A_1$ constraints from MICE are in broad agreement with the observations, given the large error bars and
taking into account that the selection of the volume limited samples in MICE differs significantly from the
selection of the observed samples. At low redshifts ($z \leq 0.3$) the luminosity dependence of the $A_1$ parameters
in MICE is consistent with a single power law, which was derived from fits to observational $A_1$ constraints
for red galaxies by \cite{fortuna21b}, while the broken power law proposed by these authors
shows clear deviations from our results.
At higher redshifts ($z > 0.3$) the luminosity dependence of $A_1$ for red galaxies in MICE decreases,
which is mainly driven by a decrease of the alignment amplitude for LRGs.
As for low redshifts, this luminosity dependence seems to be better described by a single, rather than a broken power law.
Verifying the high redshift predictions from MICE will be possible with IA measurements in upcoming spectroscopic surveys, such as DESI or PAU.

The alignment parameters for samples of blue galaxies in the simulation are consistent with zero (Fig. \ref{fig:posterior_tatt}).
This result is expected since central blue galaxies are oriented in our model
with the host halos' angular momentum, for which we find only a weak alignment
signal in Appendix \ref{app:halo_alignment}. Furthermore, we highly randomize
the orientations of these objects, to ensure that the simulation reproduces the null
detection of IA for blue galaxies in current observations.

As a last step in our analysis we investigate the contribution of IA to the angular shear correlation
$\xi_{\pm}$ in mock samples of the DES survey, taking advantage of the fact that the simulation allows us
to measure the GG, II and GI term separately from each other.
Interestingly, the II and the GI terms predicted for the DES samples by MICE are consistent with zero (Fig. \ref{fig:xipm}).
We have validated that the $w_{m+}$ measurements for the same samples show a signal that is significant
with respect to the errors. However, the amplitude of this signal is relatively low compared to
our measurements for red galaxies, which can be expected from the $\sim 50\%$ fraction of blue
galaxies that the simulation predicts for the DES-like samples (Table \ref{tab:fblue_des}).
A possible reason for the non-detection of the II and GI terms in MICE could be that the angular
shear statistics {\bf $\xi_{\pm}$} is less sensitive to IA contaminations than the projected statistics $w_{m+}$,
as it probes the integrated IA signal over a wide range of redshifts, including galaxy pairs with large physical separations
that are only weakly intrinsically aligned. The resulting decrease in the signal-to-noise ratio
could lead to a null detection for the IA terms in the angular statistics.
\newline

A weak IA signal in angular statistics lines up with findings of \citet{Wei18}, who study IA
contaminations in mock observations of the
weak lensing surveys KiDS and DLS using a semi-analytic model to implement IA in the Elucid simulation.
In contrast to our results these authors find a weak but still significant IA contribution to the GG signal. One potential reason for that difference may be the use of constraints on galaxy-halo misalignment angles
that were derived from LRGs for the entire KiDS sample in that work. This could result in an overly high signal,
as LRGs show the strongest alignment signal compared to other galaxy populations
(see e.g. Fig. \ref{fig:loglum_a1a2a3}).

Another reason for the weak alignment in the DES-like samples
could be the fact that predictions for the alignment of dim galaxies (which constitute a significant
part of the sample) are presumably affected
by the relatively low mass resolution, causing noise in the measured host halo orientations and hence
decreasing the predicted galaxy alignment (see Appendix \ref{app:halo_alignment}).
However, a weak contribution of IA in DES-like samples lines up with recent findings from \citet*{Secco21}, who
derive constraints on IA model parameters from the cosmological analysis of the cosmological weak lensing signal
and find those to be consistent with zero.

We stress here that the IA predictions from the MICE simulation for DES-like samples need to be taken with caution
since the simulation has only been tested directly against IA observations from red galaxies at redshift that
are well below those probed by DES. Furthermore, the DES samples contain a high fraction of blue
galaxies, in particular at high redshifts (Table \ref{tab:fblue_des}) for which the simulation predicts no IA
signal by construction. This lack of alignment has been observed for blue galaxies at low redshifts.
However, the alignment of blue galaxies at high redshifts remains unconstrained by observations.

Future improvements of the simulation could therefore consist in taking into account IA observations
at higher redshifts from upcoming spectrosopic surveys.
An interesting extension of our work in that regard based on current observations would be to reduce the
misalignment of blue galaxies, such that the simulation reproduces the observed null detection
for this type of galaxies (e.g. from direct measurements in a SDSS sample
\citep{Johnston19} or indirect measurements in DES Y1 data \citep{Samuroff19}) within the corresponding errors. 
Such a decrease of misalignment may lead to more significant IA contributions to the
predicted DES Y3 lensing signal.

An additional improvement of our modeling could consist in a more realistic selection of discs and ellipticals,
for instance using two different color indices as discussed in
Section \ref{sec:color_cuts}. In our current implementation all galaxies defined
as red by a single color index cut are treated as ellipticals and are therefore aligned with their
host halos principle axes. A more sophisticated color cut could identify a fraction of red objects
as discs and align them with their host halo's angular momentum vector, which might change
the predictions of our simulation. One shortcoming which is harder to address
is that a significant fraction of galaxies have an irregular morphology, which is currently not taken
into account in the modeling. Hydrodynamic simulations may help to find an effective model that describes the
intrinsic alignment of such galaxies.
More realism could further be added to the simulation by introducing a dependence of satellite alignment on the distance to the host halo center, which has been found in observations
\citep{Huang18, Georgiou19} as well as in hydrodynamic simulations \citep{Knebe20}.

There are several potential applications of our IA simulation not explored here. For instance, studies of IA in third-order lensing correlations \citep{Schmitz2018, Pyne21}, in particular because such statistics have been detected at high signal-to-noise in recent DES Y3 data \citep{Secco3pt}. Another application would be to study priors on the IA modeling, in particular on TATT parameters that are poorly constrained otherwise.

We expect that the unique size and depth of the MICE IA simulation presented here, and the improvements in the IA assignment model, will play a central role in constraining our understanding of the IA contribution in ongoing and future weak lensing observations.

\section*{Availability of data and software}

The MICE IA simulation presented in this work as well as the
halo catalog used in the modeling are publicly available
at CosmoHub\footnote{https://cosmohub.pic.es}
\citep{Carretero17,Tallada20}. A public version of our IA simulation code
is available on GitHub \footnote{https://github.com/kaidhoffmann/genIAL}.

\begin{acknowledgments}
We thank Christopher Bonnett for the initial idea to implement IA in MICE.
We thank Sukhdeep Singh and Rachel Mandelbaum for providing $w_{g+}$
measurements from the BOSS LOWZ sample.
We thank David Navarro for useful comments on the draft.
KH acknowledges support by the Swiss National Science Foundation (Grant No. 173716, 198674),
and from the Forschungskredit Grant of the University of Zurich (Projekt K-76106-01-01).
JB and SS are partially supported by NSF grant AST-2206563.
This work was also partly supported by the program "Unidad de Excelencia Mar\'ia de Maeztu CEX2020-001058-M".
CosmoHub is hosted by the Port d'Informaci\'o Cient\'ifica (PIC), maintained through a collaboration of Centro de Investigaciones Energ\'eticas, Medioambientales  y Tecnol\'ogicas (CIEMAT) and Institut de F\'isica d'Altes Energies (IFAE), with additional support from Universitat Aut\'onoma de Barcelona (UAB).
Funding for the DES Projects has been provided by the U.S. Department of Energy, the U.S. National Science Foundation,
the Ministry of Science and Education of Spain, the Science and Technology Facilities Council of the United Kingdom,
the Higher Education Funding Council for England, the National Center for Supercomputing.
Applications at the University of Illinois at Urbana-Champaign, the Kavli Institute of Cosmological Physics at
the University of Chicago, the Center for Cosmology and Astro-Particle Physics at the Ohio State University,
the Mitchell Institute for Fundamental Physics and Astronomy at Texas A\&M University, Financiadora de Estudos e Projetos,
Funda{\c c}{\~a}o Carlos Chagas Filho de Amparo {\`a} Pesquisa do Estado do Rio de Janeiro, Conselho Nacional de Desenvolvimento Cient{\'i}fico e Tecnol{\'o}gico and the Minist{\'e}rio da Ci{\^e}ncia, Tecnologia e Inova{\c c}{\~a}o, the Deutsche Forschungsgemeinschaft
and the Collaborating Institutions in the Dark Energy Survey.
The Collaborating Institutions are Argonne National Laboratory, the University of California at Santa Cruz, the University of Cambridge,
Centro de Investigaciones Energ{\'e}ticas, Medioambientales y Tecnol{\'o}gicas-Madrid, the University of Chicago, University College London,
the DES-Brazil Consortium, the University of Edinburgh, the Eidgen{\"o}ssische Technische Hochschule (ETH) Z{\"u}rich,
Fermi National Accelerator Laboratory, the University of Illinois at Urbana-Champaign, the Institut de Ci{\`e}ncies de l'Espai (IEEC/CSIC),
the Institut de F{\'i}sica d'Altes Energies, Lawrence Berkeley National Laboratory, the Ludwig-Maximilians Universit{\"a}t M{\"u}nchen
and the associated Excellence Cluster Universe, the University of Michigan, NSF's NOIRLab, the University of Nottingham,
The Ohio State University, the University of Pennsylvania, the University of Portsmouth, SLAC National Accelerator Laboratory,
Stanford University, the University of Sussex, Texas A\&M University, and the OzDES Membership Consortium.
Based in part on observations at Cerro Tololo Inter-American Observatory at NSF's NOIRLab (NOIRLab Prop. ID 2012B-0001; PI: J. Frieman),
which is managed by the Association of Universities for Research in Astronomy (AURA) under a cooperative agreement
with the National Science Foundation.
The DES data management system is supported by the National Science Foundation under Grant Numbers AST-1138766 and AST-1536171.
The DES participants from Spanish institutions are partially supported by MICINN under grants ESP2017-89838, PGC2018-094773,
PGC2018-102021, SEV-2016-0588, SEV-2016-0597, and MDM-2015-0509, some of which include ERDF funds from the European Union.
IFAE is partially funded by the CERCA program of the Generalitat de Catalunya.
Research leading to these results has received funding from the European Research
Council under the European Union's Seventh Framework Program (FP7/2007-2013) including ERC grant agreements 240672, 291329, and 306478.
We acknowledge support from the Brazilian Instituto Nacional de Ci\^encia e Tecnologia (INCT) do e-Universo (CNPq grant 465376/2014-2).
This manuscript has been authored by Fermi Research Alliance, LLC under Contract No. DE-AC02-07CH11359 with the U.S. Department of Energy,
Office of Science, Office of High Energy Physics.

\end{acknowledgments}

\appendix

\section{Dark matter halo alignment}
\label{app:halo_alignment}

\subsection{Effects of noise in halo orientations}

\begin{figure}
\centering
    \includegraphics[width=0.49\textwidth]{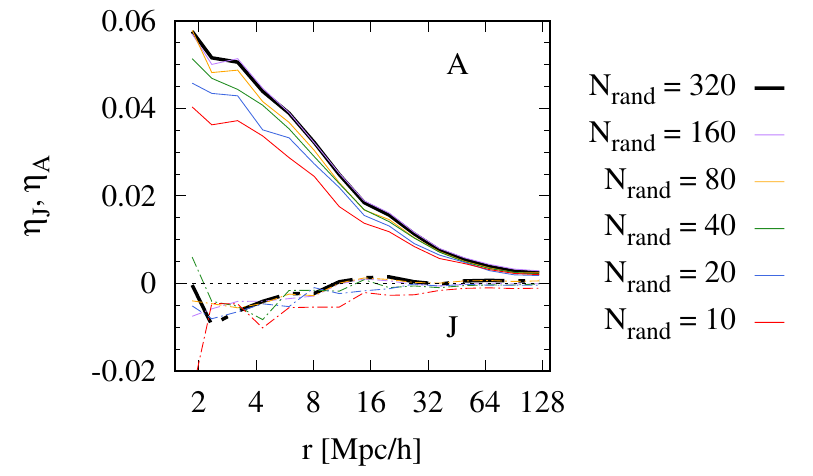}
    \caption{3D halo alignment statistics quantifying the
    average alignment of the major axis and angular
    momentum of dark matter halos in MICE
    ($\bf A$ and $\bf J$, solid and dashed lines respectively)
    and the vector $\bf r$ pointing to neighboring halos, as defined in 
    Equation \ref{eq:eta}. Results are shown for halos
    at $z=0.25$ with more than $N_p=320$ particles ($M_h \gtrsim 10^{12.97} h^{-1} \textrm{M}_{\odot}$).
    $N_{rand}$ indicates the number of random particles per halo from which $\bf A$ and $\bf J$
    were measured. $\eta=0$ corresponds to random alignment.}
    \label{fig:halo_alignment_restest}
\end{figure}

We investigate here the impact of noise resulting from low halo particle numbers
on halo alignment statistics in MICE. For that purpose we select
halos in the MICE light-cone with $N_p \gtrsim 320$ particles
($M_h \gtrsim 10^{12.97} h^{-1} \textrm{M}_{\odot}$),
then select randomly a subset of $N_{rand}$ particles from these halos without replacement and measure the halo major
axis and angular momentum vectors ({$\bf A$} and {$\bf J$} respectively) as detailed in
Section \ref{sec:halo_orientations}. Finally, we measure the 3D alignment statistics
of these halos with the large-scale structure as
%
\begin{equation}
\eta_X(r) = \langle |{\bf{\hat X}}_{1} \cdot {\bf \hat r}| \rangle (r) - 1/2,
\label{eq:eta}
\end{equation}
%
which is the inner product of the the unit vectors ${\bf{\hat X}}$ 
and ${\bf \hat r}$, where $\bf{X}$ refers to either {$\bf A$} or {$\bf J$},
$\bf{r}$ is a vector pointing to a neighboring halo and
$\langle \ldots \rangle$ denotes the average over all halo pairs
separated by the distance $r$.
Relative orientations of $\bf X$ and $\bf r$ that are random, parallel and perpendicular
to each other lead to $\eta_X=0$, $>0$ and $<0$ respectively.
In Fig. \ref{fig:halo_alignment_restest} we compare our measurements of $\eta_A$ and $\eta_J$ for
various values of $N_{rand}$. We find a positive amplitude of $\eta_A$,
which indicates that the major axes tend to point towards neighboring halos.
The amplitude starts to decrease significantly for $N_{rand} \lesssim 80$.
At $N_{rand}=10$ the amplitude is decreased by roughly $30$\%, while we still
find a clear signal.

The amplitude of $\eta_J$ is mostly negative, indicating that the
halos' angular momenta tend to be aligned perpendicular
to the vector pointing towards neighboring halos. We see no clear
dependence of $\eta_J$ on $N_{rand}$, as we saw it for $\eta_A$.
A potential explanation for hat finding could be that the dispersion
of the particles' angular momentum directions is low for the halos used in our
test, such that the average does not vary much across different random
subsets of particles.
Another interesting result is that the absolute values of
$\eta_J$ are significantly lower than those of $\eta_A$.
The lower amplitude of $\eta_J$ implies that blue galaxies,
which are aligned with $\bf J$ in our simulation, will
always show a significantly weaker IA signal than red galaxies,
which are aligned with $\bf A$, unless we set a much stronger
misalignment for red than for blue galaxies.

\subsection{Mass and redshifts dependence}

\begin{figure}
\centering
    \includegraphics[width=0.45\textwidth]{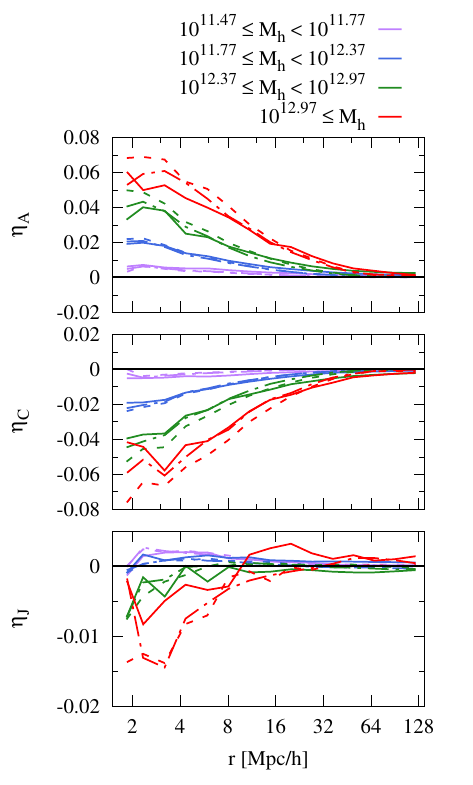}
    \caption{3D halo alignment statistics of the halo major axis, minor axis
    and angular momentum in the MICE simulation
    (top, central and bottom panels respectively). Results are shown for halos
    in the four mass ranges that are indicated on the top. The limits
    at $log_{10}(M_h) = (11.47, 11.77, 12.37, 12.97) \ [log_{10}([h^{-1}\textrm{M}_\odot)]$
    correspond to $N_p = (10,20,80,320)$ particles per halo. Solid, dashed-dotted and
    dashed lines show results at the redshifts $z=0.2$, $0.7$ and $1.2$ respectively.}
    \label{fig:halo_alignment}
\end{figure}

We consolidate this last point by comparing the amplitudes of $\eta_J$ with those
of $\eta_A$ as well as $\eta_C$ for halos in four mass samples
at three different redshifts of the MICE light-cone in Fig. \ref{fig:halo_alignment}.
For $\eta_A$ we find positive amplitudes for all mass and redshift samples.
The amplitudes are increasing with the halo mass and decrease with redshift. Overall the
redshift dependence is relatively weak compared to the mass dependence,
in particular for low halo masses. This finding explains why the large scale
amplitudes of $w_{g+}$ in Fig. \ref{fig:wgp} and the corresponding  $A_1$ parameter
in Fig. \ref{fig:loglum_a1a2a3} are decreasing more strongly with redshift
for luminous than for dim samples, since the luminous galaxies reside in more
massive halos than dim galaxies.
The fact that $\eta_A$ increases more strongly with halo mass than with redshifts
may be partially explained by the mass dependent noise on the halo orientations,
causing a weaker alignment for lower halo masses. However, the resolution effects
on $\eta_A$ that we show in Fig. \ref{fig:halo_alignment_restest} are weaker
than the mass dependence in Fig. \ref{fig:halo_alignment}.
The results for $\eta_C$ are similar to those $\eta_A$, while the amplitudes
are negative since $\bf{C} \perp \bf{A}$.

The results for $\eta_J$ line up with the results from Fig. \ref{fig:halo_alignment_restest}
as the amplitude is weaker than for $\eta_A$ and $\eta_C$ for all redshift and
mass samples. A similar finding has been reported by \citet{Forero14}, who compare the
alignment amplitude of halo shapes and angular momenta with the large-scale structure
in a cosmological simulation. This consolidates our expectation that blue galaxies
in our simulation would be weakly aligned compared to red galaxies,
even when setting a lower misalignment for blue galaxies than
used in our current model. It is further interesting to note that $\eta_J$
shows a more complex dependence on mass and scale than $\eta_A$ and $\eta_C$.
For massive objects we find the amplitude to switch from negative to positive
for increasing scales, indicating a change in the orientations of $\bf J$ from
perpendicular to parallel with respect to neighboring halo directions.
Another sign flip occurs at small scales ($r \lesssim 20 \ h^{-1} \textrm{Mpc} $)
between low and high mass samples. Such a flip has also been found in other
simulation based studies of halo spin alignment \citep[e.g.][]{Aragon14, Forero14, Lee20}.
However, a detailed discussion of these findings is beyond the scope of this work.

\section{Color, magnitude and redshift distributions}
\label{app:cmz_distributions}

\subsection{BOSS LOWZ}

\begin{figure}
    \includegraphics[width=0.45\textwidth]{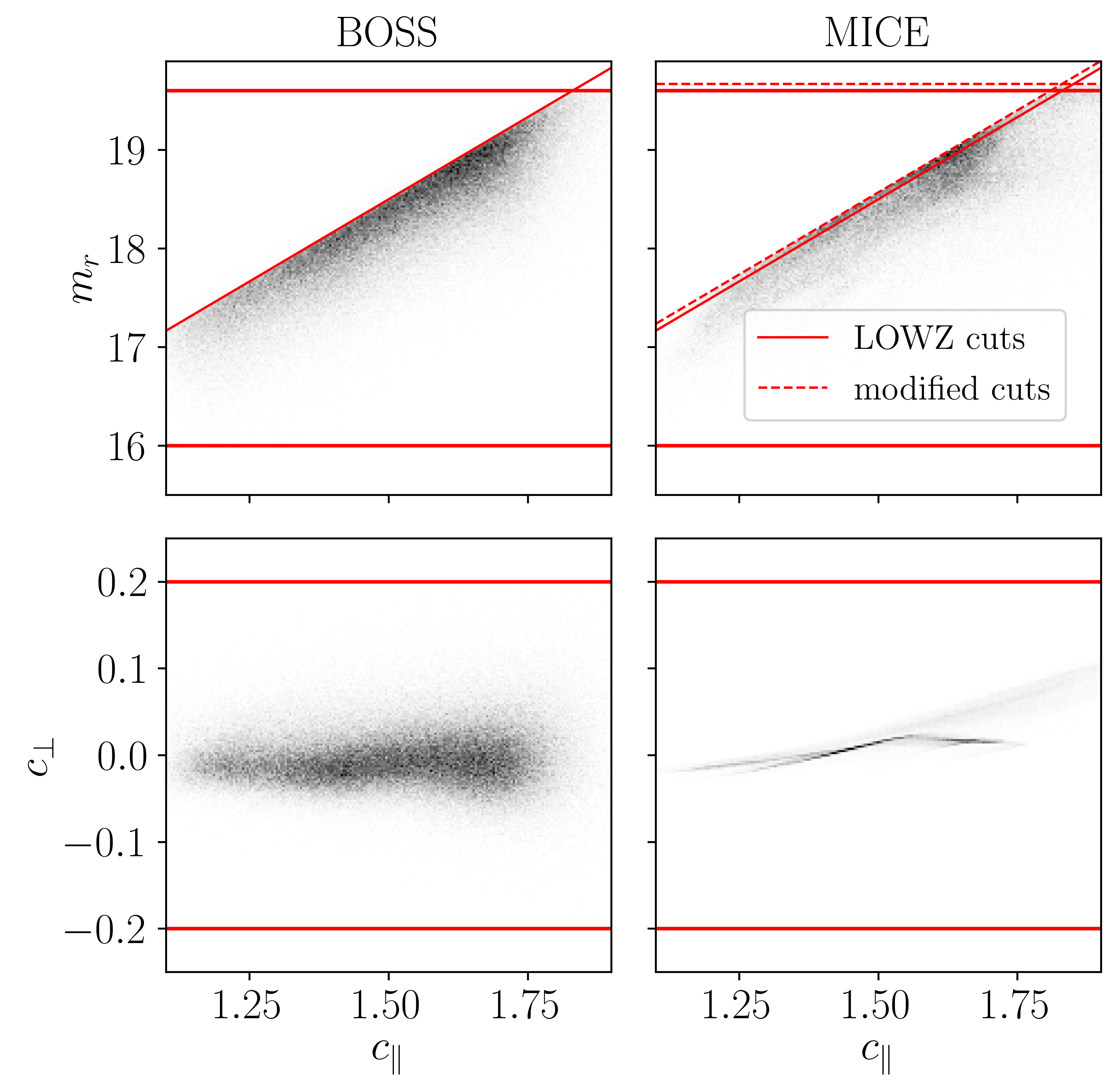}
    \caption{LOWZ selection cuts in BOSS and MICE (left right respectively)
    on the apparent SDSS $r$-band magnitude $m_r$ and the color cuts $c_\perp$ and $c_\parallel$,
    as defined in Equation (\ref{eq:lowzcuts1}).
    The $m_r$ cut for MICE is slightly shifted to dimmer magnitudes
    in order to match the BOSS LOWZ number density, as indicated by the
    red dotted line in the top right panel.}
    \label{fig:lowz_cuts}
\end{figure}

In Fig. \ref{fig:lowz_cuts} we compare the distribution of $m_r$, $c_\parallel$ and $c_\perp$\
(defined in Equation (\ref{eq:lowzcuts2}))
of galaxies in MICE to observations from the BOSS LOWZ Data Release $11$
\footnote{https://data.sdss.org/sas/dr11/boss/lss/} in the combined Southern and Northern galactic pole,
focusing on the SM16 redshift range. We find a reasonable agreement between the observed and simulated
distributions. However, when applying the LOWZ cuts on MICE with $\Delta m_r = 0$ we find a
$\sim 16 \%$ lower number density than in the observations. The latter is obtained from the number of LOWZ
Data Release 12 galaxies in the SM16 redshift range, $N_g = 249938$ and the corresponding effective area of
$A = 8,579 \ deg^2$, given in Table $2$ of \citet{Reid16}, which leads to
$n_g = N_g / A_{eff} = 29.97$. We therefore adjust the cuts on $m_r$ in MICE
$\Delta m_r = 0.085$, which results in a $\sim 0.2 \%$ agreement in the observed
number density, with $154617$ galaxies in the $5156.62 \ deg^2$ octant of the
MICE simulation.

Note that the LOWZ catalog used by SM16 is slightly
reduced in size and density compared to the original LOWZ sample
since only galaxies for which three different shape measurements
were available were used in that analysis. As a result dim
objects were excluded from their analysis and the average
brightness is increased to some degree. One can therefore expect
the clustering amplitude in the LOWZ sample to be slightly increased.
Including the exact same selection effects in the mock construction
is not feasible as the MICE simulation does not include the relevant
observational systematics. However, we do not expect these effects to
be critical, since the clustering amplitude in our mock is in good agreement
with the observations (Fig. \ref{fig:wgg}).

\begin{figure}
    \includegraphics[width=0.45\textwidth]{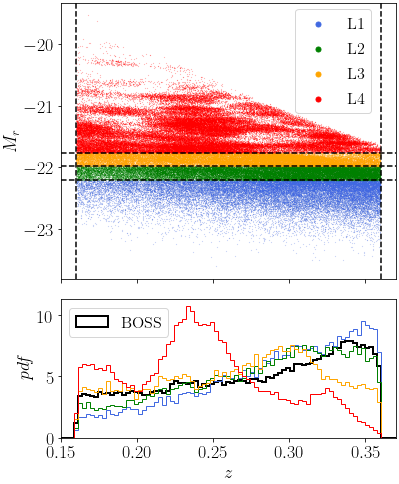}
    \caption{{\it Top}: Absolute SDSS r-band magnitude versus redshift
    of galaxies in the mock BOSS LOWZ catalog. The LOWZ sample is
    split into four luminosity sub-samples (L1-L4, from bright to faint)
    by quantiles, following SM16.
    The sample limits in redshift and magnitude are shown as dashed vertical
    and horizontal lines respectively. {\it Bottom}: Redshift distribution
    of the MICE luminosity samples in the same color coding as in the top panel.
    The black line shows the observed distribution of the entire LOWZ sample from BOSS DR11.
    }
    \label{fig:lowz_lum1234}
\end{figure}

The selection of luminosity sub-samples in the mock BOSS LOWZ sample from MICE
is illustrated in the top panel of Fig. \ref{fig:lowz_lum1234}, where we show the
joint magnitude-redshift distribution. The bottom panel of the same figure
compares the redshift distribution of the sub-samples in MICE LOWZ with the
observed redshift distribution from the BOSS DR11.
We find that the redshift distribution of the MICE sub-samples L1-L3 are 
consistent with the observations, while the dimmest sub-sample L4 
has a strong overabundance of galaxies around $z\sim0.25$.
We have validated that this overabundance is present at all angular positions. It might thus
result from an interplay of the HOD-SHAM methodology used for assigning magnitudes and colors to galaxies
and the LOWZ sample selection, leading to a preferred selection of objects at that redshift.
Further investigations are needed to fully understand this effect. However, we do not expect this
overabundance to significantly affect our measurements of the
projected  correlations $w_{gg}$ and $w_{g+}$ since it is isotropic and therefore
taken into account in the random catalogs used in the estimators (see Section \ref{sec:2pt_shear_statistics}).

\subsection{DES-like samples}

\begin{figure}
	\includegraphics[width=\columnwidth]{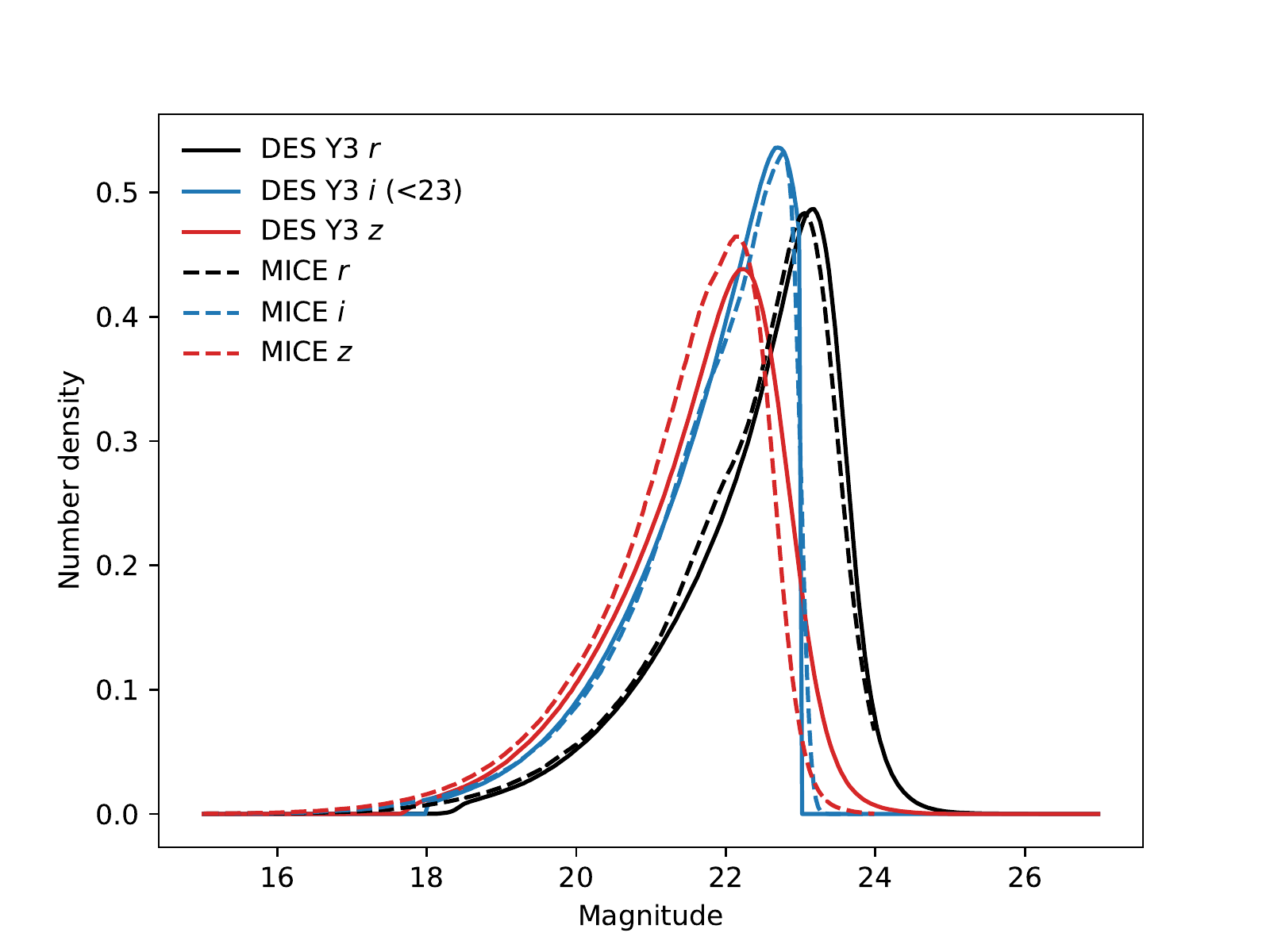}
   \caption{Comparison between magnitude distributions on the DES Y3 \metacal~ catalog and the MICE photometry in $riz$ bands. We introduce
   an extra apparent magnitude cut of $i<23$ to \metacal~ in this comparison, matching MICE specifications.}
    \label{fig:appmag_pdf_mice_des}
\end{figure}

We verify the selection of the DES-like catalog in MICE by comparing the distributions of apparent magnitudes in the
$r$, $i$ and $z$ DES broads from the DES Y3 data and the remapped MICE photometry (described in Section \ref{section:mice_intro})
in Fig. \ref{fig:appmag_pdf_mice_des}. We find a reasonable agreement between MICE and DES Y3, with MICE containing a slightly
longer tail toward bright objects. The DES Y3 distributions in that figure employ the fiducial quality cuts described
in \citet*{Gatti21}, plus an extra cut removing objects in \metacal\ with magnitude $i>23$ to enable a closer comparison
with the MICE photometry, which is limited at that magnitude\footnote{Note that the $i<23$ cut does not affect the
comparison to COSMOS data in Fig. \ref{fig:cz_cosmos_mice}, since this is based on MICE galaxies in an area, which is
complete down to $i<24$.}.

\begin{figure}
\centering
    \includegraphics[width=0.46\textwidth]{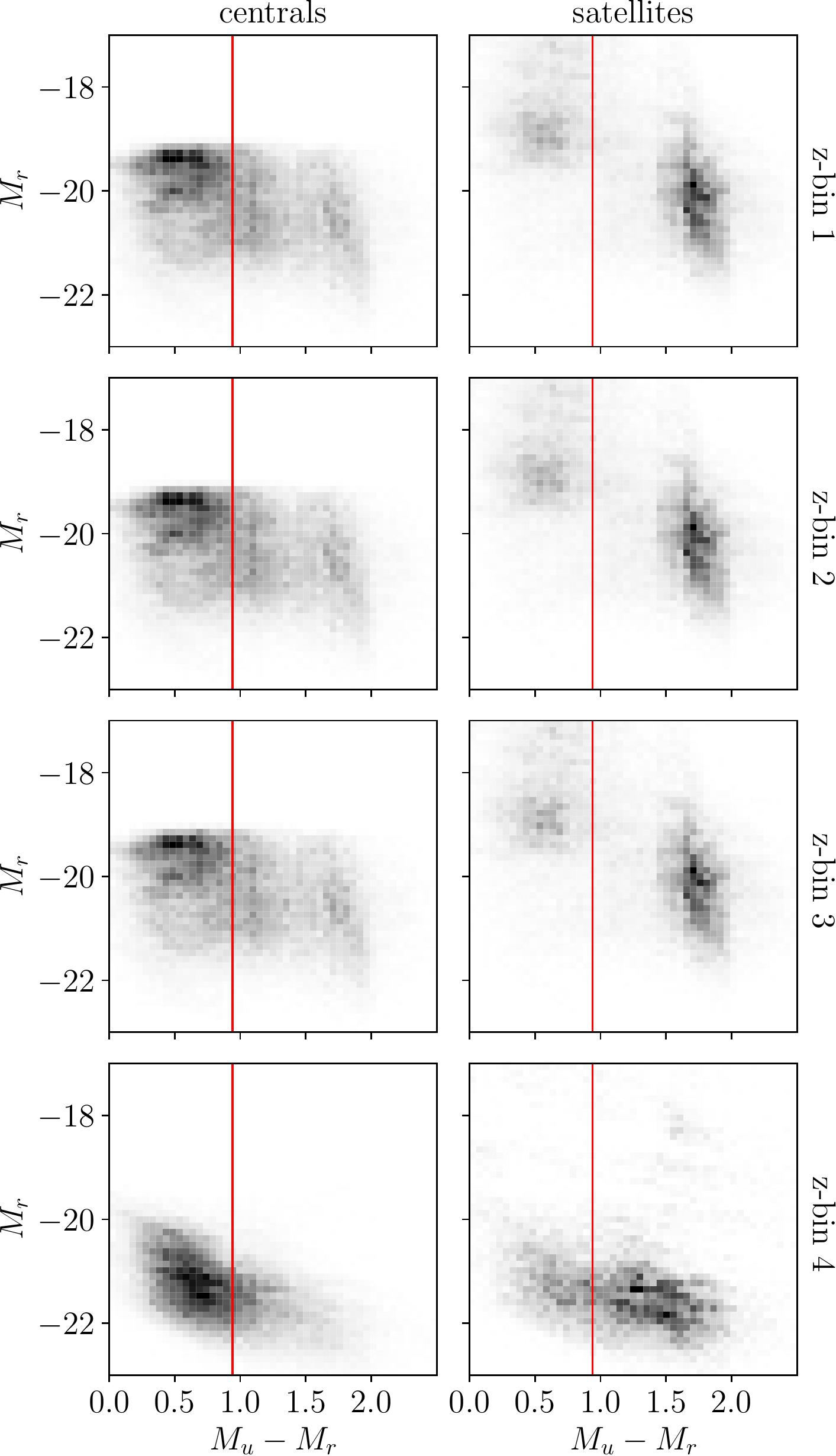}
    \caption{Normalized distribution of the $u-r$ restframe color index versus absolute r-band magnitude 
    in four redshift bins of the DES-like samples with i<23. Red lines indicate the color cut used
    to define red and blue galaxies in the IA modeling.}
    \label{fig:cmd_mice_des_samples}
\end{figure}

In Fig. \ref{fig:cmd_mice_des_samples} we display the $u-r$ color index (as defined in Section \ref{sec:color_cuts})
versus the absolute SDSS $r$-band magnitude for the DES-like redshift samples from MICE, showing how a 
significant fraction of central galaxies are defined as blue according to the color cut used in our modeling,
(see Table \ref{tab:fblue_des}). 
Fig. \ref{fig:cmd_mice_des_samples} further reveals a lack of central galaxies dimmer than $M_r\sim-19$.
This cut-off results from the fact that we require halos to have at least $10$ particles. This condition
imposes an absolute magnitude limit that affects only central galaxies as a consequence of the HOD-SHAM
model with which galaxies are generated in MICE \footnote{Note here that this lack of galaxies in low mass
halos only affects the MICE IA catalog, not the full simulated catalog, for which FoF halos down to two
particles were populated with galaxies.}. Due to the apparent magnitude limit in the DES-like samples of $i<23$,
this artefict affects mainly centrals in the two lowest redshift bins $1$ and $2$ of the DES-like samples.
The $i<23$ cut on the other hand affects a significant fraction of centrals as well as satellites in the two
highest redshift bins $3$ and $4$.

Both, the $M_r\sim-19$ and the $i=23$ cuts, can potentially boost the IA amplitude, as they remove centrals
in low mass halos as well as faint satellites.
Centrals in low mass halos should be weakly aligned due to a decrease in halo alignment with decreasing
mass \citep{Piras18}, and, moreover, the noise in measured orientations increases as the number of halo
particles decreases. Satellites with faint absolute magnitudes have more randomized orientations than
bright satellites, according to our semi-analytic IA model (Section \ref{sec:model_orientations}).
Removing such highly randomized objects from the sample could hence increase the amplitude of the IA statistics.
However, investigating the magnitude of such an increase would require a higher resolution simulation that
is complete below to $i=23$. In this work we therefore regard the IA signal predicted by MICE for the DES-like
samples as an upper bound for an IA signal that we would measure when including galaxies in halos with less
than $10$ particles and apparent magnitudes below $i=23$.

\subsection{Volume limited sample selection}

\begin{figure}
\centering
    \includegraphics[width=0.45\textwidth]{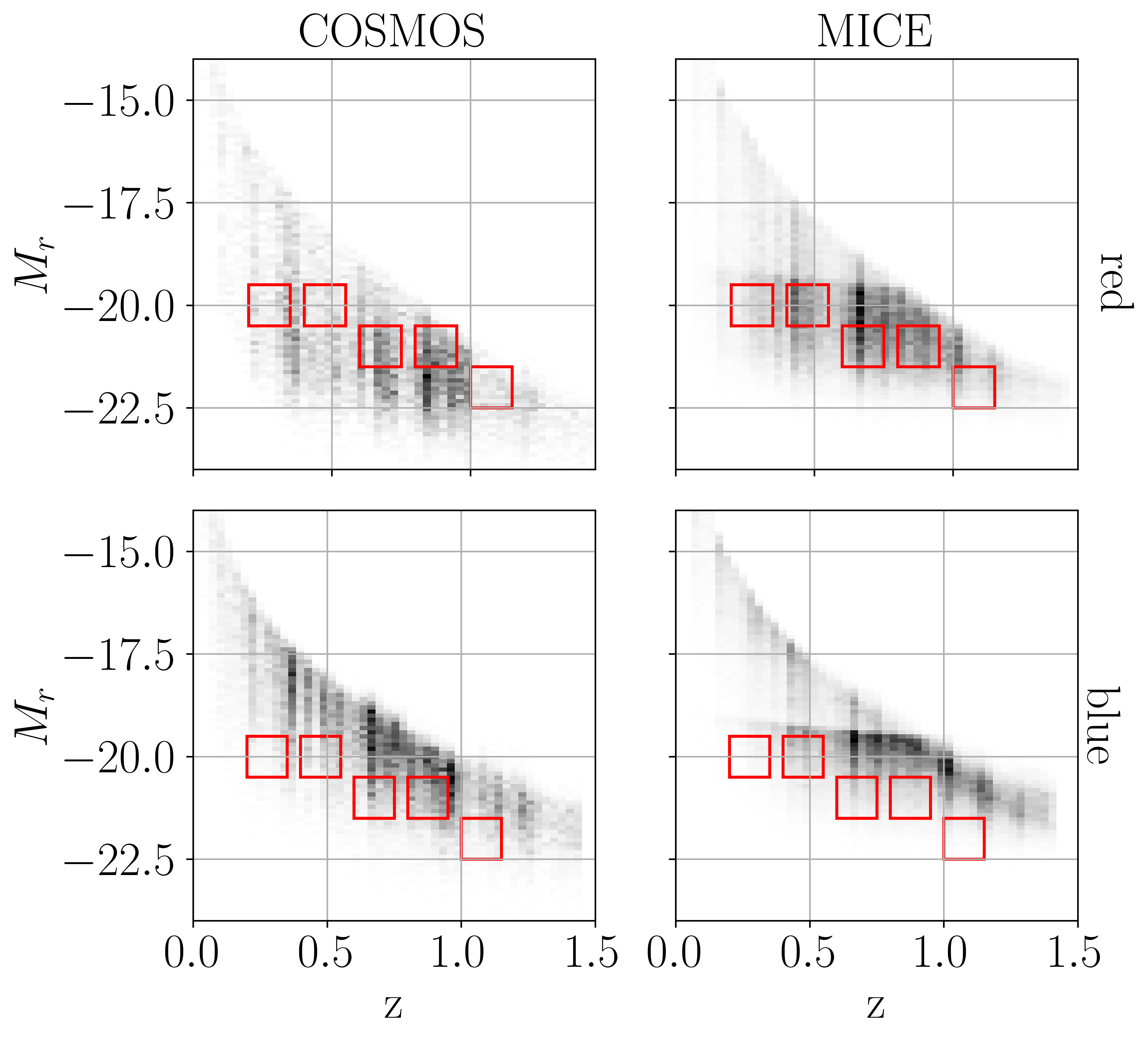}
    \caption{Normalized distribution of absolute Subaru $r$-band magnitudes in COSMOS
    and MICE for red and blue galaxies. The set of volume limited sub-samples
    used for comparing the axis ratio distributions in Fig. \ref{fig:q2d_cosmos_mice}
    are marked by red rectangles. The drop in the
    density for galaxies in MICE with $M_r>-20$ results from a lack of host halos
    with less than $10$ particles (see discussion Section \ref{sec:data:des_samples})}
    \label{fig:zmc_samples_cosmos_mice}
\end{figure}

\begin{figure}
\centering
    \includegraphics[width=0.48\textwidth]{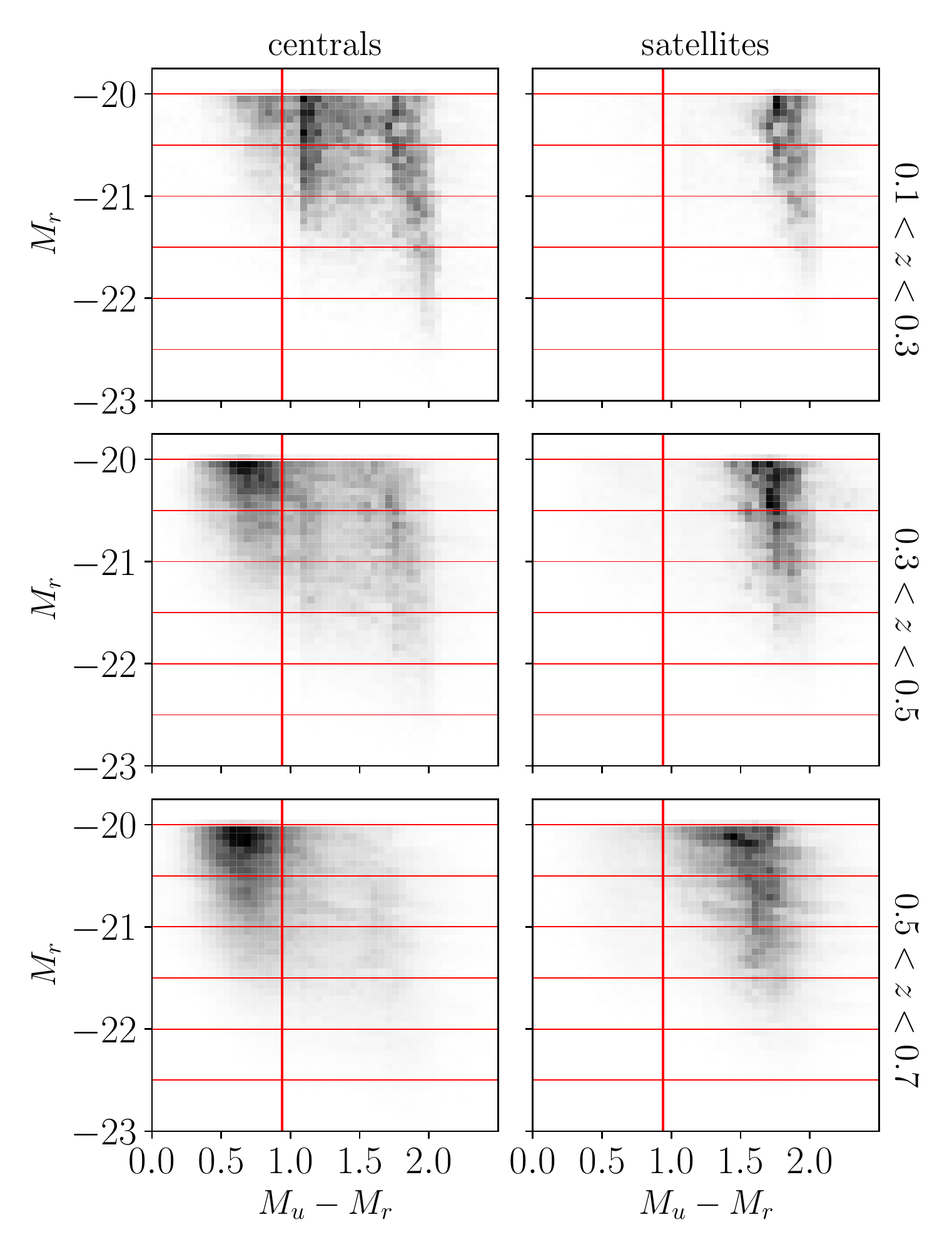}
    \caption{Normalized distribution of the $u-r$ restframe color index versus absolute r-band magnitude 
    in three redshift bins of the MICE simulation. Red dashed lines indicate the limits of the
    color-redshift samples used for the measuring predictions for $w_{m+}$
    (Fig. \ref{fig:wmp_zmc_samples}).}
    \label{fig:mice_zmc_samples}
\end{figure}

The selection of volume limited samples, used to
compare the galaxy axis ratio distributions in COSMOS and MICE
(Section \ref{sec:model_shapes}) is illustrated in
Fig. \ref{fig:zmc_samples_cosmos_mice},
separately for red and blue galaxies.
Fig. \ref{fig:mice_zmc_samples}
shows the selection of the volume limited samples that are
used to study the redshift, magnitude and color dependence of the IA statistics
in MICE (Section \ref{sec:wmp_mz_dependence}), separately for central
and satellite galaxies.

\section{Jackknife covariance}
\label{app:jk_covariance}

\begin{figure}
\centering
    \includegraphics[width=0.45\textwidth]{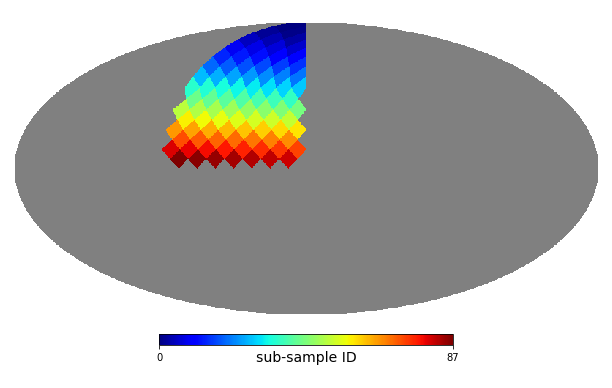}
    \caption{Healpix regions ($N_{side}=8$) in the MICE octant, used for the jackknife estimation of the covariance
    for $w_{g+}$ and $w_{m+}$.}
    \label{fig:jk_samples}
\end{figure}

\begin{figure}
\centering
    \includegraphics[width=0.45\textwidth]{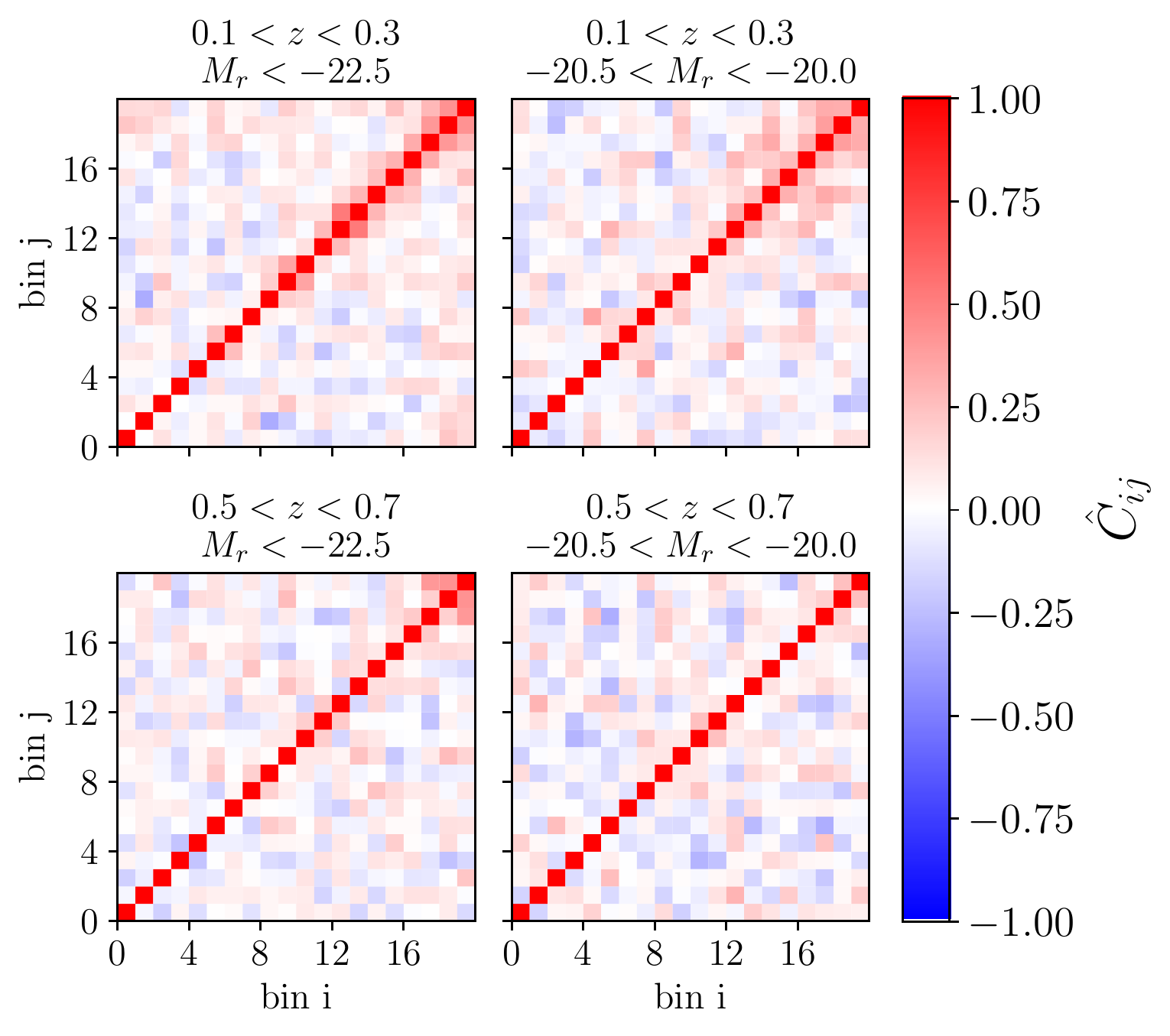}
    \caption{Jackknife estimates of the normalized covariance matrix $\hat{C}_{ij}\equiv C_{ij}/(\sigma_i\sigma_j)$
    for $w_{m+}$ measurements in the bins $i$ and $j$.
    Results are shown for red galaxies in four of our volume limited samples.}
    \label{fig:cov_wmp}
\end{figure}

Our delete-one jackknife estimates of the covariance matrix of the projected correlations,
described in Section \ref{sec:projcorr} are derived from angular sub-samples, that are defined
as healpix pixels with $N_{side}=8$. The area covered by the different sub-samples in the MICE
octant is shown in Fig. \ref{fig:jk_samples}. 
In Fig. \ref{fig:cov_wmp} we show examples of the normalized covariance estimated for
$w_{m+}$ measurements in $4$ of our $16$ volume limited samples. We find that all
covariances are dominated by the diagonal elements, while the off-diagonal elements
are noisy, which potentially affects the fits of the NLA and the TATT model
predictions for $w_{m+}$ to the measurements.

\section{Scale dependence of TATT parameters}
\label{app:tatt_params_scale}

\begin{figure*}
\centering
    \includegraphics[width=0.9\textwidth]{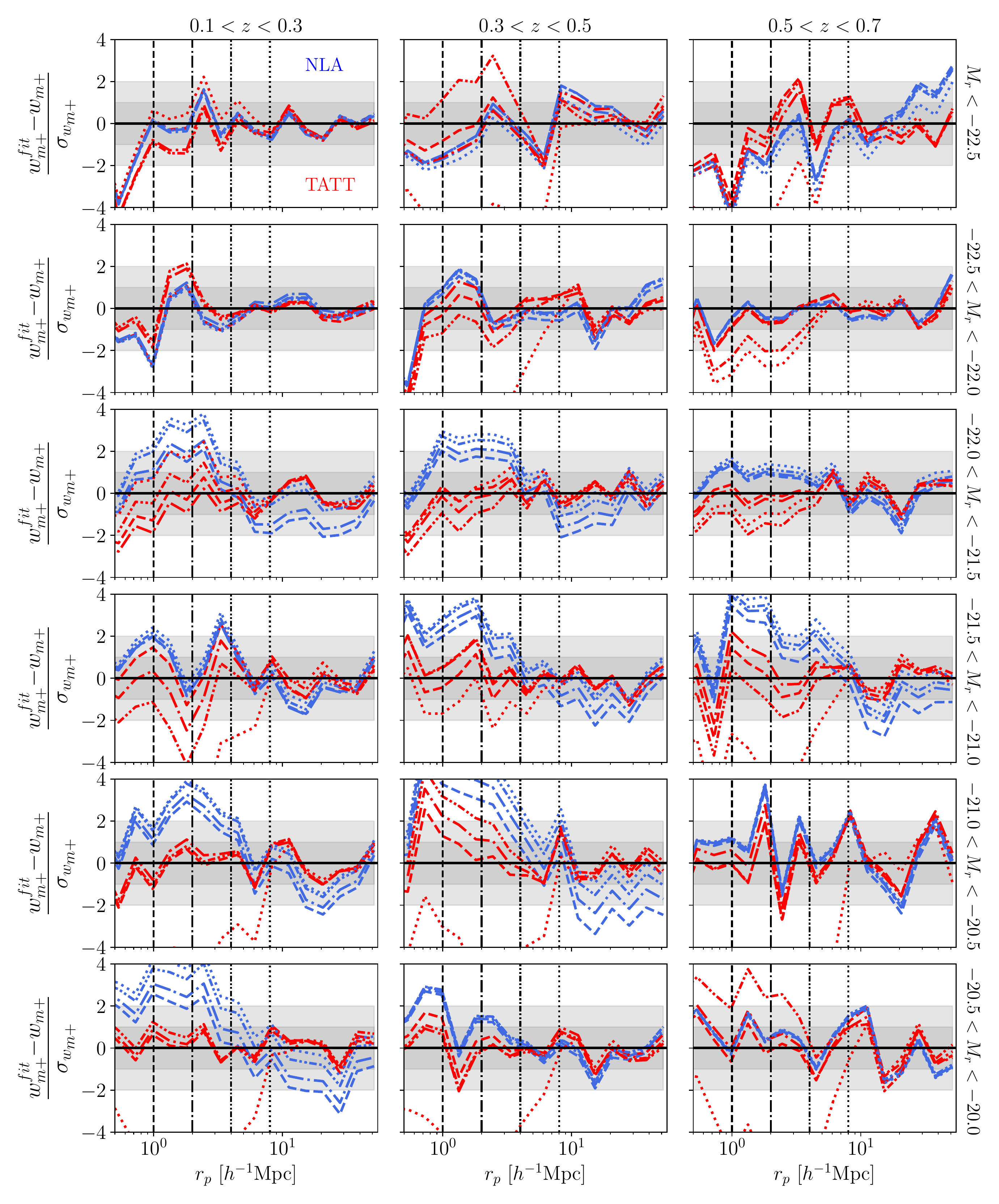}
    \caption{Significance of the deviations between the $w_{m+}$ measurements,
    shown in Fig. \ref{fig:wmp_zmc_samples} and fits to the NLA and the TATT model.
    Results are shown for $16$ volume limited samples whose redshift and magnitude ranges
    are indicated on the top and right respectively.
    Dashed, dashed-dotted, dashed-double dotted and dotted lines are results derived
    for different lower limits of the fitting range, which are set to $r_p^{min} = 1, 2, 4, 8  \ h^{-1} \textrm{Mpc}$
    respectively. These limits are marked by the vertical lines in the corresponding line types.
    The upper limit of the fitting range is set to $r_p^{max}=60\ h^{-1} \textrm{Mpc}$.
    }
    \label{fig:delta_wmp_zmc_samples}
\end{figure*}

We investigate here how well the NLA and the TATT model predictions for $w_{m+}$
fit the measurements over different ranges of the transverse distance scale $r_p$.
In addition we study how variations in the fitting range affect the inferred
model parameters. For that purpose we perform the Bayesian parameter inference,
described in Section \ref{sec:fitting}, for the lower limits of
$r_p^{min} \in [1, 2, 4, 8] \ h^{-1} \textrm{Mpc}$ and the upper limits of
$r_p^{max} \in [30, 60] \ h^{-1} \textrm{Mpc}$.
In Fig. \ref{fig:delta_wmp_zmc_samples} we show the significance of the deviations
between $w_{m+}(r_p)$ measurements and fits, defined as the absolute difference
over the standard deviation, for different $r_p^{min}$ and a fixed value of $r_p^{max}=60\ h^{-1} \textrm{Mpc}$.
Results are shown for the same $16$ volume limited sample of red galaxies as in Fig. \ref{fig:wmp_zmc_samples}.
Note that we do not investigate results for blue samples, since their alignment signal is
consistent with zero by construction.
The results in Fig. \ref{fig:delta_wmp_zmc_samples} show that the fits of the TATT model
do not deviate by more than $2 \sigma$ from the measurements within the scale range over which
the fit is performed. This is true even for $r_p^{min}=1 \ h^{-1} \textrm{Mpc}$,
which we therefore choose as lower scale cut for the fits shown in Fig. \ref{fig:wmp_zmc_samples}.
In addition we find that the fitting performance of the TATT model at large scales
($r_p>8 \ h^{-1} \textrm{Mpc}$) is only weakly affected by the small scale cut.
The results for the TATT model are contrasted by those for the NLA model. For this model we find
strong deviations of more than $4 \sigma$ when fitting between $1$ and $60 \ h^{-1} \textrm{Mpc}$.
The fitting performance at large scales is thereby strongly affected by the lower scale cut,
indicating a lack of flexibility in the model. However, when restricting the lower scale
cut to $r_p^{min}=8 \ h^{-1} \textrm{Mpc}$ we find the NLA model to fit the measurements
with a similar $2 \sigma$ uncertainty as the TATT model. This scale cut is therefore
chosen for the NLA fits shown in Fig. \ref{fig:wmp_zmc_samples}.
It is further interesting to note that there is no clear dependence of the fitting performance
on either magnitude or redshift and hence on the amplitude of $w_{m+}$.

\begin{figure*}
\centering
    \includegraphics[width=0.85\textwidth]{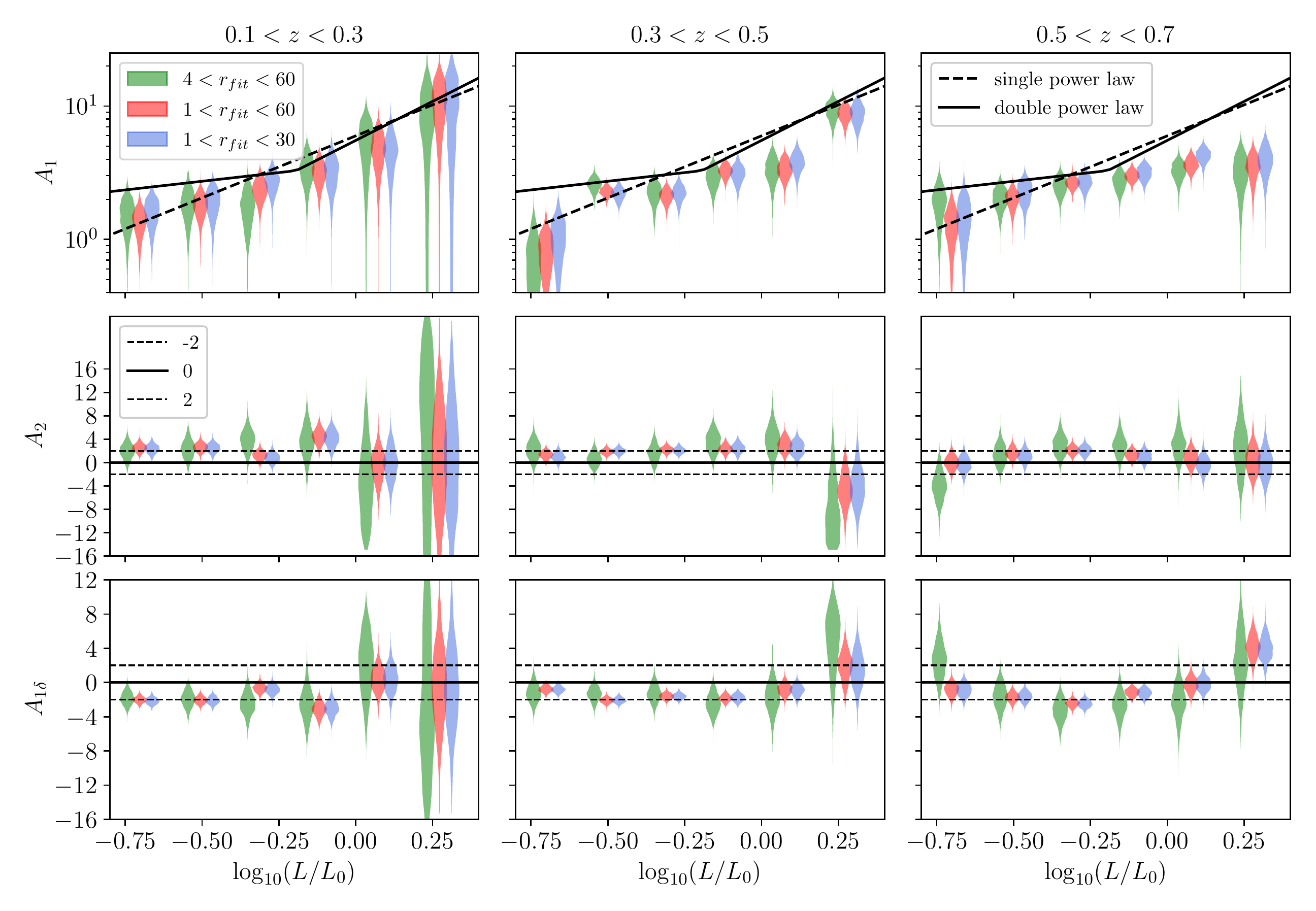}
    \caption{
    Marginalized posterior distributions of the TATT model parameters from fits to $w_{m+}$ measurements in 
    different volume limited samples of red galaxies in MICE versus each sample's
    logarithmic normalized mean $r$-band luminosity (analogous to Fig. \ref{fig:loglum_a1a2a3}).
    Results are shown for different scale ranges used for the fits, as indicated in the
    top left panel in units of $h^{-1} \textrm{Mpc}$.}
    \label{fig:loglum_a1a2a3_scale_cuts}
\end{figure*}

%
In Fig. \ref{fig:loglum_a1a2a3_scale_cuts} we show the parameters of the TATT model
versus the luminosity of the different volume limited samples in three redshift bins.
We find no significant change of the parameters when changing the upper (lower)
limit of the fiducial fitting range of $1 < r_p^{max}=60\ h^{-1} \textrm{Mpc}$
to $4$ ($30$) $h^{-1} \textrm{Mpc}$. This finding shows that the conclusions
drawn from Fig. \ref{fig:loglum_a1a2a3} are robust towards moderate
variations of the chosen fitting range.
%


\bibliographystyle{mnras_2author}
\bibliography{references}


\end{document}